%% file: svFitMEM.tex
\def\verPreprint{1}
\def\verPAPER{2}
\def\ver{1}

\ifx\ver\verPreprint
\documentclass[a4paper,english,11pt]{article}
\usepackage[bindingoffset=0.5cm,left=2.5cm,right=2.5cm,top=2.5cm,bottom=2.5cm,footskip=1.0cm]{geometry}
\usepackage{lineno,hepnames,bm,multirow,amssymb,authblk,graphicx,newclude,xspace,hyperref}
\fi
\ifx\ver\verPAPER
\documentclass[1p]{elsarticle}
\usepackage{lineno,hyperref,hepnames,bm,multirow,amssymb,xspace}
\fi

\modulolinenumbers[5]

\renewcommand{\Plepton}{\ensuremath{\ell}}
\newcommand{\Phadron}{\ensuremath{\textrm{h}}}
\newcommand{\tauh}{\ensuremath{\Pgt_{\textrm{h}}}\xspace}
\newcommand{\tauhnu}{\ensuremath{\tauh \, \Pnu_{\kern-0.10em \Pgt}}\xspace}
\newcommand{\tauhnuOne}{\ensuremath{\tauh^{(1)} \, \Pnu^{(1)}_{\kern-0.10em \Pgt}}\xspace}
\newcommand{\tauhnuTwo}{\ensuremath{\tauh^{(2)} \, \Pnu^{(2)}_{\kern-0.10em \Pgt}}\xspace}
\newcommand{\ellnunu}{\ensuremath{\Plepton \, \APnu_{\kern-0.10em \Plepton} \, \Pnu_{\kern-0.10em \Pgt}}\xspace}
\newcommand{\ellnunuOne}{\ensuremath{\Plepton^{(1)} \, \APnu^{(1)}_{\kern-0.10em \Plepton} \, \Pnu^{(1)}_{\kern-0.10em \Pgt}}\xspace}
\newcommand{\ellnunuTwo}{\ensuremath{\Plepton^{(2)} \, \APnu^{(2)}_{\kern-0.10em \Plepton} \, \Pnu^{(2)}_{\kern-0.10em \Pgt}}\xspace}
\newcommand{\ellMinusnunu}{\ensuremath{\Plepton^{-} \, \APnu_{\kern-0.10em \Plepton} \, \Pnu_{\kern-0.10em \Pgt}}\xspace}
\newcommand{\ellPlusnunu}{\ensuremath{\Plepton^{+} \, \Pnu_{\kern-0.10em \Plepton} \, \APnu_{\kern-0.10em \Pgt}}\xspace}
\newcommand{\enunu}{\ensuremath{\Pe \, \APnu_{\kern-0.10em \Pe} \, \Pnu_{\kern-0.10em \Pgt}}\xspace}
\newcommand{\mununu}{\ensuremath{\Pgm \, \APnu_{\kern-0.10em \Pgm} \, \Pnu_{\kern-0.10em \Pgt}}\xspace}

\newcommand{\phat}{\ensuremath{\bm{\hat{p}}}\xspace}
\newcommand{\pT}{\ensuremath{p_{\textrm{T}}}\xspace}
\newcommand{\pThat}{\ensuremath{\hat{p}_{\textrm{T}}}\xspace}

\newcommand{\thetahat}{\ensuremath{\hat{\theta}}\xspace}
\newcommand{\phihat}{\ensuremath{\hat{\phi}}\xspace}
\newcommand{\mhat}{\ensuremath{\hat{m}}\xspace}
\newcommand{\Ehat}{\ensuremath{\hat{E}}\xspace}
\newcommand{\ET}{\ensuremath{E_{\textrm{T}}}\xspace}

\newcommand{\pX}{\ensuremath{p_{\textrm{x}}}\xspace}
\newcommand{\pXhat}{\ensuremath{\hat{p}_{\textrm{x}}}\xspace}
\newcommand{\pY}{\ensuremath{p_{\textrm{y}}}\xspace}
\newcommand{\pYhat}{\ensuremath{\hat{p}_{\textrm{y}}}\xspace}
\newcommand{\pZ}{\ensuremath{p_{\textrm{z}}}\xspace}
\newcommand{\pZhat}{\ensuremath{\hat{p}_{\textrm{z}}}\xspace}

\newcommand{\MET}{\ensuremath{p_{\textrm{T}}^{\textrm{\kern0.10em miss}}}\xspace}
\newcommand{\vecMET}{\ensuremath{\bm{p}_{\textrm{T}}^{\textrm{\kern0.10em miss}}}\xspace}
\newcommand{\METx}{\ensuremath{p_{\textrm{x}}^{\textrm{\kern0.10em miss}}}\xspace}
\newcommand{\METy}{\ensuremath{p_{\textrm{y}}^{\textrm{\kern0.10em miss}}}\xspace}
\newcommand{\MeV}{\ensuremath{\textrm{MeV}}\xspace}
\newcommand{\GeV}{\ensuremath{\textrm{GeV}}\xspace}
\newcommand{\TeV}{\ensuremath{\textrm{TeV}}\xspace}
\newcommand{\rec}{\ensuremath{\textrm{rec}}}
\newcommand{\true}{\ensuremath{\textrm{true}}}
\newcommand{\vis}{\ensuremath{\textrm{vis}}}
\newcommand{\inv}{\ensuremath{\textrm{inv}}}
\newcommand{\eff}{\ensuremath{\textrm{eff}}}

\newcommand{\T}{\ensuremath{\textrm{T}}}
\newcommand{\cf}{cf.\xspace}
\newcommand{\ie}{i.e.\xspace}

\newcommand{\BW}{\ensuremath{\textrm{BW}}}
\newcommand{\rad}{\ensuremath{\textrm{rad}}\xspace}

\def\TReg{\textsuperscript{\textregistered}}
\usepackage{array}
\newcolumntype{C}[1]{>{\centering\arraybackslash}p{#1}}

\begin{document}

\ifx\ver\verPAPER
\begin{frontmatter}
\fi

\title{Reconstruction of the Higgs mass in events with Higgs bosons
  decaying into a pair of $\Pgt$ leptons using matrix element techniques}


\ifx\ver\verPreprint
\author[1]{Lorenzo Bianchini}
\author[2]{Betty Calpas}
\author[3]{John Conway}
\author[4]{Andrew Fowlie}
\author[2, 5]{Luca Marzola}
\author[2]{Lucia Perrini}
\author[6]{Christian Veelken}
\affil[1]{Institute for Particle Physics, ETH Zurich, 8093 Zurich, Switzerland}
\affil[2]{National Institute for Chemical Physics and Biophysics, 10143 Tallinn, Estonia}
\affil[3]{Department of Physics, University of California, Davis, CA 95616}
\affil[4]{ARC Centre of Excellence for Particle Physics at the Tera-scale, Monash University, Melbourne, Victoria 3800, Australia}
\affil[5]{Institute of Physics, University of Tartu, 50411 Tartu, Estonia}
\affil[6]{CERN, 1211 Geneva, Switzerland}
\fi
\ifx\ver\verPAPER
\author[eth]{Lorenzo Bianchini}
\ead{lorenzo.bianchini@cern.ch}
\author[tallinn]{Betty Calpas}
\ead{betty.calpas@cern.ch}
\author[ucd]{John Conway}
\ead{conway@physics.ucdavis.edu}
\author[melb]{Andrew Fowlie}
\ead{andrew.fowlie@monash.edu}
\author[tartu]{Luca Marzola}
\ead{luca.marzola@ut.ee}
\author[tallinn]{Lucia Perrini}
\ead{lucia.perrini@cern.ch}
\author[cern]{Christian Veelken}
\ead{christian.veelken@cern.ch}
\address[eth]{Institute for Particle Physics, ETH Zurich, 8093 Zurich, Switzerland}
\address[ucd]{Department of Physics, University of California, Davis, CA 95616}
\address[tallinn]{National Institute for Chemical Physics and Biophysics, 10143 Tallinn, Estonia}
\address[tartu]{Institute of Physics, University of Tartu, 51014 Tartu, Estonia}
\address[melb]{ARC Centre of Excellence for Particle Physics at the Tera-scale, Monash University, Melbourne, Victoria 3800, Australia}
\address[cern]{CERN, 1211 Geneva, Switzerland}
\fi

\ifx\ver\verPreprint
\maketitle
\fi

\begin{abstract}
We present an algorithm for the reconstruction of the Higgs mass in events with Higgs bosons decaying into a pair of $\Pgt$ leptons.
The algorithm is based on matrix element (ME) techniques and achieves
a relative resolution on the Higgs boson mass of typically $15$--$20\%$.
A previous version of the algorithm has been used in analyses of Higgs
boson production performed by the CMS collaboration during LHC Run
$1$.
The algorithm is described in detail and its performance on simulated
events is assessed.
The development of techniques to handle $\Pgt$ decays in the ME
formalism represents an important result of this paper.
\end{abstract}

\ifx\ver\verPAPER
\end{frontmatter}
\fi

\clearpage

\linenumbers

\include*{introduction}

\include*{mem}

\include*{mem_ME}
\include*{mem_hadRecoil}

\include*{mem_TF}
\include*{mem_PSintegration}

\include*{mem_xSection}

\include*{mem_numericalMaximization}
\include*{mem_logM}

\include*{classicSVfit}

\include*{performance}

\include*{discussion}

\include*{summary}

\include*{acknowledgements}


\include*{appendix}
\include*{appendix_tauToHadDecays}
\include*{appendix_tauToLepDecays}

\bibliography{svFitMEM}

\end{document}

%% file: introduction.tex
\section{Introduction}
\label{sec:introduction}

A new boson of mass $125$~\GeV has been observed by the ATLAS and CMS collaborations~\cite{Higgs-Discovery_CMS,Higgs-Discovery_ATLAS}.
The properties of the new particle are compatible with the predictions for the Standard Model (SM) 
Higgs ($\PHiggs$) boson~\cite{Englert:1964et,Higgs:1964ia,Higgs:1964pj,Guralnik:1964eu,Higgs:1966ev,Kibble:1967sv}
within the present experimental uncertainties~\cite{HIG-14-014,Chatrchyan:2014tja,Khachatryan:2014iha,HIG-14-009}.
The observation of its decay into a pair of $\Pgt$ leptons, at a rate that is compatible with the SM expectation, has been reported recently~\cite{HIG-13-004,Aad:2015vsa,HIG-15-002}.
The decay into a pair of $\Pgt$ leptons allows for the most precise measurement of the direct coupling of the SM $\PHiggs$ boson to fermions.
Decays of heavy resonances into $\Pgt$ lepton pairs furthermore provide high sensitivity to search for models with an extended Higgs sector,
constituting an important experimental signature at the LHC.

The sensitivity of the SM $\PHiggs \to \Pgt\Pgt$ analysis critically depends on
the capability to distinguish the signal from a large irreducible background, arising from $\PZ/\Pggx \to \Pgt\Pgt$ Drell--Yan (DY) production.
An important handle to separate the signal from the background is the mass of the $\Pgt$ lepton pair, which we denote by $m_{\Pgt\Pgt}$.
The signal is expected to show up as a small bump on the high mass tail of the $m_{\Pgt\Pgt}$ distribution of the background
(see e.g. Figs. 8, 9, and 11 of Ref.~\cite{HIG-13-004}).

The separation of the signal from the background improves if the mass distribution for the signal is narrow.
The SM predicts the total width of the $\PHiggs$ boson to be $\approx 4$~\MeV.
Present experimental upper limits on the total width amount to $\approx 10$ times the SM value~\cite{HIG-14-002,Aad:2015xua}.
These limits have been obtained by comparing the rates for off-shell versus on-shell $\PHiggs$ boson production and depend on certain assumptions.
Direct, model independent, upper limits on the total $\PHiggs$ width, 
obtained by analyzing the mass spectra in $\PHiggs \to \PZ\PZ \to 4 \Plepton$ ($\Plepton = \Pe, \Pgm$) and $\PHiggs \to \Pgamma\Pgamma$ events, are $\approx 1$~\GeV.
In contrast, the width of the $m_{\Pgt\Pgt}$ distribution reconstructed in SM $\PHiggs \to \Pgt\Pgt$ events typically amounts to $\approx 20$~\GeV 
and is dominated purely by the experimental resolution.

Different methods for the reconstruction of $m_{\Pgt\Pgt}$ 
have been discussed in the literature~\cite{massRecoCollinearApprox,neutrinoRecByVertexInfo,MMC,Barr:2011he,Gripaios:2012th}. 
The SVfit algorithm~\cite{SVfit} has been used to reconstruct the $\PHiggs$ boson mass in the SM $\PHiggs \to \Pgt\Pgt$ analysis
as well as in searches for further $\PHiggs$ bosons predicted by models beyond the SM 
performed by the CMS collaboration during LHC Run $1$~\cite{HIG-10-002,HIG-11-029,HIG-13-004,HIG-13-021,HIG-14-029,HIG-14-033,HIG-14-034,HIG-15-001,HIG-15-013}.
Compared to alternative mass variables,
the usage of the SVfit algorithm has improved the sensitivity of the SM $\PHiggs \to \Pgt\Pgt$ analysis for measuring the signal rate by $\approx 40\%$~\cite{HIG-13-004}.
The improvement in sensitivity corresponds to a gain by about a factor of two in integrated luminosity of the analysed dataset.

In this paper we report on the development of an improved version of the SVfit algorithm.
Two variants of the improved algorithm have been implemented.
The first variant allows to reconstruct the mass $m_{\Pgt\Pgt}$ of the $\Pgt$ lepton pair.
It has been developed within the paradigm of the matrix element (ME) method~\cite{Kondo:1988yd,Kondo:1991dw}.
Whereas the algorithm described in Ref.~\cite{SVfit} uses a likelihood function of arbitrary normalization,
the improved algorithm is based on a proper normalization within the formalism of the ME method.
The second variant of the improved algorithm uses a likelihood function of arbitrary normalization.
The algorithm allows for the reconstruction of not only the mass $m_{\Pgt\Pgt}$ of the $\Pgt$ lepton pair,
but of any kinematic function of the two $\Pgt$ leptons,
including the $\pT$, $\eta$, $\phi$, and transverse mass of the $\Pgt$ lepton pair.
A further improvement concerns the extension of the algorithm to account for the experimental resolution 
on the reconstruction of hadrons that are produced in the $\Pgt$ decays.

The development of the formalism to handle $\Pgt$ decays in the ME method constitutes an important result of this paper,
which has not been discussed in the literature so far.
The formalism described in this paper allows one to extend the ME generated by automatized tools such as CompHEP~\cite{CompHEP1,CompHEP2} or MadGraph~\cite{MadGraph},
which treat $\Pgt$ leptons as stable particles, by the capability to handle $\Pgt$ decays.

The paper is organized as follows. 
In Section~\ref{sec:mem} we describe the first variant of the improved algorithm, 
in particular the formalism that we developed to handle $\Pgt$ decays in the ME method
and our treatment of the experimental resolution on the reconstruction of hadrons that are produced in the $\Pgt$ decays.
The second variant of the algorithm is presented in Section~\ref{sec:classicSVfit}.
The performance of both variants of the improved algorithm in terms of achieved $m_{\Pgt\Pgt}$ resolution 
is compared to the previous version of the SVfit algorithm, used during LHC Run $1$, 
and to selected alternative mass observables in Section~\ref{sec:performance}.
The results are discussed in Section~\ref{sec:discussion}.
The paper concludes with a summary in Section~\ref{sec:summary}.

%% file: mem.tex
\section{The Matrix element method}
\label{sec:mem}

In the ME method, an estimate for the unknown model parameter $\Theta$
is obtained by maximizing the probability density:
\begin{align}
\mathcal{P}(\bm{y}|\Theta) = & \frac{\Omega(\bm{y})}{\sigma(\Theta) \,
\mathcal{A}(\Theta)} \, \int \, dx_{a} \, dx_{b} \,
d\Phi_{n} \, \frac{f(x_{a}) \, f(x_{b})}{2 \, x_{a} \, x_{b} \, s} \, (2\pi)^{4} \,
\delta( x_{a} \, E_{a} + x_{b} \, E_{b} - \sum_{i}^{n}
E_{(i)}) \cdot \nonumber \\
 & \quad \delta^{3}( x_{a} \, \bm{p}^{a} + x_{b} \, \bm{p}^{b} - \sum_{i}^{n}
\bm{p}^{(i)}) \, 
  \vert \mathcal{M}(\bm{p},\Theta) \vert^{2} \, W(\bm{y}|\bm{p}) \,
  \epsilon(\bm{p},\Theta) 
\label{eq:mem}
\end{align}
with respect to $\Theta$.
The symbols $E_{a}$ and $E_{b}$ and $\bm{p}^{a}$ and $\bm{p}^{b}$ denote, respectively, the energies and momenta of the two colliding protons,
$\sqrt{s}$ represents the centre-of-mass energy,
$x_{a}$ and $x_{b}$ denote the Bjorken scaling variables~\cite{Bjorkenx}
and $f(x_{a})$ and $f(x_{b})$ the corresponding parton distribution
functions (PDF).
We use the MSTW 2008 LO PDF set~\cite{MSTW} to evaluate $f(x_{a})$ and $f(x_{b})$.
We furthermore denote by $n$ the number of particles in the final state,
by $\bm{p}^{(i)}$ the momentum of the $i$-th final state particle
and by $d\Phi_{n} = \prod_{i}^{n} \,
\frac{d^{3}\bm{p}^{(i)}}{(2\pi)^{3} \, 2 E_{(i)}}$ the differential $n$-particle
phase space element.
Vector quantities are represented by bold letters.
The symbol $\vert \mathcal{M}(\bm{p},\Theta) \vert^{2}$ represents the
squared modulus of the ME for
the process.
The $\delta$-functions $\delta( x_{a} \, E_{a} + x_{b} \, E_{b} - \sum_{i}^{n} E_{(i)})$
and $\delta^{3}( x_{a} \, \bm{p}^{a} + x_{b} \, \bm{p}^{b} - \sum_{i}^{n} \bm{p}^{(i)})$ 
impose energy and momentum conservation.
The set of observables measured in the
detector is denoted by $\bm{y}$.
The function $W(\bm{y}|\bm{p})$ represents the probability density to
observe the measured values $\bm{y}$, given a point $\bm{p}$ in the
$n$-particle phase space, and
is referred to as ``transfer function'' (TF) in the
literature, while the function $\Omega(\bm{y})$ is referred to as ``indicator function''~\cite{Fiedler:2010sg,Volobouev:2011vb}.
The value of the indicator function is $1$ for events which pass the event selection criteria and $0$ otherwise.
The efficiency for an event originating at the phase space point
$\bm{p}$ to pass the event selection, \ie to end up with measured
observables $\bm{y}$ for which $\Omega(\bm{y}) = 1$,
is denoted by $\epsilon(\bm{p},\Theta)$. 
The symbol $\sigma(\Theta)$ corresponds to the inclusive cross section for the process under study,
in the case considered in this paper the production, in $\Pp\Pp$
collisions, of a $\PHiggs$ boson decaying into a pair of $\Pgt$ leptons
with subsequent decay of the $\Pgt$ pair.
The symbol $\mathcal{A}(\Theta)$ represents the acceptance of the
event selection, that is, the percentage of events which pass the event
selection criteria.
Division by $\sigma(\Theta) \cdot \mathcal{A}(\Theta)$ ensures that $\mathcal{P}(\bm{y}|\Theta)$ has
the correct normalization required for a probability density, 
i.e. $\int \, d\bm{y} \, \mathcal{P}(\bm{y}|\Theta) = 1$ for every $\Theta$, 
provided that the TF satisfy the normalization condition
$\int \, d\bm{y} \, \Omega(\bm{y}) \, W(\bm{y}|\bm{p}) = 1$
for every $\bm{p}$.

The meaning of Eq.~(\ref{eq:mem}) is as follows.
The best estimate for the unknown model parameter $\Theta$ is given by the
value $\hat{\Theta}$ which maximizes the probability density to observe precisely the 
values $\bm{y}$ that are measured in the detector. 
Within the scope of this paper, the unknown model parameter $\Theta$
corresponds to the mass $m_{\PHiggs}$ of the $\PHiggs$ boson or,
equivalently, to the true mass $m_{\Pgt\Pgt}$ of the $\Pgt$ lepton
pair in a given event.

The individual terms of Eq.~(\ref{eq:mem}) are described in
Sections~\ref{sec:mem_ME} to~\ref{sec:mem_logM}.
The ME for the process $\Pp\Pp \to \PHiggs \to \Pgt\Pgt$
with subsequent decay of the $\Pgt$ leptons 
via $\Pgt \to \enunu$, $\Pgt \to \mununu$, or $\Pgt \to \textrm{hadrons} + \Pnut$
is described in Section~\ref{sec:mem_ME}.
A complication arises from the fact that we use a leading order (LO)
ME to model the $\PHiggs$ boson production process $\Pp\Pp \to
\PHiggs$. The LO ME strictly applies only to events in which the $\PHiggs$ boson has zero $\pT$,
while the production of $\PHiggs$ bosons at the LHC typically
proceeds in association with jets.
The treatment of events with hadronic activity, in which the $\PHiggs$ boson has non-zero $\pT$,
is detailed in Section~\ref{sec:mem_hadRecoil}.
The TFs are described in Section~\ref{sec:mem_TF} and
the integration over the $n$-particle phase space is described in
Section~\ref{sec:mem_PSintegration}.
The computation of the cross section $\sigma(m_{\PHiggs})$ that is needed for a proper normalization of the probability density $\mathcal{P}(\bm{y}|m_{\PHiggs})$
in Eq.~(\ref{eq:mem}) is described in Section~\ref{sec:mem_xSection}.
The numerical maximization of the probability density $\mathcal{P}(\bm{y}|m_{\PHiggs})$
with respect to the mass $m_{\PHiggs}$ of the $\PHiggs$ boson is described in
Section~\ref{sec:mem_numericalMaximization}.
We denote this version of the SVfit algorithm by SVfitMEM.

We conclude this section with a description of an artificial term that we
choose to add to the probability density $\mathcal{P}(\bm{y}|m_{\PHiggs})$, in
order to reduce tails in the $m_{\Pgt\Pgt}$ distribution reconstructed
by the algorithm. The structure of this ``regularization'' term is described in
Section~\ref{sec:mem_logM}.
Distributions of $m_{\Pgt\Pgt}$ in simulated events,
reconstructed with and without this term, are presented in Section~\ref{sec:performance}.

As a consequence of using LO ME to model the $\PHiggs$ boson production process $\Pp\Pp \to \PHiggs$,
the efficiency $\epsilon(\bm{p},m_{\PHiggs})$ and the acceptance $\mathcal{A}(m_{\PHiggs})$ cannot be determined reliably,
because they depend on the $\PHiggs$ boson $\pT$ spectrum.
In particular for $\PHiggs$ bosons of low mass $m_{\PHiggs}$,
the probability for the particles produced in the $\Pgt$ lepton decays to pass selection criteria on $\pT$ and $\eta$, 
which are necessitated by trigger requirements at the LHC,
may vary significantly as function of $\PHiggs$ boson $\pT$.
For this reason, we will assume that $m_{\Pgt\Pgt}$ is reconstructed before any event selection criteria are applied,
\ie $\Omega(\bm{y}) = 1$, $\epsilon(\bm{p},m_{\PHiggs}) = 1$, and $\mathcal{A}(m_{\PHiggs}) = 1$
for all evaluations of Eq.~(\ref{eq:mem}).
We expect these assumptions to introduce a small bias on the reconstructed $m_{\Pgt\Pgt}$ values
and possibly a small degradation in $m_{\Pgt\Pgt}$ resolution.
The bias can be corrected with the Monte Carlo simulation, 
and we do not expect it to cause a problem in practical applications of our algorithm.

%% file: mem_ME.tex
\subsection{Matrix element}
\label{sec:mem_ME}

We decompose the squared modulus of the ME, $\vert \mathcal{M}(\bm{p},m_{\PHiggs}) \vert^{2}$, for the process $\Pp\Pp \to \PHiggs \to \Pgt\Pgt$
with subsequent decay of the $\Pgt$ leptons into electrons, muons, or
hadrons into five parts:
\begin{equation}
\vert \mathcal{M}(\bm{p},m_{\PHiggs}) \vert^{2} = 
 \vert \mathcal{M}_{\Pp\Pp \to \PHiggs \to
   \Pgt\Pgt}(\bm{p},m_{\PHiggs}) \vert^{2} 
\cdot \vert \BW_{\Pgt}^{(1)} \vert^{2} 
\cdot \vert \mathcal{M}^{(1)}_{\Pgt\to\cdots}(\bm{p}) \vert^{2} 
\cdot \vert \BW_{\Pgt}^{(2)} \vert^{2} 
\cdot \vert \mathcal{M}^{(2)}_{\Pgt\to\cdots}(\bm{p}) \vert^{2} \, ,
 \label{eq:meFactorization}
\end{equation}
where we use the superscripts $(1)$ and $(2)$ to refer to the $\Pgt$ lepton of positive and negative charge, respectively.
The first term, $\vert \mathcal{M}_{\Pp\Pp \to \PHiggs \to
  \Pgt\Pgt}(\bm{p},m_{\PHiggs}) \vert^{2}$, represents the squared
modulus of the ME for $\PHiggs$ boson production with subsequent decay of the $\PHiggs$ boson into a pair of $\Pgt$ leptons.
This term can be computed by using automatized tools such as
CompHEP or MadGraph or it can be taken from the literature.

\begin{figure}
\begin{center}
\includegraphics*[height=54mm]{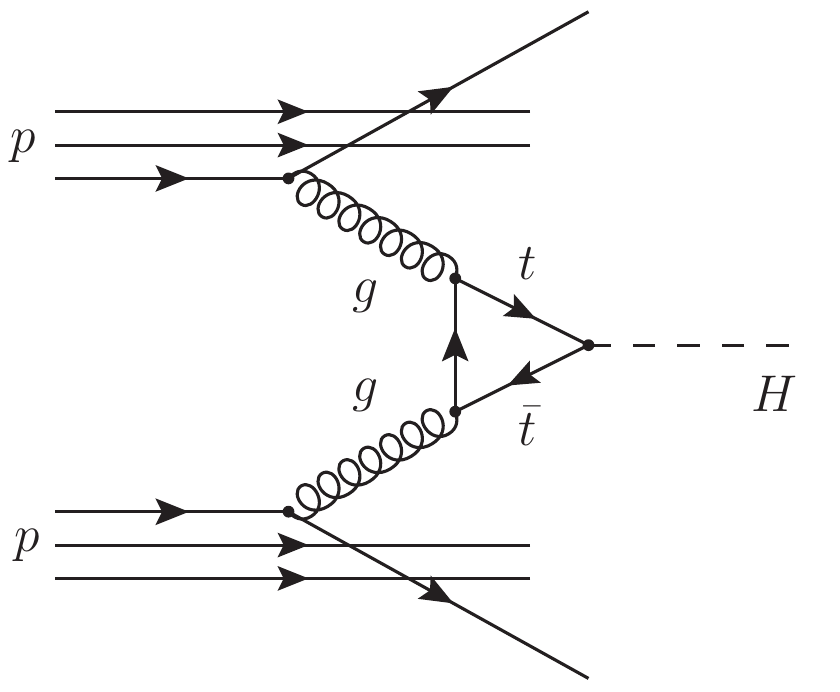}
\end{center}
\caption{
  LO Feynman diagram for $\PHiggs$ boson production in $\Pp\Pp$ collisions via the gluon fusion process.
}
\label{fig:ggH_FeynmanDiagram}
\end{figure}

We take $\vert \mathcal{M}_{\Pp\Pp \to \PHiggs \to
  \Pgt\Pgt}(\bm{p},m_{\PHiggs}) \vert^{2}$ from the literature
and decompose it into a product of three factors:
\begin{equation}
\vert \mathcal{M}_{\Pp\Pp \to \PHiggs \to \Pgt\Pgt} \vert^{2} =
 \vert \mathcal{M}_{\Pg\Pg \to \PHiggs} \vert^{2} 
\cdot \vert \BW_{\PHiggs} \vert^{2} 
\cdot \vert \mathcal{M}_{\PHiggs \to \Pgt\Pgt} \vert^{2} \, .
\label{eq:meHiggsProduction_and_Decay}
\end{equation}
We model the $\PHiggs$ boson production using the LO ME for the gluon fusion process $\Pg\Pg \to \PHiggs$,
which accounts for about $90\%$ of the total $\PHiggs$ boson production rate at the LHC.
The corresponding Feynman diagram is shown in Fig.~\ref{fig:ggH_FeynmanDiagram}.
The squared modulus of the ME reads~\cite{me_ggHprod}:
\begin{equation}
\vert \mathcal{M}_{\Pg\Pg \to \PHiggs} \vert^{2} = 
 \frac{\sqrt{2} \, G_{F}}{256 \, \pi^{2}} \, \alpha_{s}^{2} \, \zeta^{2} \, m_{\PHiggs}^{4} \vert 1 + (1 - \zeta) \, f(\zeta) \vert^{2} \, ,
\label{eq:meHiggsProduction}
\end{equation}
with $\zeta = 4\, \frac{m_{\Pqt}^{2}}{m_{\PHiggs}^{2}}$ and:
\begin{equation}
f(\zeta) = 
\begin{cases} 
\arcsin^{2} \frac{1}{\sqrt{\zeta}}  & \mbox{if } \zeta \geq 1 \, , \\
-\frac{1}{4} \, \left( \log\frac{1 + \sqrt{1 - \zeta}}{1 - \sqrt{1 - \zeta}} - i\pi \right)^{2} & \mbox{if } \zeta < 1 \, .
\end{cases}
\label{eq:meHiggsProduction_ftau}
\end{equation}
The symbols $G_{F}$ and $\alpha_{s}$ denote, respectively, the Fermi and
strong coupling constant. Their numerical values are:
\begin{equation} 
G_{F} = 1.166 \cdot 10^{-5}\mbox{~GeV}^{-2} \, \mbox{ and } \,
\alpha_{s}(m_{\PZ}) = 0.139384 \, .
\label{eq:def_G_F} 
\end{equation}
The squared modulus of the Breit-Wigner propagator $\vert
\BW_{\PHiggs} \vert^{2}$, given by:
\begin{equation}
\vert \BW_{\PHiggs} \vert^{2} = \frac{1}{(q_{\PHiggs}^{2} -
  m_{\PHiggs}^{2})^{2} + m_{\PHiggs}^{2} \, \Gamma_{\PHiggs}^{2}} 
\label{eq:meHiggsBreitWigner}
\end{equation}
associates $\PHiggs$ boson production and decay.
In Eq.~(\ref{eq:meHiggsBreitWigner}), the symbol
$q_{\PHiggs}^{2} = (E_{\Pgt(1)} + E_{\Pgt(2)})^{2} - (\bm{p}^{\Pgt(1)} + \bm{p}^{\Pgt(2)})^{2}$ denotes the mass of the $\Pgt$ lepton pair.
The squared modulus of the ME for the decay of the $\PHiggs$ boson
into a $\Pgt$ lepton pair, summed over the spin states of the two $\Pgt$ leptons, is given by:
\begin{equation}
\vert \mathcal{M}_{\PHiggs \to \Pgt\Pgt} \vert^{2} = 
 \frac{4 \, G_{F} \, m_{\Pgt}^{2}}{\sqrt{2}\pi} \, m_{\PHiggs}^{2} \left( 1 - \frac{4 \, m_{\Pgt}^{2}}{m_{\PHiggs}^{2}} \right) \, ,
\label{eq:meHiggsDecay}
\end{equation}
and is related to the branching ratio $\mathcal{B}(\PHiggs \to \Pgt\Pgt)$
by~\cite{me_HtoTauTau}:
\begin{equation}
\mathcal{B}(\PHiggs \to \Pgt\Pgt) 
 = \frac{1}{16\pi \, m_{\PHiggs} \, \Gamma_{\PHiggs}} \, \sqrt{1 - \frac{4 \, m_{\Pgt}^{2}}{m_{\PHiggs}^{2}}} \, \vert \mathcal{M}_{\PHiggs \to \Pgt\Pgt} \vert^{2} \, .
\label{eq:meHiggsDecay_by_BR}
\end{equation}
For a SM $\PHiggs$ boson,
the branching ratio $\mathcal{B}(\PHiggs \to \Pgt\Pgt)$ becomes small and the total width $\Gamma_{\PHiggs}$ becomes large
once the decay into a pair of $\PW$ bosons is kinematically possible,
\ie for $m_{\PHiggs} \gtrsim 2 \, m_{\PW}$.
We remark that in theories beyond the SM, which motivate the search
for additional heavy scalars,
the branching ratio and total width may be very different from the SM
values.
In this paper, we assume $\mathcal{B}(\PHiggs \to \Pgt\Pgt) = 100\%$
and $\Gamma_{\PHiggs} = 10^{-2} \cdot m_{\PHiggs}$, 
and we compute $\vert \mathcal{M}_{\PHiggs \to \Pgt\Pgt} \vert^{2}$ according to Eq.~(\ref{eq:meHiggsDecay_by_BR}).
The branching ratio for the decay $\PHiggs \to \Pgt\Pgt$ and the total width of the $\PHiggs$ boson
actually have no effect on the value of $m_{\Pgt\Pgt}$ reconstructed by the algorithm, 
provided that $\mathcal{B}(\PHiggs \to \Pgt\Pgt)$ and $\Gamma_{\PHiggs}$ are treated consistently
when computing the squared modulus of the ME $\vert \mathcal{M}(\bm{p},m_{\PHiggs}) \vert^{2}$ and
the normalization factor $1/\sigma(m_{\PHiggs})$ in Eq.~(\ref{eq:mem}).

\begin{figure}[h]
\setlength{\unitlength}{1mm}
\begin{center}
\begin{picture}(150,52)(0,0)
\put(-4.5, -4.0){\mbox{\includegraphics*[height=56mm]
  {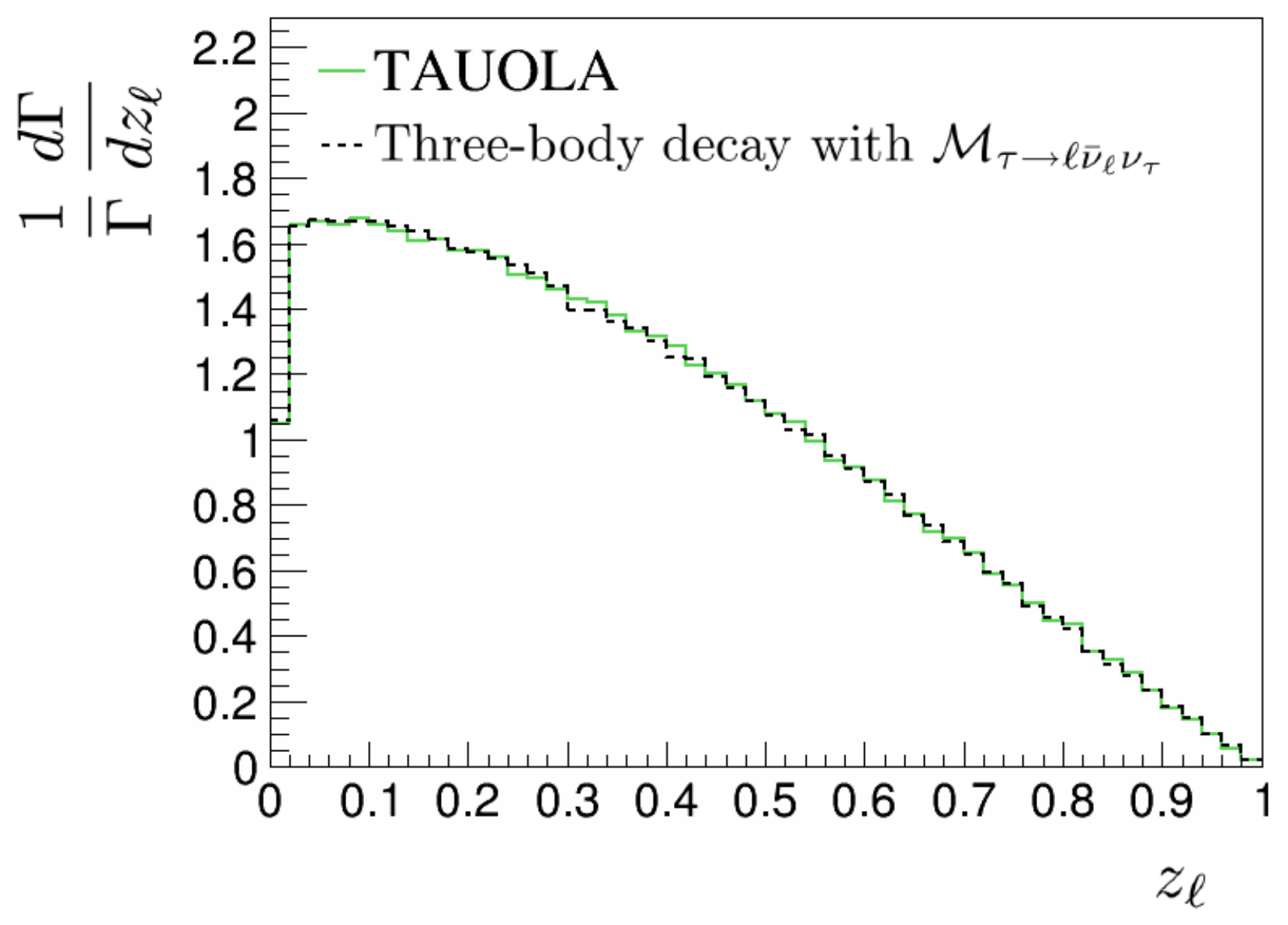}}}
\put(77.0, -4.0){\mbox{\includegraphics*[height=56mm]
  {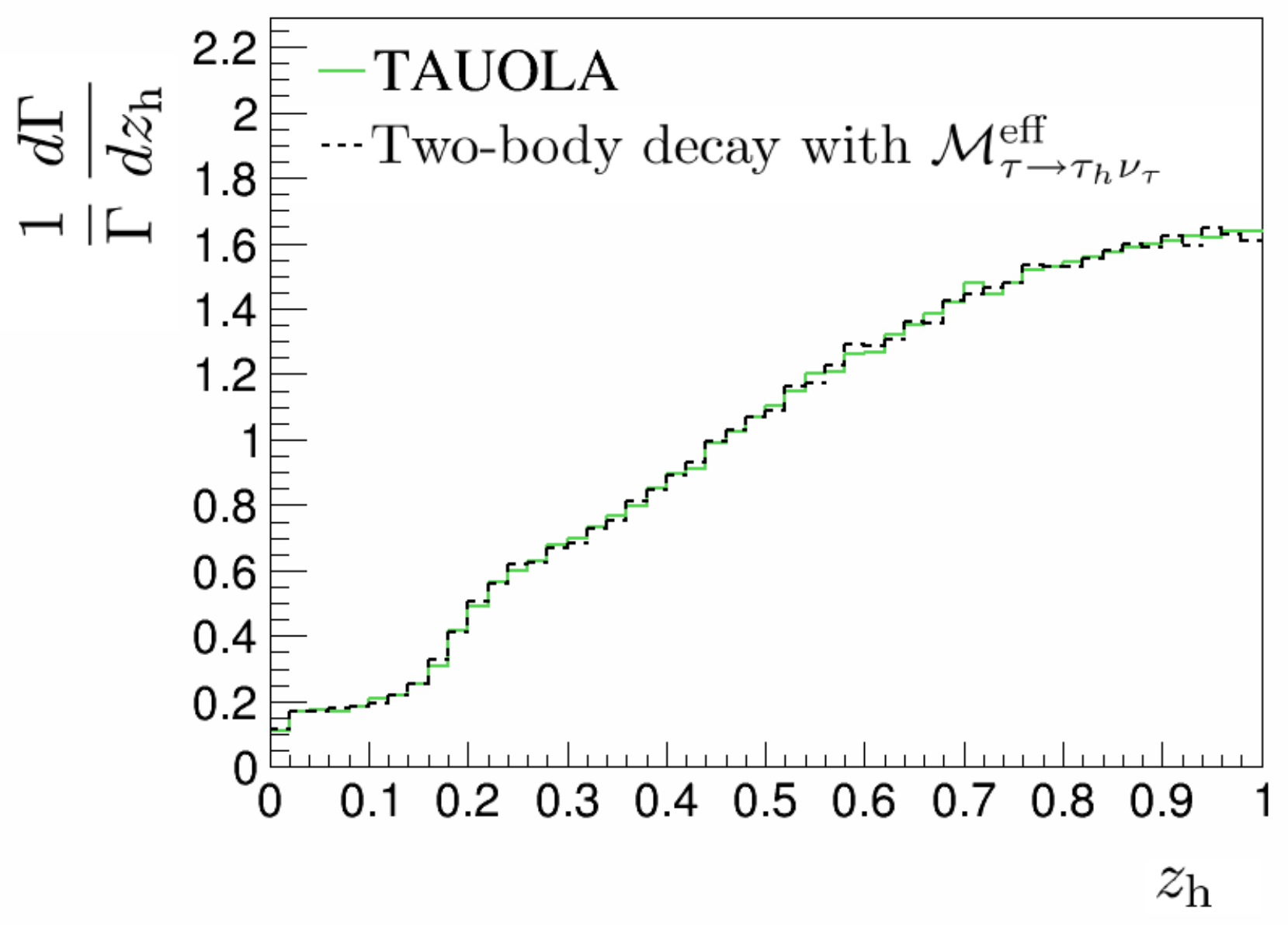}}}
\end{picture}
\end{center}
\caption{
  Fraction $z$ of the $\Pgt$ lepton energy, in the laboratory frame,
  carried by visible $\Pgt$ decay products in simulated SM $\PHiggs
  \to \Pgt\Pgt$ events of $m_{\PHiggs} = 125$~\GeV.
  The case of $\Pgt \to \mununu$ decays is shown on the left and the case of $\Pgt \to \textrm{hadrons} + \Pnut$ decays on the right.
  Our simplified model, which treats hadronic $\Pgt$ decays as
  two-body decays into a hadronic system $\tauh$ and a $\Pnut$,
  reproduces the distribution in $z$ obtained with a detailed Monte Carlo simulation of the $\Pgt$ decays, 
  based on TAUOLA~\cite{tauola}.
}
\label{fig:tauDecay_z}
\end{figure}

Concerning the decay of the $\Pgt$ leptons,
we use the narrow-width approximation (NWA) and for each $\Pgt$
lepton we take a separate and independent average over its spin
states, effectively ignoring the correlation in spin orientations between
the two $\Pgt$ leptons.
In the NWA, the squared modulus of the Breit-Wigner propagator $\vert
\BW_{\Pgt} \vert^{2}$ that associates $\Pgt$ lepton production and
decay yields a $\delta$-function:
\begin{equation}
\vert \BW_{\Pgt} \vert^{2} = \frac{\pi}{m_{\Pgt} \, \Gamma_{\Pgt}} \,
\delta ( q_{\Pgt}^{2} - m_{\Pgt}^{2} ) \, \mbox{ with } \, 
\Gamma_{\Pgt} = \frac{1}{\Delta t} =
 2.267 \cdot 10^{-12}\textrm{~\GeV} \, ,
\end{equation}
where $\Delta t = 290 \times 10^{-15}$~s denotes the lifetime of the
$\Pgt$ lepton~\cite{PDG}.
The factor $\vert \mathcal{M}^{(i)}_{\Pgt\to\cdots}
\vert^{2}$ in Eq.~(\ref{eq:meFactorization}) represents the decay of the $i$-th $\Pgt$ lepton.
For the decays $\Pgt \to \enunu$ and $\Pgt
\to \mununu$, which we refer to as ``leptonic'' $\Pgt$ decays, we take the ME from the literature.
Taking the average over the spin states of the $\Pgt$ lepton,
the squared modulus of the ME is given by~\cite{Barger:1987nn}:
\begin{equation}
\vert\mathcal{M}_{\Pgt \to \ellnunu} \vert^{2} = 64 \, G^{2}_{F} \,
\left( E_{\Pgt} \, E_{\APnu_{\Plepton}} - \bm{p}^{\Pgt} \cdot
  \bm{p}^{\APnu_{\Plepton}} \right) \, \left( E_{\Plepton} \,
  E_{\Pnut} - \bm{p}^{\Plepton} \cdot \bm{p}^{\Pnut} \right) \, .
\label{eq:leptonic_tau_decays_ME}
\end{equation}
The modelling of the decays $\Pgt \to \textrm{hadrons} + \Pnut$ 
by matrix elements is more difficult, 
due to the fact that $\Pgt$ leptons decay to a variety of hadronic
final states, and because some of the decays proceed via intermediate 
meson resonances~\cite{PDG}.
We refer to these decays as ``hadronic'' $\Pgt$ decays.
The ME for the dominant hadronic $\Pgt$ decay modes are discussed in the literature~\cite{Bullock:1992yt,Raychaudhuri:1995kv}.
In this paper, we use a simplified formalism and treat $\Pgt \to \textrm{hadrons} + \Pnut$ decays as two-body decays into a hadronic system $\tauh$ of momentum $\bm{p}^{\vis}$ and mass $m_{\vis}$ plus a $\Pnut$.
The squared modulus of the ME for the decay is taken to be constant and denoted by $\vert\mathcal{M}^{\eff}_{\Pgt \to \tauhnu}\vert^{2}$.
The value of $\vert\mathcal{M}^{\eff}_{\Pgt \to \tauhnu}\vert^{2}$ is
chosen such that it reproduces the measured branching fraction for hadronic $\Pgt$ decays.
The following relation holds for the considered case of a two-body decay:
\begin{equation}
\mathcal{B}(\Pgt \to \textrm{hadrons} + \Pnut) = \frac{1}{16 \pi \, m_{\Pgt} \, \Gamma_{\Pgt}} \cdot \frac{m_{\Pgt}^{2} - m_{\vis}^{2}}{m_{\Pgt}^{2}} \cdot \vert \mathcal{M}^{\textrm{eff}}_{\Pgt \to
  \tauhnu} \vert^{2} \, ,
\end{equation}
from which it follows that:
\begin{equation}
\vert \mathcal{M}^{\textrm{eff}}_{\Pgt \to \tauh\Pnut} \vert^{2} = 16 \pi \, m_{\Pgt} \, \Gamma_{\Pgt} 
  \cdot \frac{m_{\Pgt}^{2}}{m_{\Pgt}^{2} - m_{\vis}^{2}} \, \mathcal{B}(\Pgt \to \textrm{hadrons} + \Pnut) \, , 
\end{equation}
with $\mathcal{B}(\Pgt \to \textrm{hadrons} + \Pnut) = 0.648$~\cite{PDG}.
We have verified that the sum of all hadronic final states produced in $\Pgt$ lepton decays
is well reproduced by our simplified model.
Fig.~\ref{fig:tauDecay_z} shows the fraction of $\Pgt$ lepton energy,
in the laboratory frame, carried by the ``visible'' $\Pgt$ decay
products:
\begin{equation}
z = \frac{E_{\vis}}{E_{\Pgt}} \, .
\label{eq:def_z}
\end{equation}
We use the term ``visible'' $\Pgt$ decay products to refer to the sum
of all hadrons produced in decays of the type $\Pgt \to \textrm{hadrons} + \Pnut$ 
as well as to the electron or muon produced in the decays $\Pgt \to \enunu$ and $\Pgt \to \mununu$, respectively.

%% file: mem_hadRecoil.tex
\subsection{Treatment of hadronic activity}
\label{sec:mem_hadRecoil}

The phase-space for QCD radiation is large at the centre-of-mass
energies of the LHC
and as a consequence, particles with masses up to a few hundred GeV 
are typically produced in association with a sizeable hadronic activity~\cite{Alwall:2010cq}.
We refer to the vectorial sum of all particles in the event that do not originate from the $\PHiggs$ boson decay
as the ``hadronic recoil'' and denote the corresponding momentum by $\bm{p}^{\rec}$.
CMS, as well as ATLAS, have developed sophisticated techniques to improve the reconstruction 
of the hadronic recoil~\cite{CMS-JME-13-003,ATLAS-CONF-2014-019}.

The effect of the hadronic recoil is to modify the kinematics of the $\PHiggs$ boson decay
by boosting the $\Pgt$ leptons and their decay products,
in the transverse plane and in longitudinal direction.
The longitudinal component of $\bm{p}^{\rec}$ can be accounted for by
a small adjustement of the Bjorken scaling variables $x_{a}$ and
$x_{b}$ and does not cause a problem for modeling the $\PHiggs$ boson
production by LO ME.
However, the components of $\bm{p}^{\rec}$ in the transverse plane need to be accounted for.
The LO ME that we use to model the production of the $\PHiggs$ boson 
is adequate for events in which the $\PHiggs$ boson has zero $\pT$,
implying the transverse components of $\bm{p}^{\rec}$ to be zero.
In order to use the LO ME in analyses at the LHC,
the momenta of the visible $\Pgt$ decay products and of the neutrinos produced in the $\Pgt$ decays
need to be transformed into a frame in which the $\PHiggs$ boson has zero $\pT$.
The transformation is achieved by a Lorentz boost in the transverse plane,
using $(\hat{E}^{\rec} = \pThat^{\rec}, \pXhat^{\rec}, \pYhat^{\rec}, \pZhat^{\rec} = 0)$ as boost
vector.
We denote the true values of energies and momenta by symbols with a
hat if they are given in the laboratory frame, and by symbols a tilde in case they are given in the zero transverse momentum frame of the $\PHiggs$ boson.
Symbols with neither hat nor tilde denote values measured in the
laboratory frame.

The hadronic recoil is reconstructed with a typical resolution of $10$--$20$~\GeV at the LHC~\cite{CMS-JME-13-003,ATLAS-CONF-2014-019}.
It is demonstrated in Ref.~\cite{Alwall:2010cq} that the imperfect reconstruction of the hadronic recoil
in the detector may cause a significant bias on the $\PHiggs$ boson mass reconstructed by the ME method in case one ignores
the experimental resolution on the
hadronic recoil. We account
for the experimental resolution by introduction
of a TF $W_{\rec}( \pX^{\rec}, \pY^{\rec} | \pXhat^{\rec},
\pYhat^{\rec} )$.
We then marginalize Eq.~(\ref{eq:mem}) with respect to $\pXhat^{\rec}$ and $\pYhat^{\rec}$:
\begin{align}
& \mathcal{P}(\bm{y};\pX^{\rec},\pY^{\rec}|m_{\PHiggs}) =
\frac{1}{\sigma'(m_{\PHiggs})} \, \int \, dx_{a} \, dx_{b} \, d\Phi_{n} \,
d\pXhat^{\rec} \, d\pYhat^{\rec} \, \frac{f(x_{a}) f(x_{b})}{2 \,
  x_{a} \, x_{b} \, s} \cdot \nonumber \\
& \qquad (2\pi)^{4} \, \delta( x_{a} \, \Ehat_{a} + x_{b} \, \Ehat_{b} -
(\Ehat^{\rec} + \Ehat_{\Pgt(1)} + \Ehat_{\Pgt(2)}) ) \, \delta^{3}( x_{a} \,
\bm{\hat{p}}^{a} + x_{b} \, \bm{\hat{p}}^{b} - (\bm{\hat{p}}^{\rec} + \bm{\hat{p}}^{\Pgt(1)}
+ \bm{\hat{p}}^{\Pgt(2)}) ) \cdot \nonumber \\
& \qquad \vert \mathcal{M}(\bm{\tilde{p}},m_{\PHiggs}) \vert^{2} \, W(\bm{y}|\bm{\hat{p}}) \, W_{\rec}( \pX^{\rec}, \pY^{\rec} | \pXhat^{\rec}, \pYhat^{\rec} ) \, .
\label{eq:mem_mod_hadRecoil}
\end{align}
The cross section $\sigma(m_{\PHiggs})$ needs to be replaced by:
\begin{equation}
\sigma'(m_{\PHiggs}) = \sigma(m_{\PHiggs}) \, \int \, d\pXhat^{\rec} \, d\pYhat^{\rec} 
\end{equation}
in order for $\mathcal{P}(\bm{y};\pX^{\rec},\pY^{\rec}|m_{\PHiggs})$ to have the correct dimensions.
The symbol $\bm{y}$ refers to the measured momenta of the visible $\Pgt$ decay products, $\bm{p^{\vis(1)}}$ and $\bm{p^{\vis(2)}}$.
Using the relations $(\hat{E}_{a},\bm{\hat{p}}^{a}) = \frac{\sqrt{s}}{2} \, (1, 0,
0, 1)$ and $(\hat{E}_{b},\bm{\hat{p}}^{b}) = \frac{\sqrt{s}}{2} \, (1,
0, 0, -1)$
for the energies and momenta of the two colliding protons,
the integral over the $\delta$-function that ensures the conservation of energy and momentum can be expressed by:
\begin{align}
& \int \, dx_{a} \, dx_{b} \, d\pXhat^{\rec} \, d\pYhat^{\rec} \, \delta(
x_{a} \, \Ehat_{a} + x_{b} \, \Ehat_{b} - (\Ehat^{\rec} +
\Ehat_{\Pgt(1)} + \Ehat_{\Pgt(2)})) \cdot \nonumber \\
& \qquad \delta^{3}(
x_{a} \, \bm{p}^{a} + x_{b} \, \bm{\hat{p}}^{b} - (\bm{\hat{p}}^{\rec} +
\bm{\hat{p}}^{\Pgt(1)} + \bm{\hat{p}}^{\Pgt(2)})) \nonumber \\
= & \, \frac{2}{s} \, \int \,
d\pXhat^{\rec} \, d\pYhat^{\rec} \, 
\delta( \pXhat^{\rec} + \pXhat^{\Pgt(1)} + \pXhat^{\Pgt(2)} ) \,
\delta( \pYhat^{\rec} + \pYhat^{\Pgt(1)} + \pYhat^{\Pgt(2)} ) \, ,
\label{eq:delta4}
\end{align}
with:
\begin{align}
x_{a} = & \frac{1}{\sqrt{s}} \, \left( \Ehat^{\rec} + \Ehat_{\Pgt(1)} +
\Ehat_{\Pgt(2)} + (\pZhat^{\rec} + \pZhat^{\Pgt(1)} +
\pZhat^{\Pgt(2)}) \right)
\, , \nonumber \\
x_{b} = & \frac{1}{\sqrt{s}} \, \left( \Ehat^{\rec} + \Ehat_{\Pgt(1)}
  + \Ehat_{\Pgt(2)} - (\pZhat^{\rec} + \pZhat^{\Pgt(1)} +
  \pZhat^{\Pgt(2)}) \right) \, .
\label{eq:xa_and_xb}
\end{align}
The integration over $d\pXhat^{\rec}$ and $d\pYhat^{\rec}$ removes the $\delta$-functions 
$\delta( \pXhat^{\rec} + \pXhat^{\Pgt(1)} + \pXhat^{\Pgt(2)} )$ and
$\delta( \pYhat^{\rec} + \pYhat^{\Pgt(1)} + \pYhat^{\Pgt(2)} )$ in Eq.~(\ref{eq:delta4}).
Substituting Eq.~(\ref{eq:meFactorization}) into Eq.~(\ref{eq:mem_mod_hadRecoil}), we obtain:
\begin{align}
&
\mathcal{P}(\bm{p}^{\vis(1)},\bm{p}^{\vis(2)};\pX^{\rec},\pY^{\rec}|m_{\PHiggs})
= \frac{1}{\sigma'(m_{\PHiggs})} \, \frac{32\pi^{4}}{s} \, \int \,
 d\Phi_{n} \, \frac{f(x_{a}) f(x_{b})}{2 \, x_{a} \, x_{b} \, s} \, 
 \vert \mathcal{M}_{\Pp\Pp \to \PHiggs \to \Pgt\Pgt}(\bm{\tilde{p}},m_{\PHiggs}) \vert^{2} \cdot \hspace{2cm} \nonumber \\
& \qquad \vert \BW^{(1)}_{\Pgt} \vert^{2} \cdot \vert \mathcal{M}^{(1)}_{\Pgt\to\cdots}(\bm{\tilde{p}}) \vert^{2} 
 \cdot \vert \BW^{(2)}_{\Pgt} \vert^{2} \cdot \vert \mathcal{M}^{(2)}_{\Pgt\to\cdots}(\bm{\tilde{p}}) \vert^{2} \cdot \nonumber \\
& \qquad W(\bm{p}^{\vis(1)}|\bm{\hat{p}}^{\vis(1)}) \, W(\bm{p}^{\vis(2)}|\bm{\hat{p}}^{\vis(2)}) \, W_{\rec}( \pX^{\rec},\pY^{\rec} | \pXhat^{\rec},\pYhat^{\rec} ) \, ,
\label{eq:mem_with_hadRecoil}
\end{align}
with $x_{a}$ and $x_{b}$ given by Eq.~(\ref{eq:xa_and_xb}) and:
\begin{equation}
\pXhat^{\rec} = -(\pXhat^{\Pgt(1)} + \pXhat^{\Pgt(2)}) \, ,
\qquad \pYhat^{\rec} = -(\pYhat^{\Pgt(1)} + \pYhat^{\Pgt(2)}) \, .
\label{eq:xpXhat_and_pYhat}
\end{equation}
The form of the TF for the hadronic recoil, $W_{\rec}( \pX^{\rec},\pY^{\rec} | \pXhat^{\rec},\pYhat^{\rec} )$, is
discussed in Section~\ref{sec:mem_TF_hadRecoil}.

In practice, the experimental resolution on the momenta of electrons, muons and also $\tauh$
is typically negligible compared to the resolution on the hadronic recoil.
In good approximation, the resolution on the hadronic recoil is equivalent
to the resolution on the vectorial sum of the momenta, in the transverse plane,
of all particles reconstructed in the event.
The latter is referred to as ``missing transverse momentum'' and denoted by $\vecMET$.
The following relations hold for the components of the $\vecMET$
vector:
\begin{equation}
\METx = - \left( \pX^{\rec} + \pX^{\vis(1)} + \pX^{\vis(2)} \right)
\, \mbox { and } \,
\METy = - \left( \pY^{\rec} + \pY^{\vis(1)} + \pY^{\vis(2)} \right) \, .
\label{eq:met}
\end{equation}
These relations are valid on reconstruction level as well as for the
true values of the momenta, \ie Eq.~(\ref{eq:met}) is valid also
in case all $\pX$ and $\pY$ are replaced by $\pXhat$ and $\pYhat$.
Approximating the resolution on the hadronic recoil by the resolution
on $\vecMET$ has the advantage that the resolutions on $\vecMET$ have
been studied in detail and published by the ATLAS as well as CMS collaborations~\cite{CMS-JME-13-003,ATLAS-CONF-2014-019}.

%% file: mem_TF.tex
\subsection{Transfer functions}
\label{sec:mem_TF}

The treatment of the experimental resolution on the momenta of electrons, muons and $\tauh$ produced in the $\Pgt$ decays,
as well as of the hadronic recoil, are detailed in this section.
The resolution on the momentum $\bm{p}^{\vis(i)}$ of the visible $\Pgt$ decay products is parametrized as function of the true momentum $\phat^{\vis(i)}$
and modeled by TF $W(\bm{p}^{\vis(i)}|\phat^{\vis(i)})$.
The case of hadronic and leptonic $\Pgt$ decays is described in
Sections~\ref{sec:mem_TF_tauToHadDecays}
and~\ref{sec:mem_TF_tauToLepDecays}, respectively.
The neutrinos produced in the $\Pgt$ decays are not
included in the TF, but are handled separately, by the formalism detailed in Sections~\ref{sec:appendix_tauToHadDecays} and~\ref{sec:appendix_tauToLepDecays} of the appendix.
The TF $W_{\rec}( \pX^{\rec},\pY^{\rec} | \pXhat^{\rec},\pYhat^{\rec} )$ that we use to model the experimental resolution on the
hadronic recoil is discussed in Section~\ref{sec:mem_TF_hadRecoil}.

\subsubsection{Hadronic $\Pgt$ decays}
\label{sec:mem_TF_tauToHadDecays}

The energy, or equivalently $\pT$, of the $\tauh$ 
is reconstructed with a
resolution of $5$--$25\%$ at the
LHC~\cite{Aad:2014rga,TAU-14-001}.
The resolution typically varies as function of $\pT$ and $\eta$ and may additionally depend
also on the multiplicity of the charged and neutral hadrons produced in the $\Pgt$ decay.
The resolution on the $\pT$ of the $\tauh$ is of similar magnitude as the
resolution on $m_{\Pgt\Pgt}$ that we aim to achieve and needs to be
taken into account by suitable TF,
denoted by $W_{\Phadron}( \bm{p}_{\T}^{\vis} | \phat_{\T}^{\vis} )$,
when evaluating the integral in Eq.~(\ref{eq:mem_with_hadRecoil}).

\begin{figure}
\setlength{\unitlength}{1mm}
\begin{center}
\begin{picture}(160,54)(0,0)
\put(-2.5, 0.0){\mbox{\includegraphics*[height=54mm]{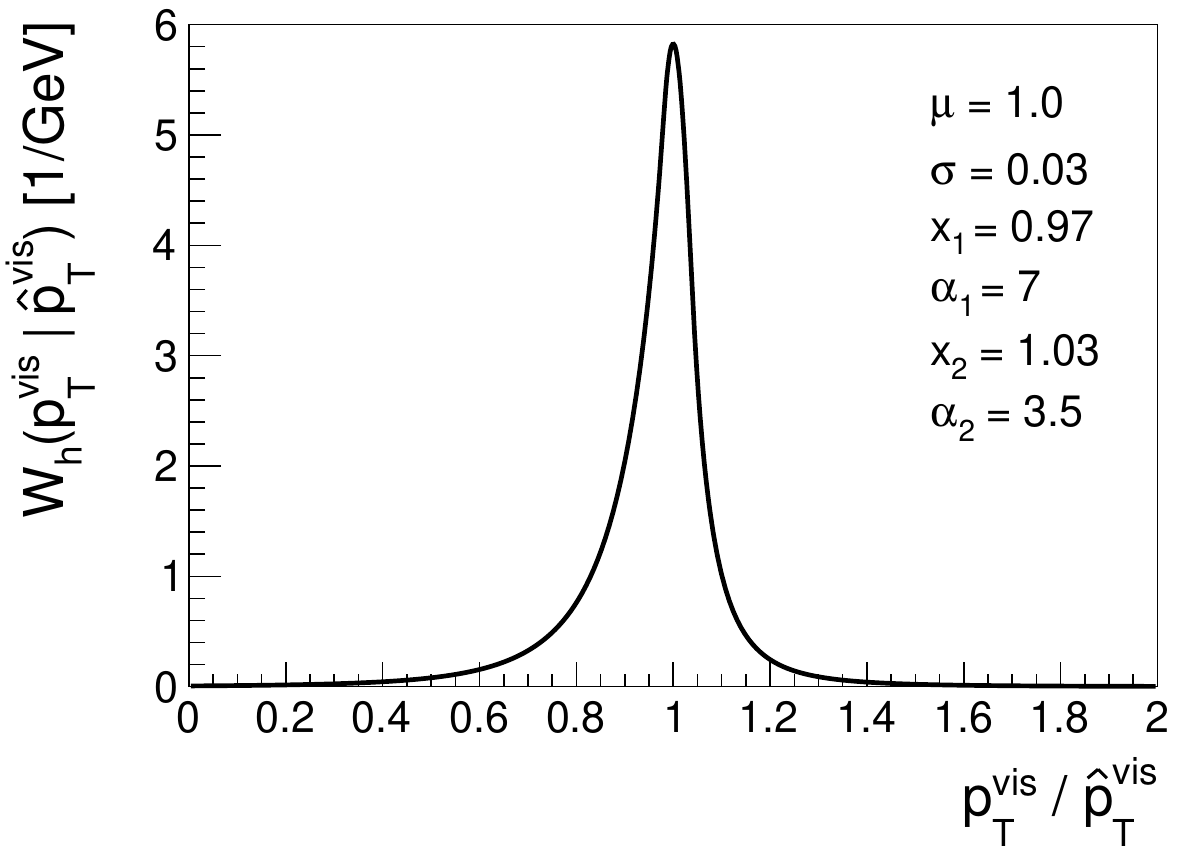}}}
\put(79.0, 0.0){\mbox{\includegraphics*[height=54mm]{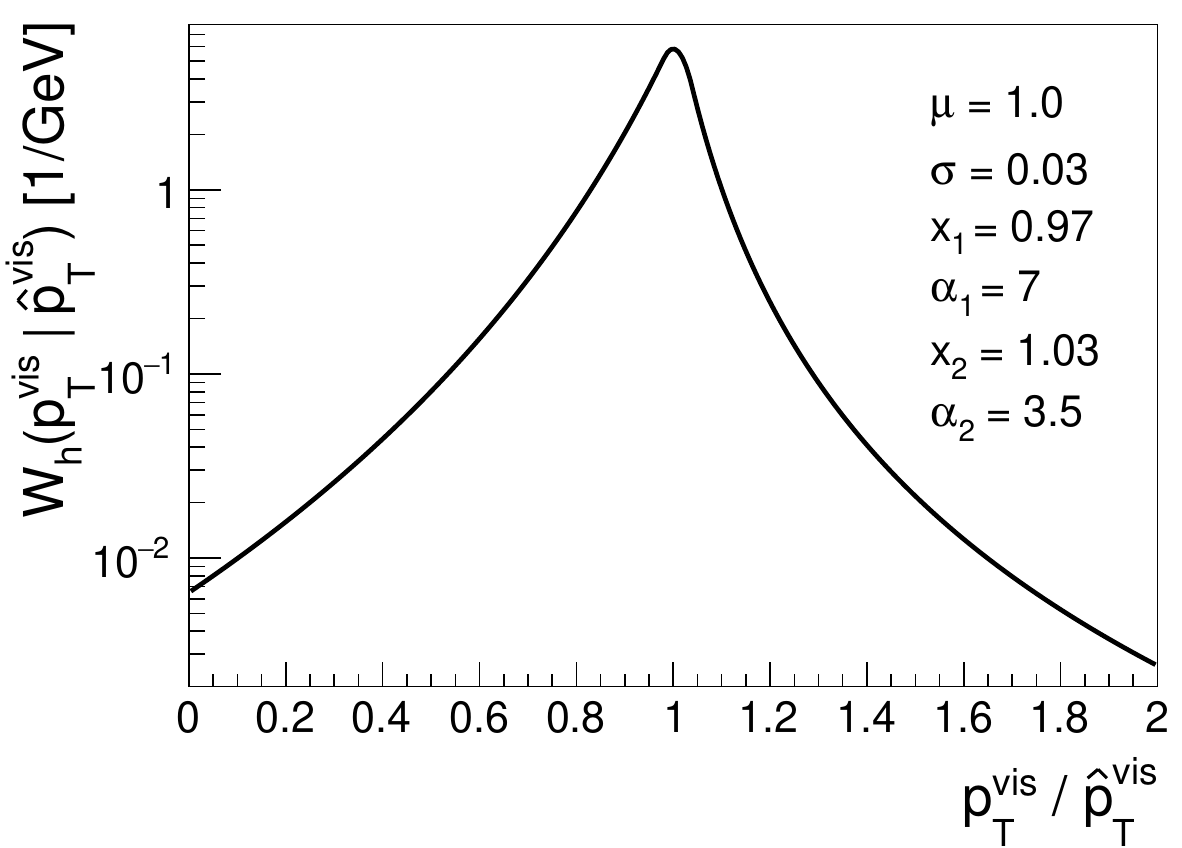}}}
\end{picture}
\end{center}
\caption{
  TF $W_{\Phadron}( \bm{p}_{\T}^{\vis} | \phat_{\T}^{\vis} )$ that models the experimental resolution on the $\pT$ of the charged and neutral hadrons produced in $\Pgt$ decays,
  given by Eq.~(\ref{eq:tf_tauToHadDecays_pT}),
  in linear (left) and logarithmic (right) scale of the ordinate.
  The values of the function parameters $\mu$, $\sigma$, $x_{1}$, $\alpha_{1}$, $x_{2}$, and $\alpha_{2}$ are given in the figure.
}
\label{fig:tf_tauToHadDecays_pT}
\end{figure}

ATLAS as well as CMS report that the energy response for hadronic
$\Pgt$ decays may be asymmetric~\cite{ATLAS:2011tfa,PRF-14-001}.
For the purpose of this paper, we assume the TF $W_{\Phadron}( \bm{p}_{\T}^{\vis} | \phat_{\T}^{\vis} )$ 
to follow a Gaussian distribution within the core region $x = \pT^{\vis}/\pThat^{\vis} \approx 1$ and to feature non-Gaussian tails,
which follow power-law functions, on both sides of the Gaussian core.
More specifically, we use the form:
\begin{equation}
W_{\Phadron}( \pT^{\vis} | \pThat^{\vis} ) = 
 \begin{cases}
   \mathcal{N} \, \xi_{1} \, \left( \frac{\alpha_{1}}{x_{1}} - x_{1} - \frac{x - \mu}{\sigma} \right)^{-\alpha_{1}} \,  
 & \mbox{if } x < x_{1} \\
   \mathcal{N} \, \exp\left( -\frac{1}{2} \, \left( \frac{x - \mu}{\sigma} \right)^{2} \right) \,
 & \mbox{if } x_{1} \leq x \leq x_{2} \\
   \mathcal{N} \, \xi_{2} \, \left( \frac{\alpha_{2}}{x_{2}} - x_{2} - \frac{x - \mu}{\sigma} \right)^{-\alpha_{2}} \,
 & \mbox{if } x > x_{2} 
 \end{cases}
\label{eq:tf_tauToHadDecays_pT}
\end{equation}
and we use the values $\mu = 1.0$, $\sigma = 0.03$, $x_{1} = 0.97$, $\alpha_{1} = 7$,
$x_{2} = 1.03$, and $\alpha_{2} = 3.5$ for its parameters.
The parameter values are chosen to approximately reproduce the $\tauh$ energy resolution expected for LHC Run $2$ in case of the CMS experiment.
The factors $\xi_{1}$ and $\xi_{2}$ are chosen such that the
function $W_{\Phadron}( \pT^{\vis} | \pThat^{\vis} )$ is continuous at
the points $x = x_{1}$ and $x = x_{2}$. 
The corresponding values are 
$\xi_{1} = \exp\left( -\frac{1}{2} \, \left( \frac{x_{1} - \mu}{\sigma} \right)^{2} \right) \cdot \left( \frac{\alpha_{1}}{x_{1}} - x_{1} - \frac{x_{1} - \mu}{\sigma} \right)^{\alpha_{1}}$
and
$\xi_{2} = \exp\left( -\frac{1}{2} \, \left( \frac{x_{2} - \mu}{\sigma} \right)^{2} \right) \cdot \left( \frac{\alpha_{2}}{x_{2}} - x_{2} - \frac{x_{2} - \mu}{\sigma} \right)^{\alpha_{2}}$.
The factor $\mathcal{N}$ is determined by the requirement that the function $W_{\Phadron}( \pT^{\vis} | \pThat^{\vis} )$ 
satisfies the normalization condition $\int \, d\pT^{\vis} \, W_{\Phadron}( \pT^{\vis} | \pThat^{\vis} ) = 1$ for any given value of $\pThat^{\vis}$.
The TF given by Eq.~(\ref{eq:tf_tauToHadDecays_pT}) is visualized in Fig.~\ref{fig:tf_tauToHadDecays_pT}.

The resolution on the direction of the $\tauh$ is on the level
of a few milliradians and is negligible in practice.
We hence model the TF for the momentum of the $\tauh$ by the product of Eq.~(\ref{eq:tf_tauToHadDecays_pT}) and two
$\delta$-functions:
\begin{equation}
W_{\Phadron}( \bm{p}^{\vis} | \phat^{\vis} ) =
 \frac{\sin^{2}\theta_{\vis}}{(\pT^{\vis})^{2}} \, 
  W_{\Phadron}( \pT^{\vis} | \pThat^{\vis} ) \,
  \delta( \theta_{\vis} - \thetahat_{\vis} ) \, 
  \delta( \phi_{\vis} - \phihat_{\vis} ) \, .
\label{eq:tf_tauToHadDecays}
\end{equation}
The factor $\sin^{2}\theta_{\vis}/(\pT^{\vis})^{2}$ is needed
to ensure the correct normalization of the TF:
\begin{align}
& \int \, d^{3}\bm{p}  \, W_{\Phadron}( \bm{p}^{\vis} | \phat^{\vis} ) = \int \, d\pX^{\vis} \, d\pY^{\vis} \, d\pZ^{\vis} \, W_{\Phadron}( \bm{p}^{\vis} | \phat^{\vis} ) \nonumber \\
& \qquad = \int \, d\pT^{\vis} \, d\theta_{\vis} \, d\phi_{\vis} \,
\frac{(\pT^{\vis})^{2}}{\sin^{2} \theta_{\vis}} \, W_{\Phadron}( \bm{p}^{\vis} | \phat^{\vis} ) \, 
  = 1 \, ,
\end{align}
where the factor $(\pT^{\vis})^{2}/\sin^{2} \theta_{\vis}$ corresponds to the Jacobian of the variable transformation 
from $(\pX^{\vis} = \pT^{\vis} \, \cos\phi_{\vis}, \pY^{\vis} = \pT^{\vis} \, \sin\phi_{\vis}, \pZ^{\vis} = \pT^{\vis}/\tan\theta_{\vis})$ 
to $(\pT^{\vis}, \theta_{\vis}, \phi_{\vis})$.

A few words of explanation are in order concerning the treatment of
the mass $m_{\vis}$ of the hadronic system $\tauh$ produced in $\Pgt
\to \textrm{hadrons} + \Pnut$ decays.
Within the scope of this paper, we will assume that $m_{\vis}$ can be
reconstructed with negligible experimental resolution.
The $\tauh$ reconstruction algorithm of the CMS experiment~\cite{TAU-14-001} allows to
reconstruct the mass of $\tauh$. The experimental resolution on $m_{\vis}$ can be taken into
account by adding, to Eq.~(\ref{eq:mem_with_hadRecoil}), a suitable TF $W_{\Phadron}( m_{\vis} | \mhat_{\vis}
)$ and marginalization with respect to the true mass $\mhat_{\vis}$ of the $\tauh$.
In case of the CMS experiment the effect of the resolution on $m_{\vis}$ on the reconstruction of $m_{\Pgt\Pgt}$ is found to be small.
The $\tauh$ reconstruction algorithm used by the ATLAS experiment during LHC Run $1$~\cite{Aad:2014rga} does not
allow to reconstruct the mass of $\tauh$.
In case $m_{\vis}$ cannot be reconstructed, one needs to add, to Eq.~(\ref{eq:mem_with_hadRecoil}),
a marginalization with respect to $m_{\vis}$.

\subsubsection{Leptonic $\Pgt$ decays}
\label{sec:mem_TF_tauToLepDecays}

We assume that the $\pT^{\vis}$, $\eta_{\vis}$, and $\phi_{\vis}$
of the electrons and muons produced in $\Pgt$ decays are measured with negligible experimental resolution.
Correspondingly, we model the TF by a product of three $\delta$-functions:
\begin{equation}
W_{\Plepton}( \bm{p}^{\vis} | \phat^{\vis} ) =  
 \frac{\sin^{2} \theta_{\vis}}{(\pT^{\vis})^{2}} \, 
  \delta( \pT^{\vis} - \pThat^{\vis} ) \, 
  \delta( \theta_{\vis} - \thetahat_{\vis} ) \, 
  \delta( \phi_{\vis} - \phihat_{\vis} ) \, .
\label{eq:tf_tauToLepDecays}
\end{equation}

\subsubsection{Hadronic recoil}
\label{sec:mem_TF_hadRecoil}

We use a two-dimensional normal distribution:
\begin{align}
W_{\rec}( \pX^{\rec},\pY^{\rec} | \pXhat^{\rec},\pYhat^{\rec} ) = & 
  \frac{1}{2\pi \, \sqrt{\vert V \vert}} \, \exp \left( -\frac{1}{2}
  \left( \begin{array}{c} \Delta\pX^{\rec} \\ \Delta\pY^{\rec} \end{array} \right)^{T}
  \cdot V^{-1} \cdot
   \left( \begin{array}{c} \Delta\pX^{\rec} \\ \Delta\pY^{\rec} \end{array} \right)
  \right) \, ,
\label{eq:tf_hadRecoil}
\end{align}
to model the experimental resolution on the momentum, in the transverse plane, 
of the hadronic recoil.
The symbols:
\begin{equation}
\Delta\pX^{\rec} = \pX^{\rec} - \pXhat^{\rec} \quad \mbox{ and } \quad
\Delta\pY^{\rec} = \pY^{\rec} - \pYhat^{\rec} 
\label{eq:tf_hadRecoil_delta}
\end{equation}
refer to the difference between reconstructed and true values of the
momentum in $x$- and $y$-direction,
and the covariance matrix:
\begin{equation}
V = \left( \begin{array}{cc} \sigma_{x}^{2} & \rho \, \sigma_{x} \, \sigma_{y} \\ \rho \, \sigma_{x} \, \sigma_{y} & \sigma_{y}^{2} \end{array} \right) 
\label{eq:tf_hadRecoil_V}
\end{equation}
accounts for the fact that the resolutions $\sigma_{x}$ and $\sigma_{y}$ in $x$- and $y$-direction may be correlated,
with the correlation coefficient denoted by $\rho$.
The symbol $\vert V \vert$ denotes the determinant of $V$.
Two-dimensional normal distributions have been demonstrated to model well the resolution on $\vecMET$ in case of the CMS experiment~\cite{CMS-JME-13-003,CMS-JME-10-009}.
As explained in Section~\ref{sec:mem_hadRecoil}, the resolution on
$\pX^{\rec}$ and $\pY^{\rec}$ is very similar to the resolution on $\METx$ and $\METy$,
motivating the use of TF of the same type for the modelling of the experimental resolution on the hadronic recoil.

For the purpose of this paper,
we assume that the components $\pX^{\rec}$ and $\pY^{\rec}$ are reconstructed with a resolution of $\sigma_{x} = \sigma_{y} = 10$~\GeV each
and that the differences $\Delta\pX^{\rec}$ and $\Delta\pY^{\rec}$ 
between reconstructed and true values of both components are uncorrelated, \ie $\rho = 0$.
In the real experiment, the covariance matrix $V$ is computed on an
event-by-event basis, based on the reconstructed hadronic activity in a given event,
using resolution functions obtained from the Monte Carlo simulation~\cite{CMS-JME-13-003,CMS-JME-10-009}.

%% file: mem_PSintegration.tex
\subsection{Computation of phase space integral}
\label{sec:mem_PSintegration}

The integration over the differential $n$-particle phase space element
$d\Phi_{n}$ in Eq.~(\ref{eq:mem_with_hadRecoil}) needs to be done differently for
events in which both $\Pgt$ leptons decay hadronically (``hadronic'' $\Pgt$ pair decays),
events in which one $\Pgt$ lepton decays hadronically and one $\Pgt$ lepton decays leptonically (``semi-leptonic'' $\Pgt$ pair decays),
and events in which both $\Pgt$ leptons decay leptonically (``leptonic'' $\Pgt$ pair decays).
With the approximation that we treat hadronic $\Pgt$ decays as
two-body decays, the integral
over $d\Phi_{n}$ is of dimension $12$ in case of hadronic $\Pgt$ pair decays,
$15$ in case of semi-leptonic $\Pgt$ pair decays,
and $18$ in case of leptonic $\Pgt$ pair decays.
The differential phase space element reads:
\begin{equation}
d\Phi_{n}= 
 \begin{cases} 
   d\Phi^{(1)}_{\tauhnu} \, d\Phi^{(2)}_{\tauhnu} \, 
 & \mbox{if } \Pgt^{+} \to \textrm{hadrons} + \APnut \mbox{ and } \Pgt^{-} \to \textrm{hadrons} + \Pnut \\
   d\Phi^{(1)}_{\ellnunu} \, d\Phi^{(2)}_{\tauhnu} \, 
 & \mbox{if } \Pgt^{+} \to \ellPlusnunu \mbox{ and } \Pgt^{-} \to \textrm{hadrons} + \Pnut \\
   d\Phi^{(1)}_{\tauhnu} \, d\Phi^{(2)}_{\ellnunu} \, 
 & \mbox{if } \Pgt^{+} \to \textrm{hadrons} + \APnut \mbox{ and } \Pgt^{-} \to \ellMinusnunu \\
   d\Phi^{(1)}_{\ellnunu} \, d\Phi^{(2)}_{\ellnunu} \, 
 & \mbox{if } \Pgt^{+} \to \ellPlusnunu \mbox{ and } \Pgt^{-} \to \ellMinusnunu \, ,
 \end{cases}
\label{eq:PSintegration_taupair}
\end{equation}
where:
\begin{align}
d\Phi^{(i)}_{\tauhnu} = & \frac{d^{3}\bm{\hat{p}}^{\vis(i)}}{(2\pi)^{3} \, 2
  \hat{E}_{\vis(i)}} \, \frac{d^{3}\bm{\hat{p}}^{\Pnu(i)}}{(2\pi)^{3} \, 2 \hat{E}_{\Pnu(i)}} \nonumber \\
d\Phi^{(i)}_{\ellnunu} = &
\frac{d^{3}\bm{\hat{p}}^{\vis(i)}}{(2\pi)^{3} \, 2 \hat{E}_{\vis(i)}} \, 
\frac{d^{3}\bm{\hat{p}}^{\APnu(i)}}{(2\pi)^{3} \, 2 \hat{E}_{\APnu(i)}} \, 
\frac{d^{3}\bm{\hat{p}}^{\Pnu(i)}}{(2\pi)^{3} \, 2 \hat{E}_{\Pnu(i)}} \, .
\label{eq:PSintegration_onetau}
\end{align}

The integrand in Eq.~(\ref{eq:mem_with_hadRecoil}) depends on the
four-momentum of the two $\Pgt$ leptons via the 
product $f(x_{a}) \, f(x_{b})$ of the PDF,
the factor $1/(2 \, x_{a} \, x_{b} \, s)$, referred to as ``flux factor'' in the literature,
and via the squared modulus $\vert \mathcal{M}_{\Pp\Pp \to \PHiggs \to \Pgt\Pgt}(\bm{\tilde{p}},m_{\PHiggs}) \vert^{2}$ 
of the ME for $\PHiggs$ boson production and subsequent decay into a $\Pgt$ lepton pair.
In addition, it depends on the four-momentum of the visible $\Pgt$ decay products via the
squared moduli $\vert \mathcal{M}^{(1)}_{\Pgt\to\cdots}(\bm{\tilde{p}}) \vert^{2}$ and $\vert \mathcal{M}^{(2)}_{\Pgt\to\cdots}(\bm{\tilde{p}}) \vert^{2}$ of the ME for the $\Pgt$ lepton decays
and via the transfer functions $W( \bm{p}^{\vis(1)} | \phat^{\vis(1)} )$ and $W( \bm{p}^{\vis(2)} | \phat^{\vis(2)} )$.
The $\Pgt$ lepton energies and momenta need to be computed as function of the integration variables.

The dimension of the integration over the phase-space elements
$d^{3}\bm{p}^{\Pnu(i)}$ and $d^{3}\bm{p}^{\APnu(i)} \, d^{3}\bm{p}^{\Pnu(i)}$
can be reduced by means of analytic transformations.
Two variables are sufficient to fully parametrize the kinematics of hadronic $\Pgt$ decays.
In case of leptonic $\Pgt$ decays, three variables are sufficient.
We choose to parametrize hadronic $\Pgt$ decays by the variables $z$ and $\phi_{\inv}$.
The variable $z$ represents the fraction of $\Pgt$ lepton energy, in the laboratory frame,
that is carried by the visible $\Pgt$ decay products (\cf Eq.~(\ref{eq:def_z})).
We denote the energy and momentum of the $\Pgt$ neutrino
produced in hadronic $\Pgt$ decays as well as of the neutrino pair produced in leptonic $\Pgt$
decays by the symbols $E_{\inv}$ and $\bm{p}^{\inv}$.
The energy component $E_{\inv}$ is related to the variable $z$ via:
\begin{equation}
E_{\inv} = \frac{1 - z}{z} \, E_{\vis} \, .
\label{eq:E_inv}
\end{equation}
The angle $\theta_{\inv}$ between the $\bm{p}^{\inv}$ vector and the $\bm{p}^{\vis}$ vector
is related to the variable $z$ as well, 
and is given by Eq.~(\ref{eq:hadTauDecaysCosTheta}) in case of hadronic $\Pgt$ decays 
and by Eq.~(\ref{eq:lepTauDecaysCosTheta}) in case of leptonic $\Pgt$ decays.
The variable $\phi_{\inv}$ specifies the orientation of the
$\bm{p}^{\inv}$ vector relative to the direction of the visible $\Pgt$ decay products.
In case of hadronic $\Pgt$ decays, the magnitude of the $\bm{p}^{\inv}$ vector is equal to $E_{\inv}$.
We choose the mass $m_{\inv}$ of the neutrino pair as third variable to parametrize the kinematics of leptonic $\Pgt$ decays,
so that the magnitude of the $\bm{p}^{\inv}$ vector is given by $\sqrt{\left( \frac{1 - z}{z} \right)^{2} \, E_{\vis}^{2} - m^{2}_{\inv}}$.
With the convention that $m_{\inv} = 0$ for hadronic $\Pgt$ decays,
Eqs.~(\ref{eq:hadTauDecaysCosTheta})
and~(\ref{eq:lepTauDecaysCosTheta}) can be expressed by a common form
that is valid for hadronic as well as for leptonic $\Pgt$ decays:
\begin{equation}
\cos\theta_{\inv} = \frac{\frac{1 - z}{z} \, E_{\vis}^{2} - \frac{1}{2}(m^{2}_{\Pgt} - (m^{2}_{\vis} + m^{2}_{\inv}))}{\vert\bm{p}^{\vis}\vert \, 
  \sqrt{\left( \frac{1 - z}{z} \right)^{2} \, E_{\vis}^{2} - m^{2}_{\inv}}} \, .
\label{eq:theta_inv}
\end{equation}
The $\Pgt$ lepton momentum vector is given by the sum of the
$\bm{p}^{\vis}$ and $\bm{p}^{\inv}$ vectors.

\begin{figure}[h]
\begin{center}
\includegraphics*[height=58mm]{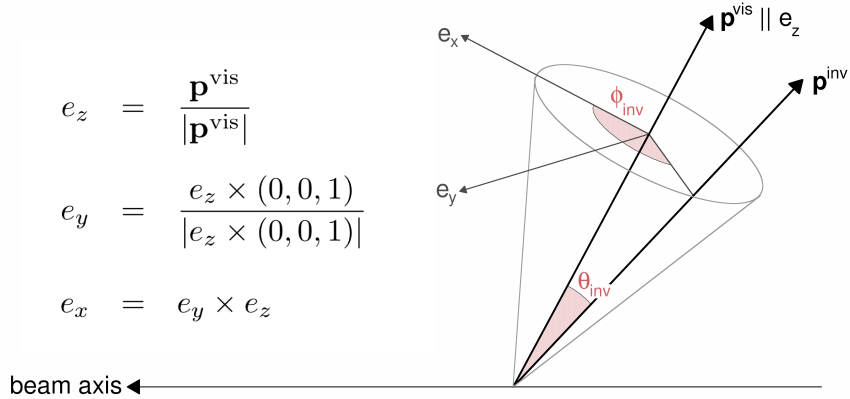}
\end{center}
\caption{
  Illustration of the variables $\theta_{\inv}$ and $\phi_{\inv}$ that specify the orientation of the $\bm{p}^{\inv}$ vector
  relative to the momentum vector $\bm{p}^{\vis}$ of the visible $\Pgt$ decay products.
}
\label{fig:tauDecayParametrization}
\end{figure}

The angles $\theta_{\inv}$ and $\phi_{\inv}$ are illustrated in Fig.~\ref{fig:tauDecayParametrization}.
The $\bm{p}^{\inv}$ vector is located on the surface of a cone,
the axis of which is given by the $\bm{p}^{\vis}$ vector and the
opening angle of which is given by Eq.~(\ref{eq:theta_inv}).
The variable $\phi_{\inv}$ represents the angle of rotation, in
counter-clockwise direction, around the
axis of the cone.
The value $\phi_{\inv} = 0$ is chosen to correspond to the case that
the $\bm{p}^{\inv}$ vector is within the plane spanned by the
$\bm{p}^{\vis}$ vector and the beam direction.

The parametrization of the $\Pgt$ decay kinematics by $\bm{p}^{\vis}$
and the variables $z$ and $\phi_{\inv}$, respectively by $z$, $\phi_{\inv}$, and $m_{\inv}$,
allows one to simplify the evaluation of the integral in Eq.~(\ref{eq:mem_with_hadRecoil}) considerably.
Expressions for the product of the differential phase space elements
$d\Phi^{(i)}_{\tauhnu}$  and $d\Phi^{(i)}_{\ellnunu}$ with the squared
modulus of the ME $\vert \BW_{\Pgt} \vert^{2} \cdot \vert \mathcal{M}^{(i)}_{\Pgt\to\cdots}(\bm{p}) \vert^{2}$ are derived in Sections~\ref{sec:appendix_tauToHadDecays} and~\ref{sec:appendix_tauToLepDecays} of the appendix.
The results are:
\begin{align}
\vert \BW_{\Pgt} \vert^{2} \cdot \vert \mathcal{M}^{(i)}_{\Pgt\to\cdots}(\bm{\tilde{p}}) \vert^{2} \, d\Phi^{(i)}_{\tauhnu} 
 = & \, \frac{\pi}{m_{\Pgt}\Gamma_{\Pgt}} \,
 f_{\Phadron}\left(\bm{\hat{p}}^{\vis(i)}, m^{\vis(i)},
   \bm{\hat{p}}^{\inv(i)}\right) \, \frac{d^{3}\bm{\hat{p}}^{\vis}}{2 \hat{E}_{\vis}} \, dz \, d\phi_{\inv} \nonumber \\
\vert \BW_{\Pgt} \vert^{2} \cdot \vert \mathcal{M}^{(i)}_{\Pgt\to\cdots}(\bm{\tilde{p}}) \vert^{2} \, d\Phi^{(i)}_{\ellnunu} 
 = & \, \frac{\pi}{m_{\Pgt}\Gamma_{\Pgt}} \, f_{\ell}\left(\bm{\hat{p}}^{\vis(i)},
 m^{\vis(i)}, \bm{\hat{p}}^{\inv(i)}\right) \, \frac{d^{3}\bm{\hat{p}}^{\vis}}{2 \hat{E}_{\vis}} \, dz \, dm^{2}_{\inv} \, d\phi_{\inv}
 \, .
\label{eq:PSint}
\end{align}
The functions $f_{\Phadron}$ and $f_{\Plepton}$ are given by
Eqs.~(\ref{eq:hadTauDecays_f})
and~(\ref{eq:lepTauDecays_f}).

Eq.~(\ref{eq:PSint}) represents the quintessence of what is needed 
in order to extend the ME generated by automatized tools such as
CompHEP or MadGraph
by the capability to handle the $\Pgt$ decays.
Instead of performing an integration over $d^{3}\bm{p}^{\Pgt(1)} \,
d^{3}\bm{p}^{\Pgt(2)}$, which treats the $\Pgt$ leptons as stable particles,
one needs to perform the integration over $\vert \BW^{(1)}_{\Pgt} \vert^{2} \cdot \vert
\mathcal{M}^{(1)}_{\Pgt\to\cdots}(\bm{\tilde{p}}) \vert^{2} \, d\Phi^{(1)} \, \vert \BW^{(2)}_{\Pgt} \vert^{2} \cdot \vert
\mathcal{M}^{(2)}_{\Pgt\to\cdots}(\bm{\tilde{p}}) \vert^{2} \, d\Phi^{(2)}$ according to
Eq.~(\ref{eq:PSint}).
The momenta of both $\Pgt$ leptons need to be computed as
function of the integration variables $z_{(1)}$, $\phi_{\inv}^{(1)}$,
$m_{\inv}^{(1)}$ and $z_{(2)}$, $\phi_{\inv}^{(2)}$,
$m_{\inv}^{(2)}$, using Eqs.~(\ref{eq:E_inv})
and~(\ref{eq:theta_inv}),
where $m_{\inv}^{(i)}$ is equal to zero in case the $i$-th $\Pgt$ lepton
decays hadronically.
The $\Pgt$ lepton momenta can then be used to evaluate the product of the PDF, the flux factor, 
and the squared modulus $\vert \mathcal{M}_{\Pp\Pp \to \PHiggs \to \Pgt\Pgt}(\bm{\tilde{p}},m_{\PHiggs}) \vert^{2}$ 
of the ME for $\PHiggs$ boson production and subsequent decay into a $\Pgt$ lepton pair in Eq.~(\ref{eq:mem_with_hadRecoil}).
 
In order to improve the accuracy of the numerical integration,
we perform a further variable transformation, replacing $z_{(2)}$ by the variable $t_{\PHiggs}$, defined below.
The transformation is executed in two steps. 
First, we replace $z_{(2)}$ by:
\begin{equation}
q_{\PHiggs}^{2} = \frac{q^{2}_{\vis}}{z_{(1)} \, z_{(2)}} \, ,
\label{eq:varTransform_z2_to_tHiggs_1}
\end{equation}
with $q_{\vis}$ denoting the mass of the visible decay products of both
$\Pgt$ leptons.
Following Eq.~(8) in Ref.~\cite{Alwall:2010cq}, we then parametrize $q_{\PHiggs}^{2}$ by:
\begin{equation}
q_{\PHiggs}^{2} = m_{\PHiggs}^{2} + m_{\PHiggs} \, \Gamma_{\PHiggs}
\tan t_{\PHiggs} \, .
\label{eq:varTransform_z2_to_tHiggs_2}
\end{equation}
The form of the variable transformation in Eqs.~(\ref{eq:varTransform_z2_to_tHiggs_1}) and~(\ref{eq:varTransform_z2_to_tHiggs_2}) 
is chosen such that the Jacobi factor of the transformation from
$z_{(2)}$ to $t_{\PHiggs}$ is proportional to the inverse of $\vert \BW_{\PHiggs} \vert^{2}$, 
the squared modulus of the Breit-Wigner propagator of the $\PHiggs$ boson in
Eq.~(\ref{eq:meHiggsBreitWigner}).
The Jacobi factor is given by:
\begin{equation}
\left\lvert \frac{\partial z_{(2)}}{\partial q_{\PHiggs}^{2}} \cdot \frac{\partial
  q_{\PHiggs}^{2}}{\partial t_{\PHiggs}} \right\rvert =
\frac{q^{2}_{\vis}}{q_{\PHiggs}^{4} \, z_{(1)}} \cdot \frac{(q_{\PHiggs}^{2}
  - m_{\PHiggs}^{2})^{2} + m_{\PHiggs}^{2} \,
  \Gamma_{\PHiggs}^{2}}{m_{\PHiggs} \, \Gamma_{\PHiggs}} \, .
\label{eq:JacobiFactor_z2_to_tHiggs}  
\end{equation}
Compared to Ref.~\cite{Alwall:2010cq} we differ by a factor $\frac{1}{\pi}$ in the derivative 
$\frac{\partial q_{\PHiggs}^{2}}{\partial t_{\PHiggs}}$. We have verified that Eq.~(\ref{eq:JacobiFactor_z2_to_tHiggs}) is correct.
This transformation improves the numerical precision of evaluating the integral 
in Eq.~(\ref{eq:mem_with_hadRecoil}) by reducing the variance of the integrand.

%% file: mem_xSection.tex
\subsection{Computation of $\sigma'(m_{\PHiggs})$}
\label{sec:mem_xSection}

According to the paradigm of the ME method, the normalization factor
$1/\sigma'(m_{\PHiggs})$ in Eq.~(\ref{eq:mem_with_hadRecoil}) is to be computed
by evaluating the integral:
\begin{align}
\sigma'(m_{\PHiggs}) = &
  \frac{32\pi^{4}}{s} \, \int \, d^{3}\bm{p}^{\vis(1)} \, d^{3}\bm{p}^{\vis(2)} \, d\pX^{\rec} \, d\pY^{\rec} \, 
  d\Phi_{n} \, \frac{f(x_{a}) f(x_{b})}{2 \, x_{a} \, x_{b} \, s} \cdot \hspace{2cm} \nonumber \\
& \qquad \vert \mathcal{M}_{\Pp\Pp \to \PHiggs \to \Pgt\Pgt}(\bm{\tilde{p}},m_{\PHiggs}) \vert^{2} 
  \cdot \vert \BW^{(1)}_{\Pgt} \vert^{2} \cdot \vert \mathcal{M}^{(1)}_{\Pgt\to\cdots}(\bm{\tilde{p}}) \vert^{2} 
  \cdot \vert \BW^{(2)}_{\Pgt} \vert^{2} \cdot \vert
  \mathcal{M}^{(2)}_{\Pgt\to\cdots}(\bm{\tilde{p}}) \vert^{2} \cdot \nonumber \\
& \qquad W(\bm{p}^{\vis(1)}|\bm{\hat{p}}^{\vis(1)}) \, W(\bm{p}^{\vis(2)}|\bm{\hat{p}}^{\vis(2)}) \, W_{\rec}( \pX^{\rec},\pY^{\rec} | \pXhat^{\rec},\pYhat^{\rec} ) \, .
\label{eq:xSection}
\end{align}
Of the factors in the integrand of Eq.~(\ref{eq:xSection}),
only the TF depend on the measured momenta $\bm{p}^{\vis(1)}$ and $\bm{p}^{\vis(2)}$ of the visible decay products of the two $\Pgt$ leptons
and on the measured momentum components $\pX^{\rec}$ and $\pY^{\rec}$ of the hadronic recoil.
All other factors depend solely on the true values of the momenta.
According to the TF normalization conditions
\begin{align}
& \int \, d^{3}\bm{p}^{\vis(1)} \, W(\bm{p}^{\vis(1)}|\bm{\hat{p}}^{\vis(1)}) = 1 \nonumber \\
& \int \, d^{3}\bm{p}^{\vis(2)} \, W(\bm{p}^{\vis(2)}|\bm{\hat{p}}^{\vis(2)}) = 1 \nonumber \\
& \int \, d\pX^{\rec} \, d\pY^{\rec} \, W_{\rec}( \pX^{\rec},\pY^{\rec} | \pXhat^{\rec},\pYhat^{\rec} ) = 1 \, ,
\end{align}
so that Eq.~(\ref{eq:xSection}) becomes:
\begin{align}
\sigma'(m_{\PHiggs}) = &
  \frac{32\pi^{4}}{s} \, \int \, 
  d\Phi_{n} \, \frac{f(x_{a}) f(x_{b})}{2 \, x_{a} \, x_{b} \, s} \,
  \hspace{2cm} \nonumber \\
& \qquad \vert \mathcal{M}_{\Pp\Pp \to \PHiggs \to
    \Pgt\Pgt}(\bm{\tilde{p}},m_{\PHiggs}) \vert^{2} 
  \cdot \vert \BW^{(1)}_{\Pgt} \vert^{2} \cdot \vert \mathcal{M}^{(1)}_{\Pgt\to\cdots}(\bm{\tilde{p}}) \vert^{2} 
  \cdot \vert \BW^{(1)}_{\Pgt} \vert^{2} \cdot \vert \mathcal{M}^{(2)}_{\Pgt\to\cdots}(\bm{\tilde{p}}) \vert^{2} \, .
\label{eq:xSection2}
\end{align}

The integral in Eq.~(\ref{eq:xSection2}) is computed numerically,
for $\PHiggs$ boson masses $m_{\PHiggs}$ ranging from $50$ to $5000$~\GeV in steps of $1$~\GeV.
The numeric integration is performed using the VAMP algorithm~\cite{VAMP},
an improved implementation of the VEGAS algorithm~\cite{VEGAS}.
The result is used for the purpose of normalizing the probability density $\mathcal{P}(\bm{p}^{\vis(1)},\bm{p}^{\vis(2)};\pX^{\rec},\pY^{\rec}|m_{\PHiggs})$
in Eq.~(\ref{eq:mem_with_hadRecoil}).

\begin{figure}
\begin{center}
\includegraphics*[height=74mm]{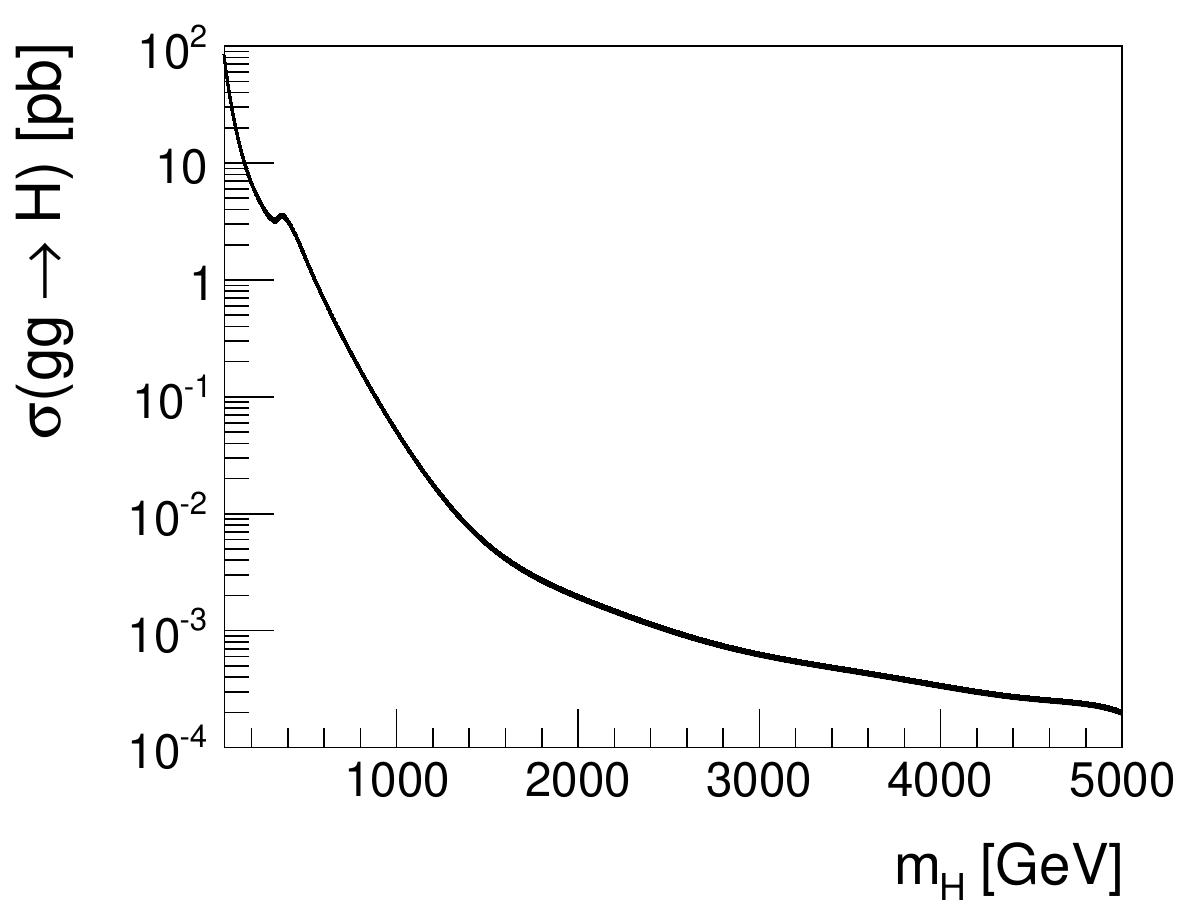}
\end{center}
\caption{
  Cross section $\sigma(m_{\PHiggs})$ as function of the $\PHiggs$ boson mass $m_{\PHiggs}$,
  computed for proton-proton collisions at $\sqrt{s} = 13$~\TeV centre-of-mass energy.
}
\label{fig:xSection}
\end{figure}

The cross sections $\sigma'(m_{\PHiggs})$ computed in this way cannot 
be directly compared to literature values,
due to the fact that we are applying the LO ME to events in which the $\PHiggs$ boson has non-zero $\pT$.
For the purpose of comparing $\sigma(m_{\PHiggs})$ to literature
values for the LO $\Pg\Pg \to \PHiggs$ cross section,
we compute the integral in Eq.~(\ref{eq:xSection2}) for the case that the $\PHiggs$ boson has zero $\pT$,
by inserting two $\delta$-functions, $\delta(\pX^{\Pgt(1)} + \pX^{\Pgt(2)})$ and $\delta(\pY^{\Pgt(1)} + \pY^{\Pgt(2)})$, into the integrand.
We then use the $\delta$-functions to remove two integration variables
analytically before evaluating the integral numerically.
The values of $\sigma(m_{\PHiggs})$ obtained in this way are shown as function of
$m_{\PHiggs}$ in Fig.~\ref{fig:xSection}.
The values agree with the literature values for the LO $\Pg\Pg \to
\PHiggs$ cross section within $\approx 10\%$. 
The level of agreement is sufficient for our purposes.

%% file: mem_numericalMaximization.tex
\subsection{Determination of $m_{\Pgt\Pgt}$}
\label{sec:mem_numericalMaximization}

The best estimate, $m_{\Pgt\Pgt}$, for the mass of the $\Pgt$ lepton pair in a given event
is obtained by computing the probability density $\mathcal{P}(\bm{p}^{\vis(1)},\bm{p}^{\vis(2)};\pX^{\rec},\pY^{\rec}|m_{\PHiggs}^{\textrm{test}(i)})$ 
for a series of mass hypotheses $m_{\PHiggs}^{\textrm{test}(i)}$, using Eq.~(\ref{eq:mem_with_hadRecoil}), and determining the value of $m_{\PHiggs}$ that maximizes this probability density.

\begin{figure}
\setlength{\unitlength}{1mm}
\begin{center}
\begin{picture}(160,68)(0,0)
\put(-2.5, 0.0){\mbox{\includegraphics*[height=68mm]{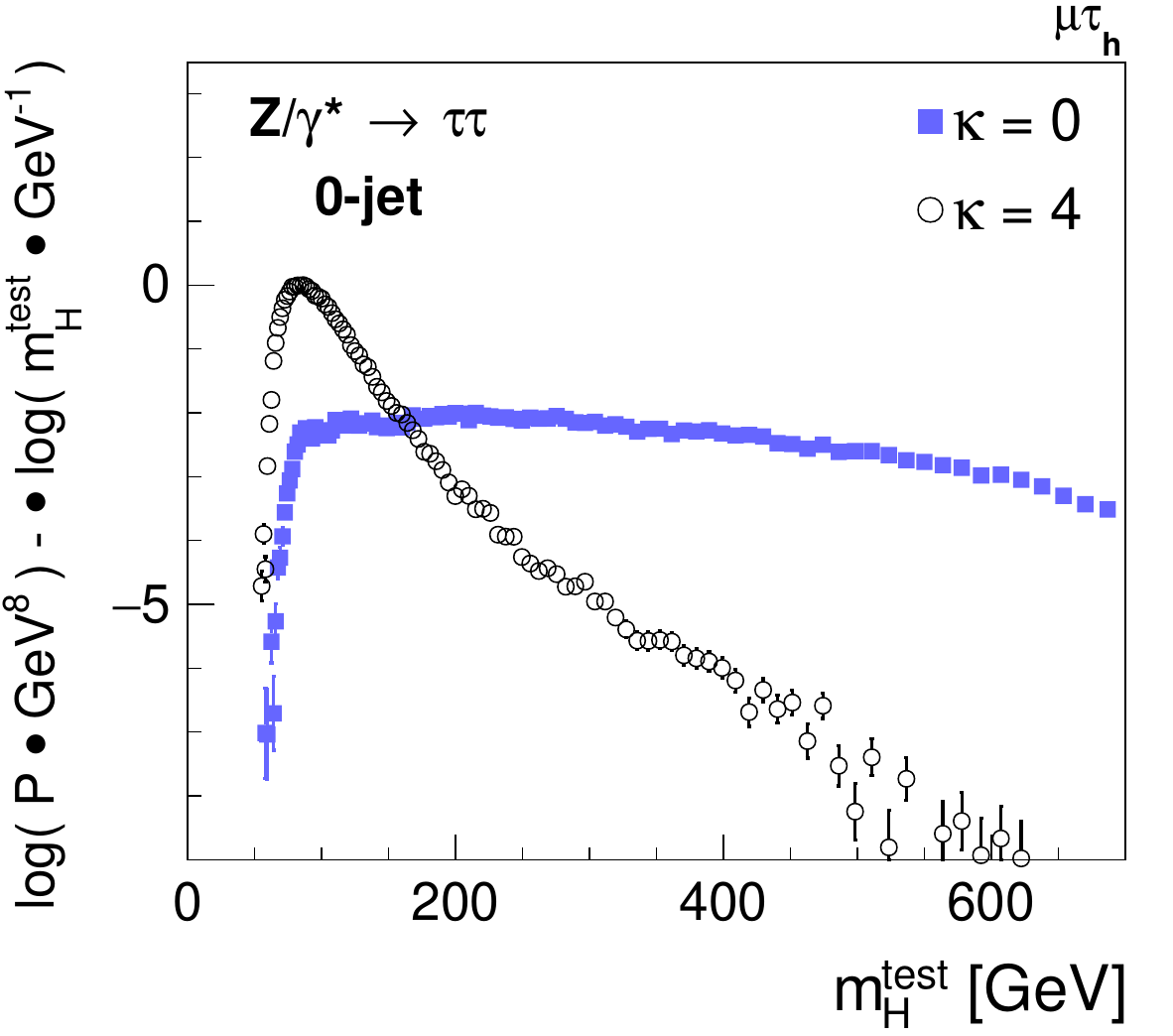}}}
\put(81.0, 0.0){\mbox{\includegraphics*[height=68mm]{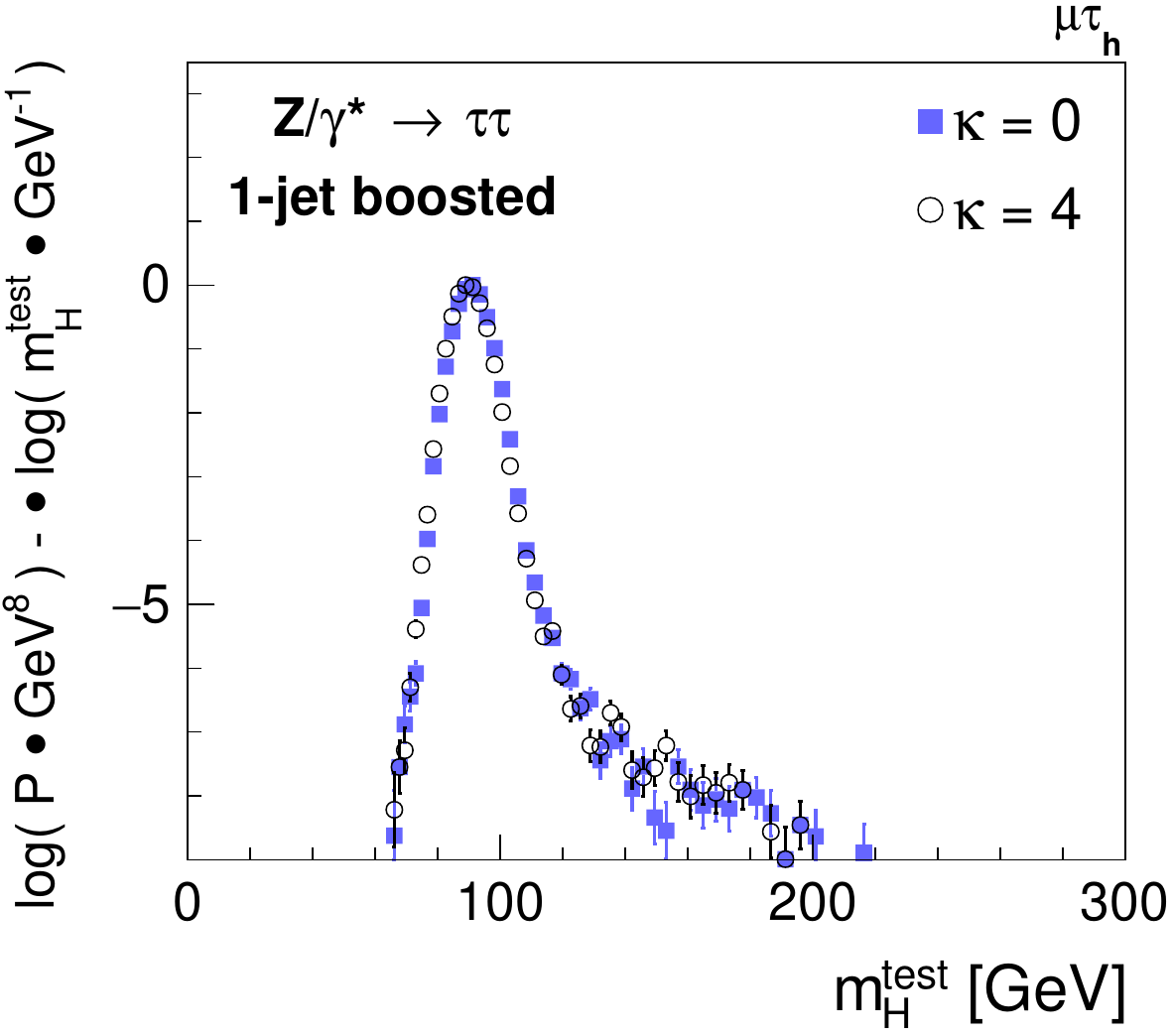}}}
\end{picture}
\end{center}
\caption{
  Graphs of the probability density $\mathcal{P}(\bm{p}^{\vis(1)},\bm{p}^{\vis(2)};\pX^{\rec},\pY^{\rec}|m_{\PHiggs}^{\textrm{test}(i)})$, 
  computed for a series of mass hypotheses $m_{\PHiggs}^{\textrm{test}(i)}$, in two exemplary simulated $\PZ/\Pggx \to \Pgt\Pgt$ background events.
  In the event shown on the left (``$0$-jet'') the $\PZ$ boson has little $\pT$, while in the event shown on the right (``$1$-jet boosted'') the $\PZ$ boson recoils against a high $\pT$ jet.
  The scale of the ordinate is adjusted such that the probability density is equal to one
  for the value of $m_{\PHiggs}^{\textrm{test}(i)}$ that maximises $\mathcal{P}(\bm{p}^{\vis(1)},\bm{p}^{\vis(2)};\pX^{\rec},\pY^{\rec}|m_{\PHiggs}^{\textrm{test}(i)})$.
}
\label{fig:likelihoodGraphs}
\end{figure}

The integral in Eq.~(\ref{eq:mem_with_hadRecoil}) is evaluated numerically using the VAMP algorithm.
For each mass hypotheses $m_{\PHiggs}^{\textrm{test}(i)}$ the integrand is evaluated $20\,000$ times.
The series of mass hypotheses is defined by a recursive relation: 
\begin{equation}
m_{\PHiggs}^{\textrm{test}(i + 1)} = (1 + \delta) \,  m_{\PHiggs}^{\textrm{test}(i)} \, \mbox{ with } \, m_{\PHiggs}^{\textrm{test}(0)} = m_{\vis} \, ,
\label{eq:mTauTau_step_size}
\end{equation}
where $m_{\vis}$ denotes the mass of the visible $\Pgt$ decay products.
The step size $\delta$ is chosen such that it is small compared to the
expected resolution on $m_{\PHiggs}$,
which typically amounts to $15$--$20\%$ relative to the true mass of the $\Pgt$ lepton pair (\cf Section~\ref{sec:performance}).

In order to reduce the computing time, the series is computed in two passes.
The purpose of the first pass, which uses a step size $\delta = 0.10$, is to find an approximate value of $m_{\Pgt\Pgt}$
that maximizes the probability density $\mathcal{P}$.
The series of mass hypotheses that is used in the first pass stops when $\mathcal{P}$ falls below one per mille 
of the maximal probability density $\mathcal{P}^{\textrm{max}}$
computed for any $m_{\PHiggs}^{\textrm{test}(i)}$ so far in a given event.
In the second pass, which uses a step size of $\delta = 0.01$,
further $\mathcal{P}$ values 
are computed for mass hypotheses $m_{\PHiggs}^{\textrm{test}(i)}$ that
are within a region around the maximum
for which the probability density computed in the first pass exceeds
$0.10 \cdot \mathcal{P}^{\textrm{max}}$.

Finally, the graph of $\log(\mathcal{P} \, \cdot \, \mbox{\GeV}^{8})$ 
versus $m_{\PHiggs}^{\textrm{test}(i)}$ is fitted by a second order polynomial
in the region around the maximum,
and $m_{\Pgt\Pgt}$ is taken to be the point at which the polynomial reaches its maximum.

Graphs of the probability density $\mathcal{P}(\bm{p}^{\vis(1)},\bm{p}^{\vis(2)};\pX^{\rec},\pY^{\rec}|m_{\PHiggs}^{\textrm{test}(i)})$ 
as function of $m_{\PHiggs}^{\textrm{test}(i)}$ are shown in Fig.~\ref{fig:likelihoodGraphs} for two exemplary events.
The events are drawn from a simulated sample of $\PZ/\Pggx \to \Pgt\Pgt$ background events, produced as described in Section~\ref{sec:performance}.
In the first event (``$0$-jet'') the $\PZ$ boson has little $\pT$, while in the second event (``$1$-jet boosted'') the $\PZ$ boson recoils against a high $\pT$ jet.
The scale of the ordinate is adjusted such that the probability density is equal to one
for the value of $m_{\PHiggs}^{\textrm{test}(i)}$ that maximises $\mathcal{P}$.
The width of the graph reflects the experimental resolution on $m_{\Pgt\Pgt}$.
As will be explained in more detail in Section~\ref{sec:performance}, the resolution on $m_{\Pgt\Pgt}$ improves considerably in case the $\PZ$ boson has high $\pT$.
The graphs of the probability density are superimposed for two cases:
for the case that the artificial regularization term described in Section~\ref{sec:mem_logM} is used and for the case that it is not used.
In the $\PZ/\Pggx \to \Pgt\Pgt$ background event in which the $\PZ$ boson has little $\pT$
the probability density decreases slowly as function of the mass hypothesis in case no artificial regularization term is used
with the effect that such events have a sizeable probability for the reconstructed $m_{\Pgt\Pgt}$ to populate, due to resolution effects, 
the region in which the Higgs boson signal is searched for.

%% file: mem_logM.tex
\subsection{Artificial regularization term}
\label{sec:mem_logM}

The SM $\PHiggs \to \Pgt\Pgt$ signal is produced at a rate about three orders of magnitude smaller 
than the irreducible $\PZ/\Pggx \to \Pgt\Pgt$ background.
As illustrated in Fig.~\ref{fig:xSection},
the production rate for hypothetical heavy resonances, such as heavy pseudoscalar Higgs bosons or heavy spin $1$ resonances,
typically decreases steeply with mass.
Physics analyses at the LHC will soon start to probe signal cross sections of order $1$~fb.
In order to maintain high sensitivity for the SM $\PHiggs \to \Pgt\Pgt$ signal as well as for hypothetical heavy resonances
it is imperative to reduce, as much as possible, high mass tails in the $m_{\Pgt\Pgt}$ distribution reconstructed in $\PZ/\Pggx \to \Pgt\Pgt$ background events,
as tails that are on or below the per mille level may yet compete with or even dwarf potential signals.

As exemplified in Fig.~\ref{fig:likelihoodGraphs},
high mass tails in the $m_{\Pgt\Pgt}$ distribution for the $\PZ/\Pggx \to \Pgt\Pgt$ background predominantly arise from events in which the $\PZ$ boson has little $\pT$
and the probability density $\mathcal{P}$ 
decreases slowly as function of $m_{\PHiggs}^{\textrm{test}(i)}$.

Penalized maximum likelihood (ML) estimation is an established method for circumventing problems 
that arise in the stability of parameter estimates in case the likelihood function is relatively flat~\cite{penalizedMaximumLikelihood1}.
Instead of maximising the log-likelihood function $\log \mathcal{L}(\Theta|\bm{y})$,
the penalized ML method finds the best estimate $\hat{\Theta}$ for the unknown model parameter $\Theta$ by maximising the sum $\log \mathcal{L}(\Theta|\bm{y}) + r(\Theta)$.
The penalty function $r(\Theta)$ is added to the log-likelihood function in order to pull the best estimate $\hat{\Theta}$ for the parameter $\Theta$
towards a value that has some rationale as good guess for $\Theta$ in information outside of the likelihood function.
The penalized ML approach can be viewed as a method for introducing some tolerable degree of bias 
in exchange for a reduction in the sampling variability of parameter estimates~\cite{penalizedMaximumLikelihood2}.

From a Bayesian perspective, the penalty function $r(\Theta)$ can be interpreted as prior distribution that describes the information that one has on the parameter $\Theta$
outside of any information conveyed by the measured observables $\bm{y}$.

In the context of the SVfit algorithm, we choose a penalty function $r(m_{\PHiggs}^{\textrm{test}(i)})$
which gives preference to low values of $m_{\Pgt\Pgt}$ 
in case the probability density $\mathcal{P}$ 
is relatively flat.
Recall that the cross section for producing resonances decaying into $\Pgt$ lepton pairs 
decreases steeply with the mass of the resonance (\cf Fig.~\ref{fig:xSection}).
The information that large mass values are less likely is not contained within the probability density defined by Eq.~(\ref{eq:mem_with_hadRecoil})
and motivates the use of a penalized ML approach with a penalty function $r(m_{\PHiggs}^{\textrm{test}(i)})$ 
that increases (decreases) for small (large) $m_{\PHiggs}^{\textrm{test}(i)}$.

In previous applications of the SVfit algorithm for data analyses performed by the CMS collaboration,
regularization functions of the following type were considered:
\begin{equation} 
r(m_{\PHiggs}^{\textrm{test}(i)}) = -\kappa \cdot \log(m_{\PHiggs}^{\textrm{test}(i)} \cdot \mbox{~\GeV}^{-1}) \, .
\label{eq:logM}
\end{equation}
We will focus on this type of regularization functions in this paper,
but remark that from a Bayesian perspective a well motivated alternative choice 
would be to use as penalty function the logarithm of the cross section as function of mass.

Our choice of the parameter $\kappa$ is performed with the objective of achieving an
optimal compromise between reducing the high mass tail in the $m_{\Pgt\Pgt}$ distribution reconstructed for 
$\PZ/\Pggx \to \Pgt\Pgt$ background events on the one hand and 
causing no or at most a small bias on the $m_{\Pgt\Pgt}$ distribution
reconstructed in signal events on the other hand.
We find that the optimal value of $\kappa$ depends on the experimental
resolution on the $\pT$ of $\tauh$ and on $\vecMET$ and hence needs to be adjusted to the experimental conditions.
Higher (lower) experimental resolution favors a small (large) value of $\kappa$. 
The optimal value of $\kappa$ may furthermore differ 
for events in which both $\Pgt$ leptons decay hadronically,
events in which one $\Pgt$ lepton decays into hadrons and the other into an electron or muon,
and events in which both $\Pgt$ leptons decay into electrons or muons,
with larger (smaller) values of $\kappa$ being favored in case both (none) of the $\Pgt$ leptons decay hadronically.

\begin{figure}
\setlength{\unitlength}{1mm}
\begin{center}
\begin{picture}(160,72)(0,0)
\put(-3.5, 0.0){\mbox{\includegraphics*[height=72mm]{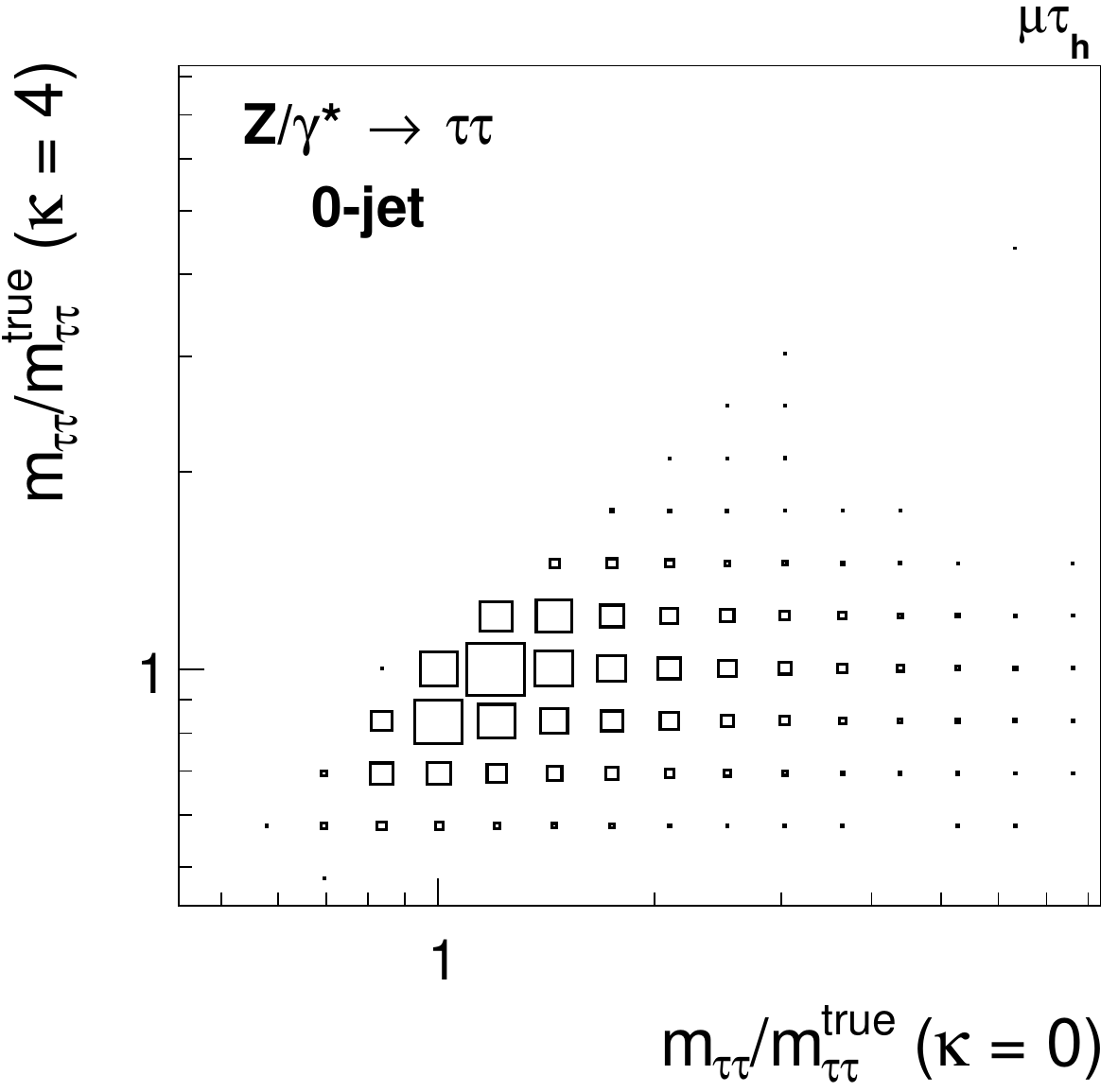}}}
\put(82.0, 0.0){\mbox{\includegraphics*[height=72mm]{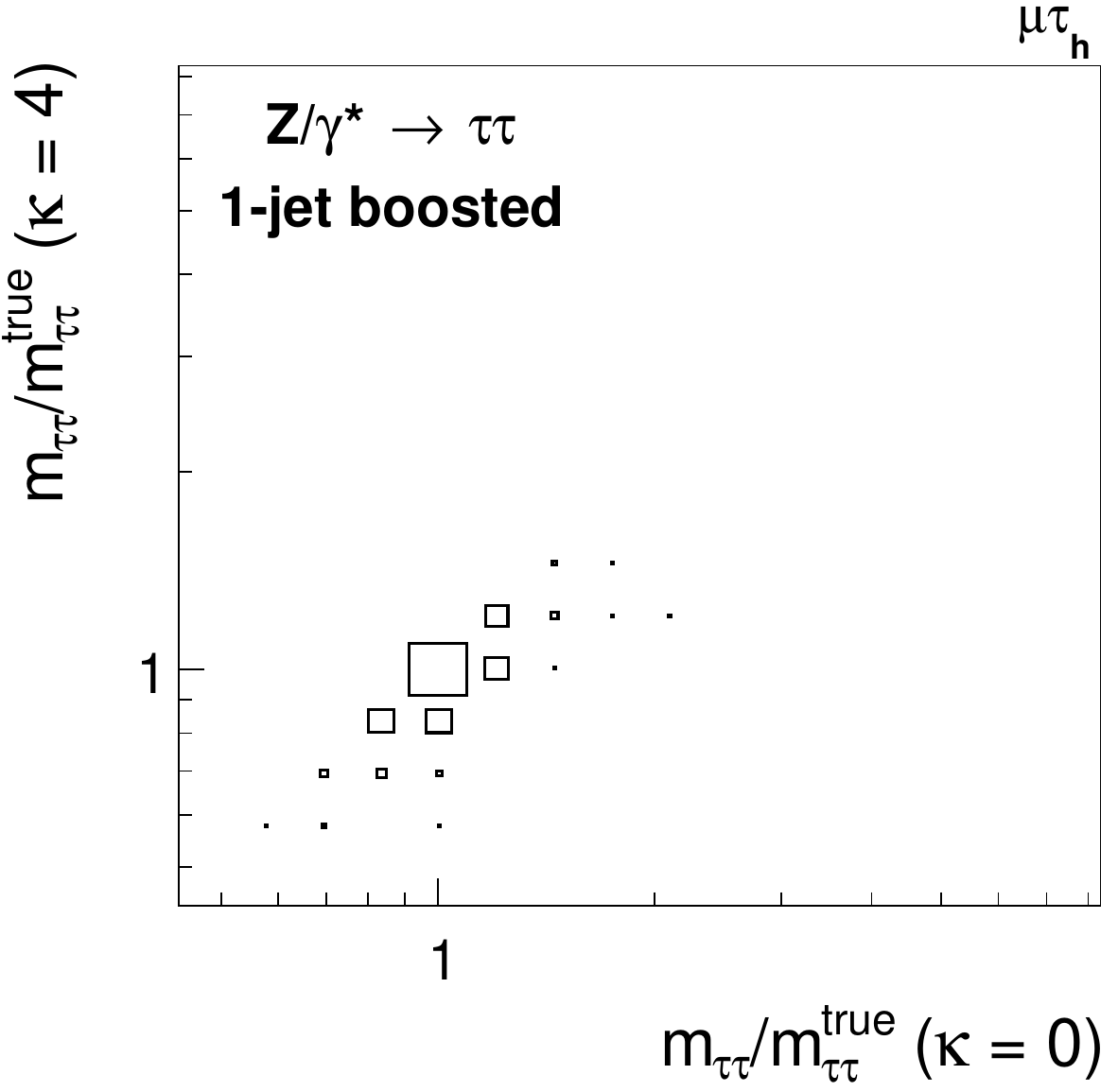}}}
\end{picture}
\end{center}
\caption{
  Correlation between $m_{\Pgt\Pgt}$ values reconstructed in simulated $\PZ/\Pggx \to \Pgt\Pgt$ background events
  in case the artificial regularization term defined by Eq.~(\ref{eq:logM}) is used (ordinate) respectively not used (axis of abscissae).
  The correlation is shown separately for events in which the the $\PZ$ boson has little $\pT$ (left) and for events in which the $\PZ$ boson recoils against a high $\pT$ jet (right).
}
\label{fig:logMscatterPlots}
\end{figure}

The effect of adding a regularization term of the kind 
$r(m_{\PHiggs}^{\textrm{test}(i)}) = -\kappa \cdot \log(m_{\PHiggs}^{\textrm{test}(i)} \cdot \mbox{~\GeV}^{-1})$ 
to the probability density $\mathcal{P}$ given by Eq.~(\ref{eq:mem_with_hadRecoil})
on the distribution in $m_{\Pgt\Pgt}$ reconstructed in $\PZ/\Pggx \to \Pgt\Pgt$ background events
is visualized in Fig.~\ref{fig:logMscatterPlots}.
The effect on events in which the $\PZ$ boson recoils against a high $\pT$ jet, for which the probability density typically exhibits a narrow maximum, is small, 
while the addition of the regularization term effectively reduces the high mass tail in the $m_{\Pgt\Pgt}$ distribution reconstructed in
events in which the $\PZ$ boson has little $\pT$,
for which the probability density decreases more slowly as function of $m_{\PHiggs}^{\textrm{test}(i)}$ (\cf Fig.~\ref{fig:likelihoodGraphs}).

%% file: classicSVfit.tex
\section{``Classic'' SVfit algorithm}
\label{sec:classicSVfit}

A variant of the SVfit algorithm without proper normalization of the likelihood function is maintained for analyses of LHC Run $2$ data.
The algorithm differs from the SVfitMEM algorithm, described in Section~\ref{sec:mem}, in two respects,
altering the integrand in Eq.~(\ref{eq:mem}) as well as the procedure for computing the integral.
The indicator function $\Omega(\bm{y})$, the PDF factor $f(x_{a}) \, f(x_{b})$, the flux factor $\frac{1}{2 \, x_{a} \, x_{b} \, s}$,
and the efficiency factor $\epsilon(\bm{p},\theta)$ of Eq.~(\ref{eq:mem}) are omitted,
as is the normalization to inclusive cross section times acceptance, given by the factor $\frac{1}{\sigma(\theta) \, \mathcal{A}(\theta)}$.
Instead of the complete ME $\vert \mathcal{M}(\bm{p},m_{\PHiggs}) \vert^{2}$ given by Eq.~(\ref{eq:meFactorization}),
only the terms 
$\vert \BW_{\Pgt}^{(1)} \vert^{2} \cdot \vert \mathcal{M}^{(1)}_{\Pgt\to\cdots}(\bm{p}) \vert^{2} \cdot \vert \BW_{\Pgt}^{(2)} \vert^{2} \cdot \vert \mathcal{M}^{(2)}_{\Pgt\to\cdots}(\bm{p}) \vert^{2}$
for the decay of the $\Pgt$ leptons are retained.
The equivalent to Eq.~(\ref{eq:mem_with_hadRecoil}) reads:
\begin{align}
&
\mathcal{P}(\bm{p}^{\vis(1)},\bm{p}^{\vis(2)};\pX^{\rec},\pY^{\rec}|m_{\PHiggs})
= \frac{32\pi^{4}}{s} \, \int \, d\Phi_{n} \, \cdot \hspace{2cm} \nonumber \\
& \qquad \vert \BW^{(1)}_{\Pgt} \vert^{2} \cdot \vert \mathcal{M}^{(1)}_{\Pgt\to\cdots}(\bm{\tilde{p}}) \vert^{2} 
 \cdot \vert \BW^{(2)}_{\Pgt} \vert^{2} \cdot \vert \mathcal{M}^{(2)}_{\Pgt\to\cdots}(\bm{\tilde{p}}) \vert^{2} \cdot \nonumber \\
& \qquad W(\bm{p}^{\vis(1)}|\bm{\hat{p}}^{\vis(1)}) \, W(\bm{p}^{\vis(2)}|\bm{\hat{p}}^{\vis(2)}) \, W_{\rec}( \pX^{\rec},\pY^{\rec} | \pXhat^{\rec},\pYhat^{\rec} ) \, .
\label{eq:likelihood_with_hadRecoil}
\end{align}
Instead of computing probability densities $\mathcal{P}(\bm{p}^{\vis(1)},\bm{p}^{\vis(2)};\pX^{\rec},\pY^{\rec}|m_{\PHiggs})$ for a series of mass hypotheses $m_{\PHiggs}^{\textrm{test}(i)}$ 
and determining the value of $m_{\PHiggs}$ that maximizes the probability density,
the following integral is computed:
\begin{align}
& \mathcal{L}(\bm{p}^{\vis(1)},\bm{p}^{\vis(2)};\pX^{\rec},\pY^{\rec}) 
= \frac{32\pi^{4}}{s} \, \int \, dm_{\PHiggs} \, d\Phi_{n} \, \cdot \hspace{2cm} \nonumber \\
& \qquad \vert \BW^{(1)}_{\Pgt} \vert^{2} \cdot \vert \mathcal{M}^{(1)}_{\Pgt\to\cdots}(\bm{\tilde{p}}) \vert^{2} 
 \cdot \vert \BW^{(2)}_{\Pgt} \vert^{2} \cdot \vert \mathcal{M}^{(2)}_{\Pgt\to\cdots}(\bm{\tilde{p}}) \vert^{2} \cdot \nonumber \\
& \qquad W(\bm{p}^{\vis(1)}|\bm{\hat{p}}^{\vis(1)}) \, W(\bm{p}^{\vis(2)}|\bm{\hat{p}}^{\vis(2)}) \, W_{\rec}( \pX^{\rec},\pY^{\rec} | \pXhat^{\rec},\pYhat^{\rec} ) \cdot \mathcal{F}(\bm{p}) \, .
\label{eq:cSVfit_with_hadRecoil}
\end{align}
The function $\mathcal{F}(\bm{p})$ in the integrand may be an arbitrary function of the momenta $\bm{p}^{(1)}$ and $\bm{p}^{(2)}$ of the two $\Pgt$ leptons.
The integral is evaluated numerically, using a custom implementation of the Markov chain Monte Carlo integration method with the Metropolis--Hastings algorithm~\cite{Metropolis_Hastings}.
The actual value $\mathcal{L}(\bm{y})$ of the integral is irrelevant.
The reconstruction of the mass $m_{\Pgt\Pgt}$ of the $\Pgt$ lepton pair is based on choosing 
$\mathcal{F}(\bm{p}) \equiv (\Ehat_{\Pgt(1)} + \Ehat_{\Pgt(2)})^{2} 
 - \left( (\pXhat^{\Pgt(1)} + \pXhat^{\Pgt(2)})^{2} + (\pYhat^{\Pgt(1)} + \pYhat^{\Pgt(2)})^{2} + (\pZhat^{\Pgt(1)} + \pZhat^{\Pgt(2)})^{2} \right)$,
recording the values of $\mathcal{F}(\bm{p})$ for each evaluation of the integrand in Eq.~(\ref{eq:cSVfit_with_hadRecoil}) by the Markov chain
and taking the median of the series of $\mathcal{F}(\bm{p})$ values
as the best estimate $m_{\Pgt\Pgt}$ for the mass of the $\Pgt$ lepton pair in a given event.
The total number of evaluations of the integrand, referred to as Markov chain ``states'',  
amounts to $100\,000$ per event. The first $10\,000$ evaluations of the integrand are used as ``burn-in'' period and are excluded from the computation of the median.
The transitions between subsequent states of the Markov chain are computed for the case that $\mathcal{F}(\bm{p}) \equiv 1$.
This choice has the advantage that the sequence of Markov chain states does not depend on $\mathcal{F}(\bm{p})$,
which allows for $\mathcal{F}(\bm{p})$ to consist of multiple components that can be evaluated by the same Markov chain.
Each component may be an arbitrary function of the momenta $\bm{p}^{(1)}$ and $\bm{p}^{(2)}$ of the two $\Pgt$ leptons.
In the default implementation of the algorithm,
the $\pT$, $\eta$, $\phi$ and transverse mass, $m_{T\Pgt\Pgt} = (\ET^{\Pgt(1)} + \ET^{\Pgt(2)})^{2} 
 - \left( (\pX^{\Pgt(1)} + \pX^{\Pgt(2)})^{2} + (\pY^{\Pgt(1)} + \pY^{\Pgt(2)})^{2} \right)$, of the $\Pgt$ lepton pair
are reconstructed in addition to the mass $m_{\Pgt\Pgt}$.
We refer to this version of the SVfit algorithm as the ``classic'' SVfit (cSVfit) algorithm.

The cSVfit algorithm covers two use cases:
data analyses use it either because of its capability to reconstruct kinematic observables of the $\Pgt$ lepton pair other than the mass $m_{\Pgt\Pgt}$
or because of its significantly reduced requirement on computing resources compared to the SVfitMEM algorithm.
The $\pT$ of the $\PHiggs$ boson was used for the purpose of categorizing events in the SM $\PHiggs \to \Pgt\Pgt$ analysis 
performed by the CMS collaboration during LHC Run $1$~\cite{HIG-13-004}.
The $\pT$ was reconstructed by computing the vectorial sum of the momenta $\bm{p}^{\vis}$ of the visible $\Pgt$ decay products and of the missing transverse momentum $\vecMET$.
We will demonstrate in Section~\ref{sec:performance} that reconstructing the $\pT$ of the $\PHiggs$ boson candidate by the cSVfit algorithm
improves the resolution compared to taking the sum of the $\bm{p}^{\vis}$ and $\vecMET$ vectors.
The transverse mass $m_{T\Pgt\Pgt}$ of the $\PHiggs$ boson candidate has been used as observable to discriminate signal from background
in the CMS search for heavy $\PHiggs$ bosons in the first LHC Run $2$ data~\cite{HIG-16-006}, 
performed in the context of the minimal supersymmetric extension of the Standard Model (MSSM)~\cite{Fayet:1974pd,Fayet:1977yc}.

Compared to the cSVfit algorithm,
the version of the SVfit algorithm used for analyses of data recorded by the CMS experiment during LHC Run $1$
used an incomplete expression for the product of the differential phase space element and for the ME modelling the $\Pgt$ lepton decays.
In particular the factor $\frac{1}{z^{2}}$ in the functions $f_{\Phadron}$ and $f_{\Plepton}$, 
given by Eqs.~(\ref{eq:hadTauDecays_f}) and~(\ref{eq:lepTauDecays_f}), were missing by mistake.
The effect of the missing factors $\frac{1}{z^{2}}$ 
is equivalent to adding an artificial regularization term of the type described in Section~\ref{sec:mem_logM} with $\kappa = 4$ to the likelihood function.
This can be seen in the limit that the angles $\theta_{\inv}$ between the $\bm{p}^{\inv}$ and $\bm{p}^{\vis}$ vectors are zero for both $\Pgt$ leptons:
In this limit, $m_{\PHiggs}^{\textrm{test}} \approx \frac{m_{\vis}}{\sqrt{z_{(1)}z_{(2)}}}$,
with $m_{\vis}$ denoting the measured mass of the visible $\Pgt$ decay products
and $\log(\mathcal{P} \cdot \textrm{~\GeV}^{8}) + 4 \cdot \log(m_{\PHiggs}^{\textrm{test}(i)} \cdot \textrm{~\GeV}^{-1}) \approx \log(\mathcal{P} \cdot \textrm{~\GeV}^{8}) + \log(m_{\vis}^{4} \cdot \textrm{~\GeV}^{-4}) - \log(\frac{1}{z_{(1)}^{2} \, z_{(2)}^{2}}) \approx \log(\frac{\mathcal{P} \cdot \textrm{~\GeV}^{8}}{z_{(1)}^{2} \, z_{(2)}^{2}})$.
The term $\log(m_{\vis}^{4} \cdot \textrm{~\GeV}^{-4})$ has been omitted from the sum in the last step.
As the term $\log(m_{\vis}^{4} \cdot \textrm{~\GeV}^{-4})$ does not depend on $m_{\PHiggs}^{\textrm{test}(i)}$,
it has no effect on the reconstruction of $m_{\Pgt\Pgt}$.

%% file: performance.tex
\section{Performance}
\label{sec:performance}

The performance of the $m_{\Pgt\Pgt}$ reconstruction is studied in simulated events.
Samples of SM $\PHiggs \to \Pgt\Pgt$ signal events
are generated with the next-to-leading-order (NLO) program POWHEG v2~\cite{POWHEG1,POWHEG2,POWHEG3}
for a $\PHiggs$ boson of mass $m_{\PHiggs} = 125$~\GeV 
and for the gluon fusion ($\Pg\Pg \to \PHiggs$) and vector boson fusion ($\Pquark\APquark \to \PHiggs$) production processes.
We also study the $m_{\Pgt\Pgt}$ reconstruction in events containing
heavy pseudoscalar Higgs bosons $\PHiggsps$ of mass $m_{\PHiggsps} = 200$, $300$, $500$, $800$, $1200$, $1800$, and $2600$~\GeV,
produced via gluon fusion, and in events containing heavy spin $1$
resonances, of mass $2500$~\GeV, that decay into $\Pgt$ pairs.
We denote the latter by the symbol $\PZ'$. 
The $\PHiggsps \to \Pgt\Pgt$ and $\PZ' \to \Pgt\Pgt$ signal samples are generated with the LO generator PYTHIA 8.2~\cite{pythia8}.
The $\PZ/\Pggx \to \Pgt\Pgt$ background sample is generated with the LO MadGraph program, in the version MadGraph\_aMCatNLO 2.2.2~\cite{MadGraph_aMCatNLO}.
The sample contains $\Pgt$ lepton pairs of true mass $m_{\Pgt\Pgt}^{\true} > 50$~\GeV.
Most of the $\Pgt$ lepton pairs have a mass near the $\PZ$ peak at $m_{\PZ} = 91.2$~GeV~\cite{PDG}, 
but the sample also contains events of significantly higher mass.
Drell--Yan events of mass $m_{\Pgt\Pgt}^{\true} < 50$~\GeV are not relevant for this study, 
because they do not pass the selection criteria on $\pT$ and $\eta$ that are applied on analysis level.
All events are generated for proton-proton collisions at $\sqrt{s} = 13$~\TeV centre-of-mass energy.
The samples produced by MadGraph and POWHEG are generated with the NNPDF3.0 set of parton distribution functions,
while the samples produced by PYTHIA use the NNPDF2.3LO set~\cite{NNPDF1,NNPDF2,NNPDF3}.
Parton shower and hadronization processes are modelled using the generator PYTHIA with the tune CUETP8M1~\cite{PYTHIA_CUETP8M1tune_CMS}.
The latter is based on the Monash tune~\cite{PYTHIA_MonashTune}.
The decays of $\Pgt$ leptons, including polarization effects, are modelled by PYTHIA.

The samples are normalized according to cross section for the purpose of comparing different mass reconstruction algorithms in terms of signal-to-background separation.
The cross section for the irreducible $\PZ/\Pggx \to \Pgt\Pgt$ background is computed at NNLO accuracy and amounts to $1.92 \times 10^{3}$~pb~\cite{FEWZ}.
The cross sections for the SM $\PHiggs \to \Pgt\Pgt$ signal have been computed as detailed in Ref.~\cite{Dittmaier:2011ti},
with the updates described in Ref.~\cite{Heinemeyer:2013tqa} included.
The product of cross section times the branching fraction for the decay into a $\Pgt$ lepton pair
amounts to $2.77$~pb for SM $\PHiggs$ bosons produced via gluon fusion 
and to $2.37 \cdot 10^{-1}$~pb for SM $\PHiggs$ bosons produced via vector boson fusion.
The $\PHiggsps \to \Pgt\Pgt$ and $\PZ' \to \Pgt\Pgt$ samples 
are scaled to a product of cross section times branching fraction, for the decay into a pair of $\Pgt$ leptons, of $1$~pb.

The events are studied on generator level.
The $\pT$, $\eta$, and $\phi$ of the electrons and muons produced in the $\Pgt$ lepton decays
are assumed to be reconstructed perfectly.
The $\tauh$ are built by adding the visible decay products of a given $\Pgt$ lepton, excluding the neutrinos.
The experimental resolution on the $\pT$ of the $\tauh$ is simulated by sampling from the TF described in Section~\ref{sec:mem_TF_tauToHadDecays},
while the $\eta$, $\phi$, and $m_{\vis}$ of the $\tauh$ are assumed to be reconstructed perfectly.
Jets are reconstructed by the anti-$k_{T}$ algorithm~\cite{AntiKT} with a distance parameter $R = 0.4$,
using all stable visible generator level particles as input.
The components $\pX^{\rec}$ and $\pY^{\rec}$ of the hadronic recoil are computed according to Eq.~(\ref{eq:met}):
$\pX^{\rec} = -( \pX^{\vis(1)} + \pX^{\vis(2)} + \METx )$
and
$\pY^{\rec} = -( \pY^{\vis(1)} + \pY^{\vis(2)} + \METy )$,
where $\METx$ and $\METy$ correspond to, respectively, the $x$ and $y$ components of the sum of the momenta of all neutrinos produced in the $\Pgt$ lepton decays.
The resolution on the hadronic recoil is simulated by sampling from the TF described in Section~\ref{sec:mem_TF_hadRecoil}.

Distributions in $m_{\Pgt\Pgt}$ are computed separately for events in which 
both $\Pgt$ leptons decay hadronically ($\tauh\tauh$), 
events in which one $\Pgt$ lepton decays hadronically and the other into a muon ($\Pgm\tauh$),
and events in which one $\Pgt$ lepton decays into a muon and the other into an electron ($\Pe\Pgm$).
The visible $\Pgt$ decay products are required to pass selection criteria on $\pT$ and $\eta$,
which are motivated by the SM $\PHiggs \to \Pgt\Pgt$ analysis performed by the CMS collaboration during LHC Run $1$~\cite{HIG-13-004}.
Events in the $\tauh\tauh$ decay channel are required to contain
two $\tauh$ of $\pT > 45$~\GeV and $\vert\eta\vert < 2.1$.
Events in the $\Pgm\tauh$ channel
are required to contain a muon of $\pT > 20$~\GeV and $\vert\eta\vert < 2.1$ plus a $\tauh$ of $\pT > 30$~\GeV and $\vert\eta\vert < 2.3$.
Events selected in the $\Pe\Pgm$ channel are required to contain a muon with $\vert\eta\vert < 2.1$ and an electron with $\vert\eta\vert < 2.4$.
The lepton of higher $\pT$ (either the electron or the muon) is required to satisfy the condition $\pT > 20$~\GeV,
while the lepton of lower $\pT$ is required to satisfy $\pT > 10$~\GeV.
Similar selection criteria on $\pT$ and $\eta$ of the visible $\Pgt$ decay products were applied in the $\PHiggs \to \Pgt\Pgt$
analyses performed by the ATLAS
collaboration during LHC Run $1$~\cite{ATLAS_HiggsTauTau_SM,ATLAS_HiggsTauTau_MSSM}.

The $m_{\Pgt\Pgt}$ reconstruction in SM $\PHiggs \to \Pgt\Pgt$ signal events is studied in event categories motivated by the
SM $\PHiggs \to \Pgt\Pgt$ analysis performed by the CMS collaboration during LHC Run $1$~\cite{HIG-13-004}:
\begin{itemize}
\item $0$-jet: 
  Events containing no jets of $\pT > 30$~\GeV and $\lvert \eta \rvert < 4.5$.
\item $1$-jet non-boosted:
  Events containing one or more jets of $\pT > 30$~\GeV and $\lvert \eta \rvert < 4.5$
  in which the $\PHiggs$ (respectively $\PZ$) boson satisfies $\pT < 100$~\GeV.
  Events in the $1$-jet non-boosted category are required not to be selected in the $2$-jet VBF category.
\item $1$-jet boosted:
  Events containing one or more jets of $\pT > 30$~\GeV and $\lvert \eta \rvert < 4.5$
  in which the $\PHiggs$ (respectively $\PZ$) boson satisfies $\pT > 100$~\GeV.
  Events in the $1$-jet boosted category are required not to be selected in the $2$-jet VBF category.
\item $2$-jet VBF:
  Events containing two or more jets of $\pT > 30$~\GeV and $\lvert \eta \rvert < 4.5$,
  with at least one pair of jets satisfying $m_{jj} > 500$~\GeV and $\Delta\eta_{jj} > 3.5$.
\end{itemize}
The event categorization is based on generator level quantities.
We do not expect that migrations between the event categories due to resolution effects 
significantly affect the conclusions of the $m_{\Pgt\Pgt}$ resolution studies.

\begin{figure}
\setlength{\unitlength}{1mm}
\begin{center}
\begin{picture}(160,212)(0,0)
\put(-4.0, -4.0){\mbox{\includegraphics*[height=216mm]
{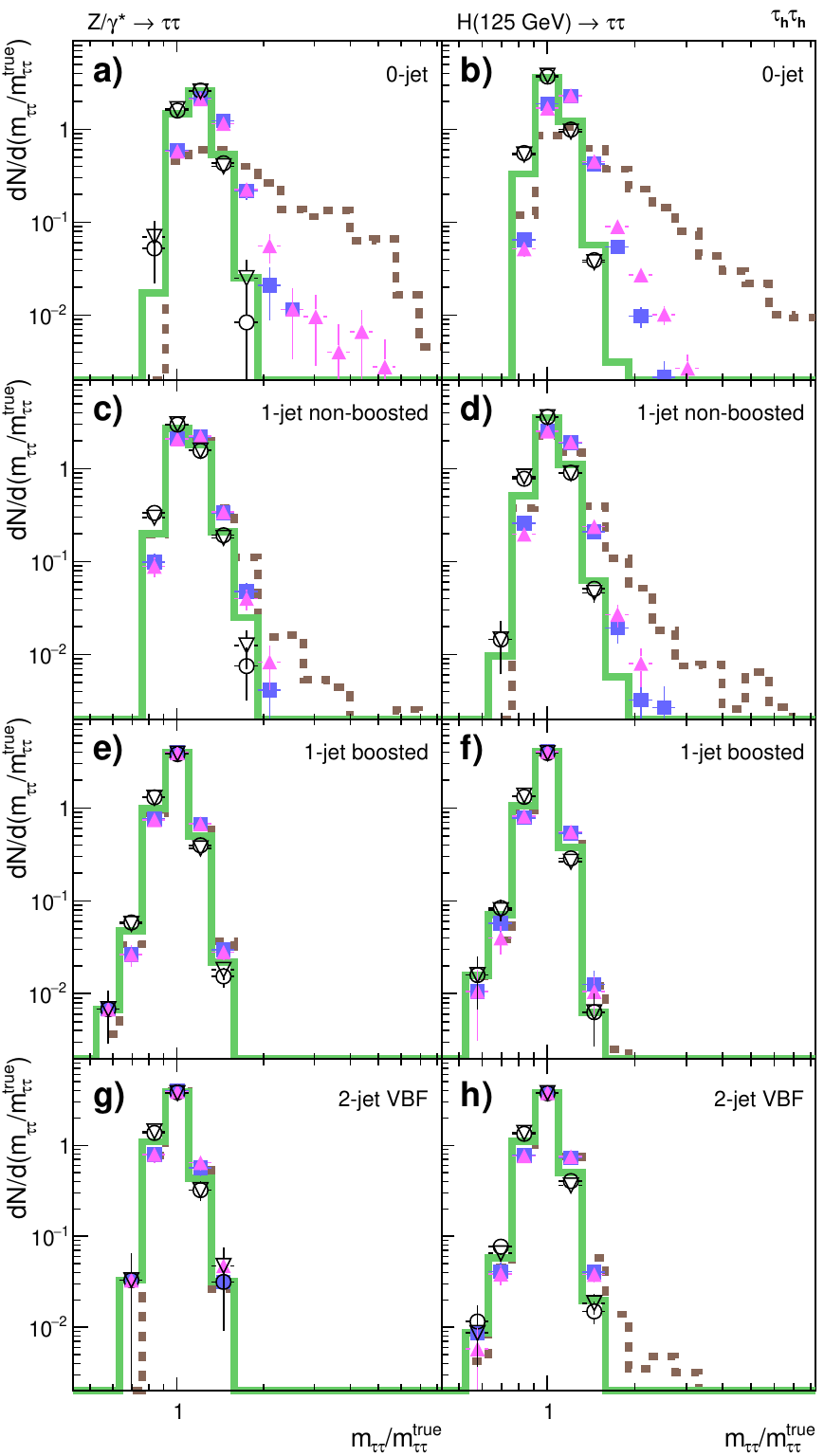}}}
\put(117.5, 145.5){\mbox{\includegraphics*[width=52mm]
{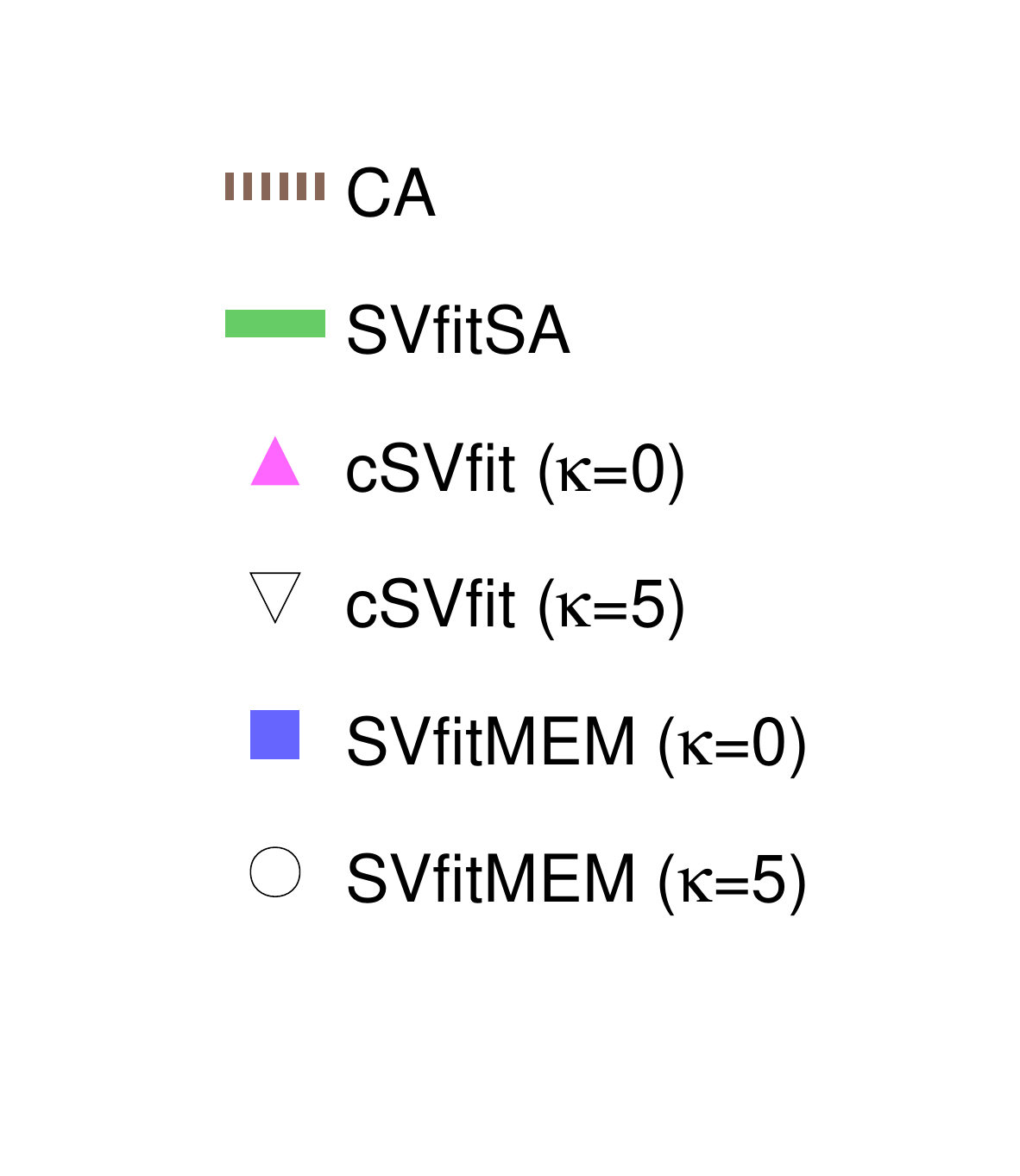}}}
\end{picture}
\end{center}
\caption{
  Distributions in $m_{\Pgt\Pgt}$ reconstructed by the CA method and different versions of the SVfit algorithm in simulated $\PZ/\Pggx \to \Pgt\Pgt$ background events (a,c,e,g)
  and in SM $\PHiggs \to \Pgt\Pgt$ signal events produced via the $\Pg\Pg \to \PHiggs$ (b,d,f) and $\Pquark\APquark \to \PHiggs$ (h) production processes
  in different event categories: $0$-jet (a,b), $1$-jet non-boosted (c,d), $1$-jet boosted (e,f),
  and $2$-jet VBF (g,h).
  The events are selected in the $\tauh\tauh$ decay channel. 
  The axis of abscissae ranges from $0.4$ to $8$.
}
\label{fig:massDistributions_sm_tautau}
\end{figure}

\begin{figure}
\setlength{\unitlength}{1mm}
\begin{center}
\begin{picture}(160,212)(0,0)
\put(-4.0, -4.0){\mbox{\includegraphics*[height=216mm]
{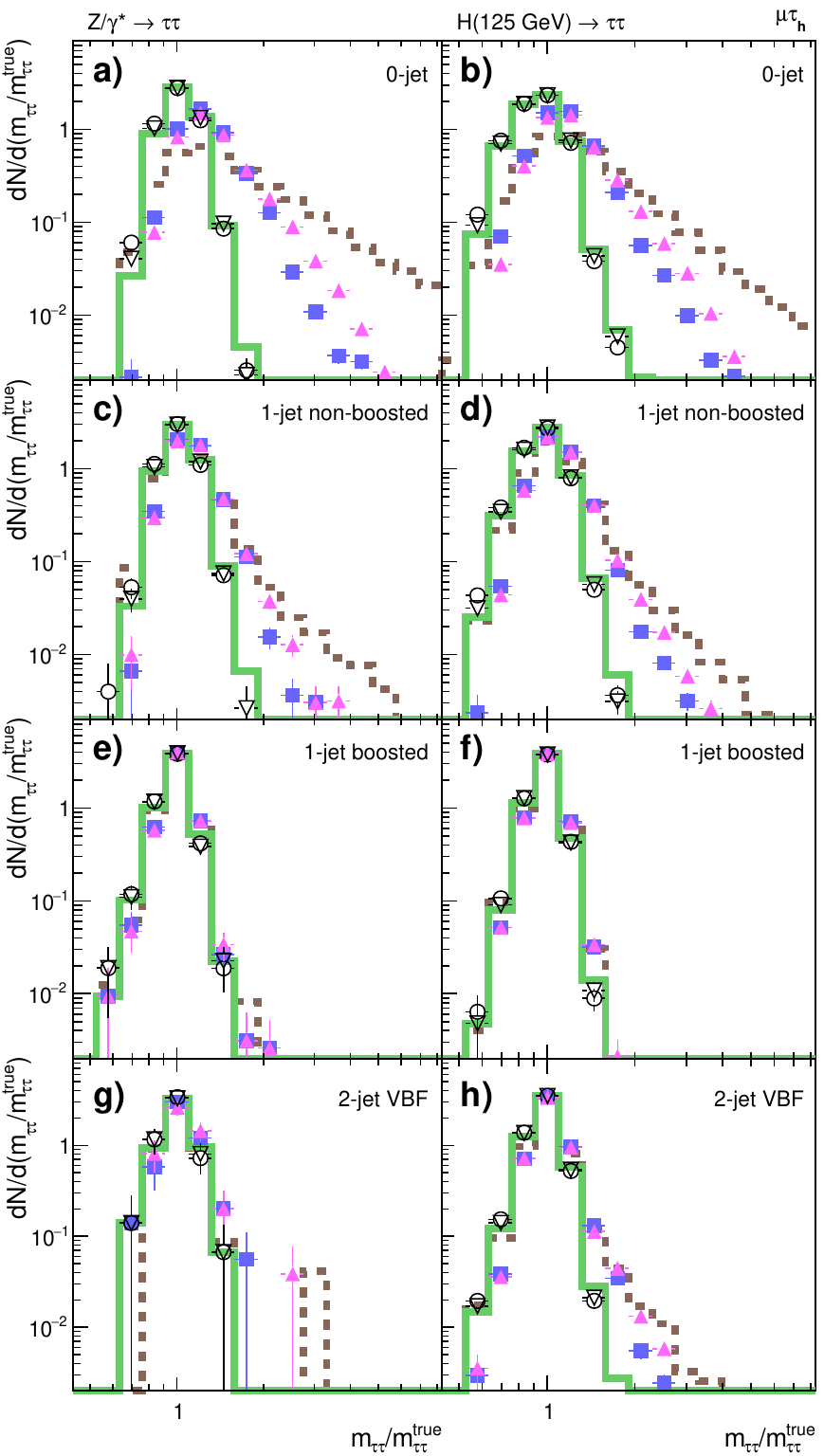}}}
\put(117.5, 145.5){\mbox{\includegraphics*[width=52mm]
{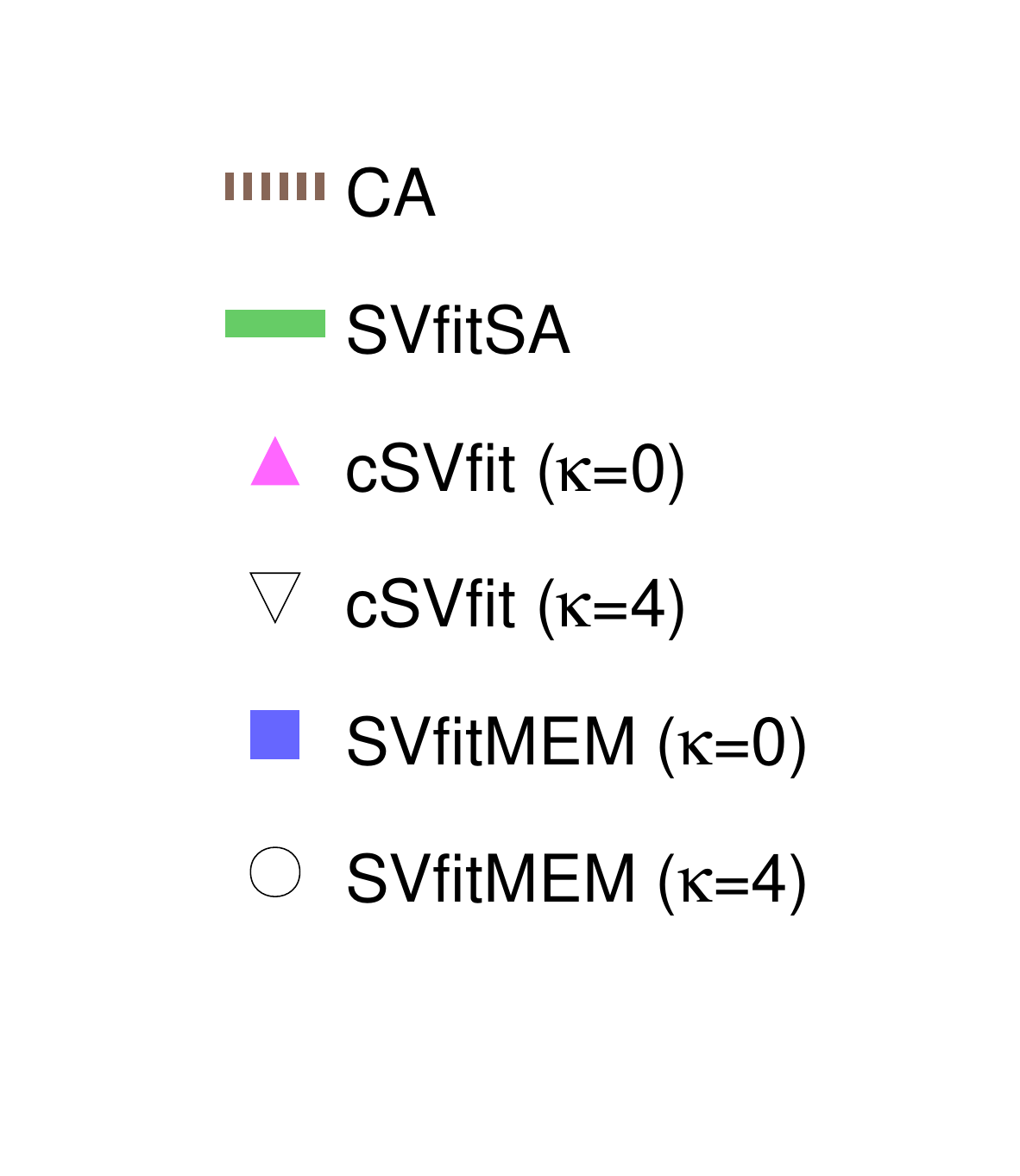}}}
\end{picture}
\end{center}
\caption{
  Distributions in $m_{\Pgt\Pgt}$ reconstructed by the CA method and different versions of the SVfit algorithm in simulated $\PZ/\Pggx \to \Pgt\Pgt$ background events (a,c,e,g)
 and in SM $\PHiggs \to \Pgt\Pgt$ signal events produced via the $\Pg\Pg \to \PHiggs$ (b,d,f) and $\Pquark\APquark \to \PHiggs$ (h) production processes
  in different event categories: $0$-jet (a,b), $1$-jet non-boosted (c,d), $1$-jet boosted (e,f),
  and $2$-jet VBF (g,h).
  The events are selected in the $\Pgm\tauh$ decay channel. 
  The axis of abscissae ranges from $0.4$ to $8$.
}
\label{fig:massDistributions_sm_mutau}
\end{figure}

\begin{figure}
\setlength{\unitlength}{1mm}
\begin{center}
\begin{picture}(160,212)(0,0)
\put(-4.0, -4.0){\mbox{\includegraphics*[height=216mm]
{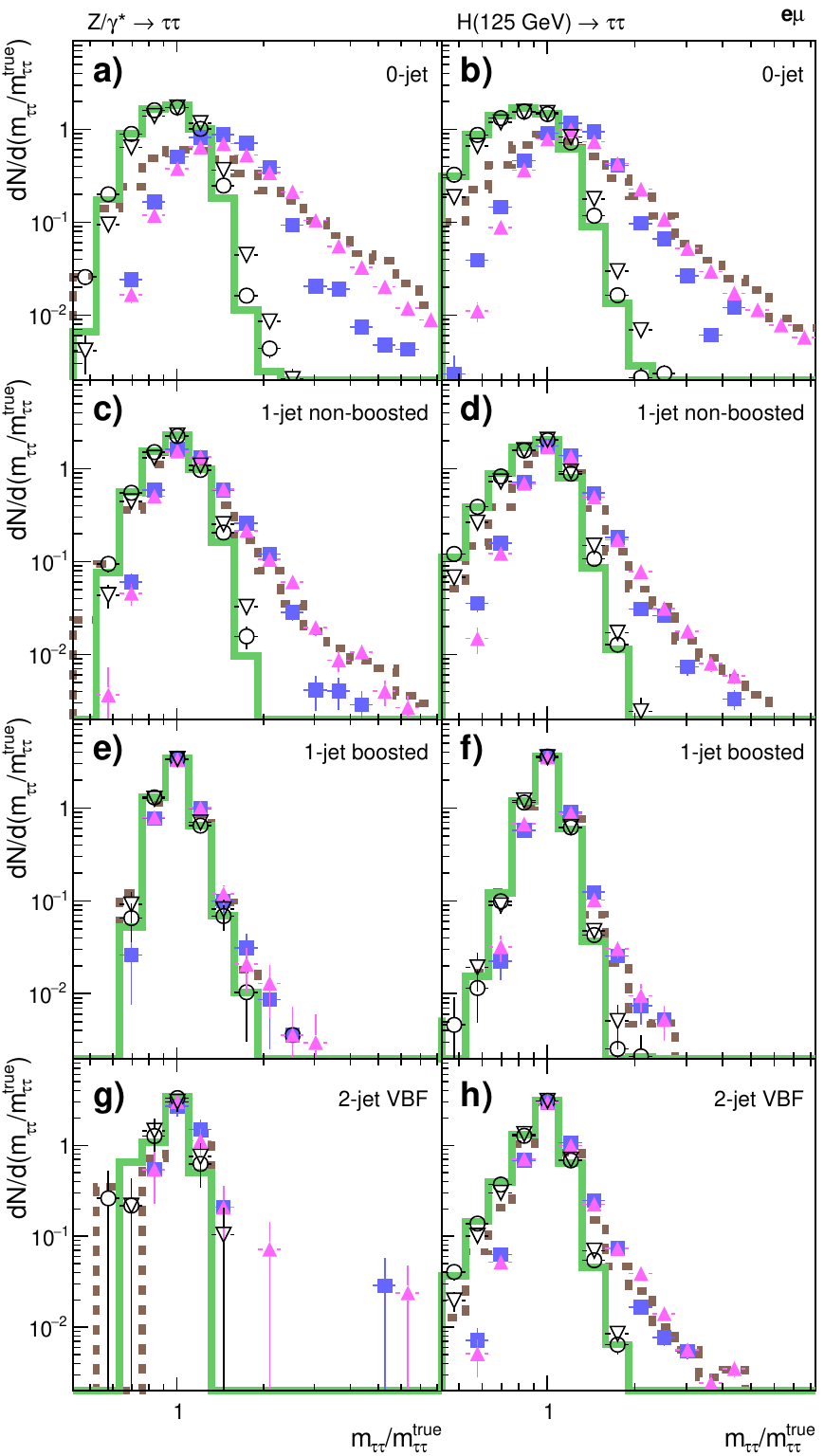}}}
\put(117.5, 145.5){\mbox{\includegraphics*[width=52mm]
{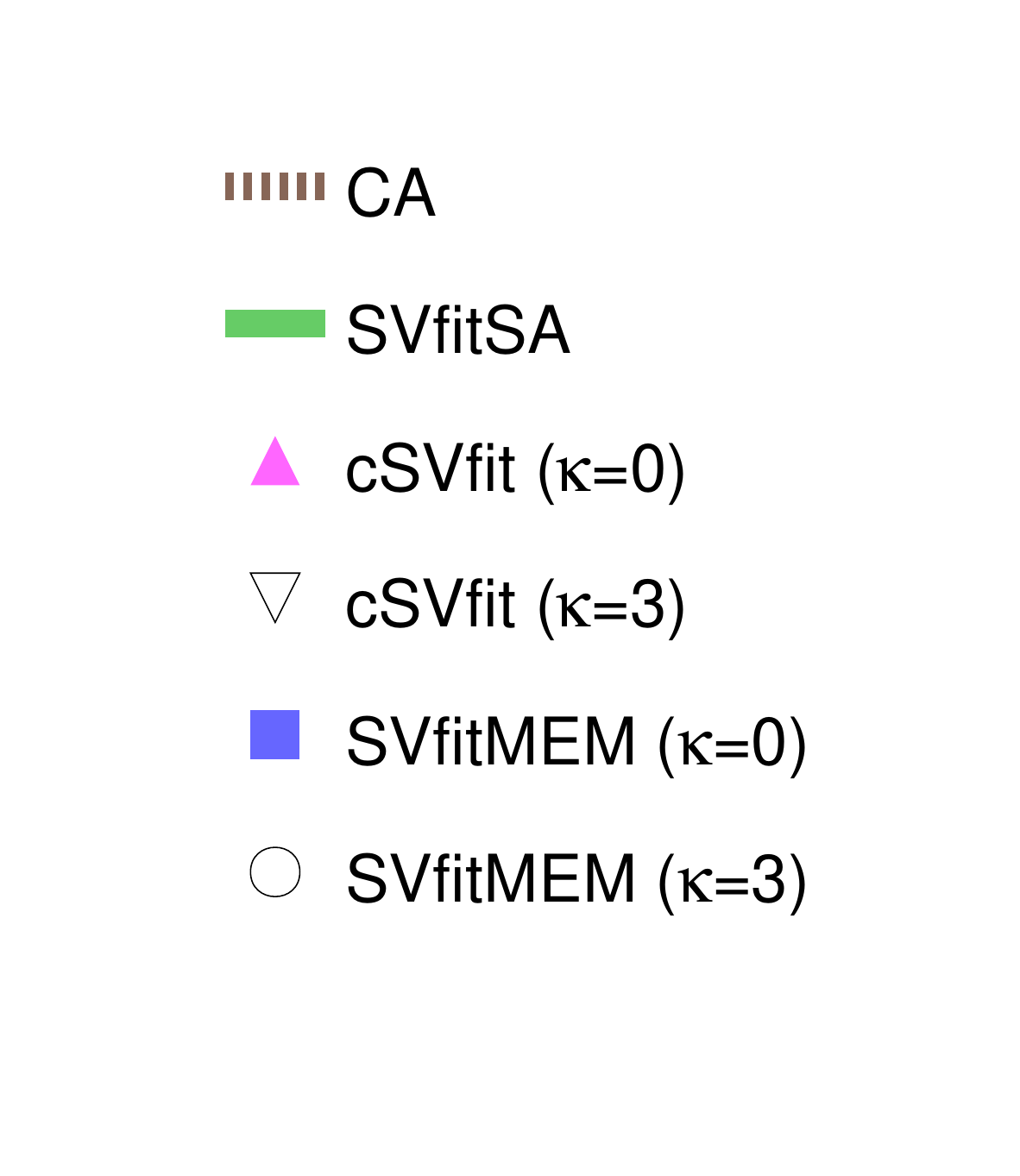}}}
\end{picture}
\end{center}
\caption{
  Distributions in $m_{\Pgt\Pgt}$ reconstructed by the CA method and different versions of the SVfit algorithm in simulated $\PZ/\Pggx \to \Pgt\Pgt$ background events (a,c,e,g)
  and in SM $\PHiggs \to \Pgt\Pgt$ signal events produced via the $\Pg\Pg \to \PHiggs$ (b,d,f) and $\Pquark\APquark \to \PHiggs$ (h) production processes
  in different event categories: $0$-jet (a,b), $1$-jet non-boosted (c,d), $1$-jet boosted (e,f),
  and $2$-jet VBF (g,h).
  The events are selected in the $\Pe\Pgm$ decay channel.
  The axis of abscissae ranges from $0.4$ to $8$.
}
\label{fig:massDistributions_sm_emu}
\end{figure}

The $m_{\Pgt\Pgt}$ distributions reconstructed by the SVfitMEM and cSVfit algorithms in these event categories 
in the $\tauh\tauh$, $\Pgm\tauh$, and $\Pe\Pgm$ decay channels are shown in Figs.~\ref{fig:massDistributions_sm_tautau} to~\ref{fig:massDistributions_sm_emu}.
They are compared to the distributions in $m_{\Pgt\Pgt}$ reconstructed by the ``collinear-approximation'' (CA) method~\cite{massRecoCollinearApprox}
and by the previous version of the
SVfit algorithm, described in Ref.~\cite{SVfit}, to which we refer to as the SVfit ``standalone'' (SVfitSA) algorithm.
The SVfitMEM and cSVfit algorithms are utilized with and without the artificial regularization term described in Section~\ref{sec:mem_logM}.
The value of $\kappa$ used for each channel is chosen such that the optimal resolution on $m_{\Pgt\Pgt}$, 
quantified in terms of the ratio of the root-mean-square (RMS) to the median $\textrm{M}$ of the distribution,
is attained.
We find that the choice of $\kappa = 5$ for the $\tauh\tauh$ channel, $\kappa = 4$ for the $\Pgm\tauh$ channel,
and $\kappa = 3$ for the $\Pe\Pgm$ channel performs well for the SVfitMEM as well as for the cSVfit algorithm.

%
%

The distributions in $m_{\Pgt\Pgt}$ reconstructed by the cSVfit and SVfitMEM algorithms with artificial regularization term
and by the SVfitSA algorithm are very similar.
The SVfitSA algorithm performs well without adding an artificial regularization term of the type described in Section~\ref{sec:mem_logM} to its likelihood function.
This is because the effect of the missing factor $\frac{1}{z^{2}}$ in the likelihood function used by the SVfitSA algorithm (\cf Section~\ref{sec:classicSVfit})
is equivalent to using an artificial regularization term with $\kappa = 4$.
Regardless of the choice of $\kappa$, the peaks of the $m_{\Pgt\Pgt}$ distributions reconstructed by the different versions of the SVfit algorithm 
are close to the true value of the mass of the $\Pgt$ lepton pair in the $1$-jet boosted and $2$-jet VBF categories.
In the $0$-jet and $1$-jet non-boosted categories,
the distributions in $m_{\Pgt\Pgt}$ reconstructed by the cSVfit and SVfitMEM algorithms with $\kappa = 0$
exhibit a tendency to overestimate the mass of the $\Pgt$ lepton pair,
resulting in pronounced high mass tails.
The choice of a small positive $\kappa$ reduces this tendency, 
makes the distributions in $m_{\Pgt\Pgt}$ peak close to the true mass of the $\Pgt$ lepton pair
and significantly reduce the high mass tails.

The effect of adding the artificial regularization term is largest for events reconstructed in the $0$-jet category,
which are the most difficult to reconstruct.
This is because in events in which the $\PHiggs$ or $\PZ$ boson has low $\pT$
the $\Pgt$ leptons are typically ``back-to-back'' in the transverse plane ($\Delta\phi_{\Pgt\Pgt} \approx \pi$), 
with the effect that the neutrinos produced in the $\Pgt$ lepton decays are emitted in opposite hemispheres and their contribution to $\MET$ cancels.
The cancellation of neutrino momenta causes mass hypotheses of low $m_{\PHiggs}^{\textrm{test}(i)} \approx m_{\vis}$
and of high $m_{\PHiggs}^{\textrm{test}(i)} \gg m_{\vis}$
to be degenerate in terms of the probability density $\mathcal{P}$,
computed according to Eq.~(\ref{eq:mem_with_hadRecoil}).
The best estimate for the mass of the $\Pgt$ lepton pair in a given event may fluctuate 
due to mismeasurements, within the experimental resolution, of the components $\pX^{\rec}$ and $\pY^{\rec}$ of the hadronic recoil or, to a lesser extent, of the $\pT$ of $\tauh$,
degrading the resolution on $m_{\Pgt\Pgt}$.

\begin{figure}
\setlength{\unitlength}{1mm}
\begin{center}
\begin{picture}(160,153.5)(0,0)
\put(-1.5, 81.5){\mbox{\includegraphics*[height=72mm]
  {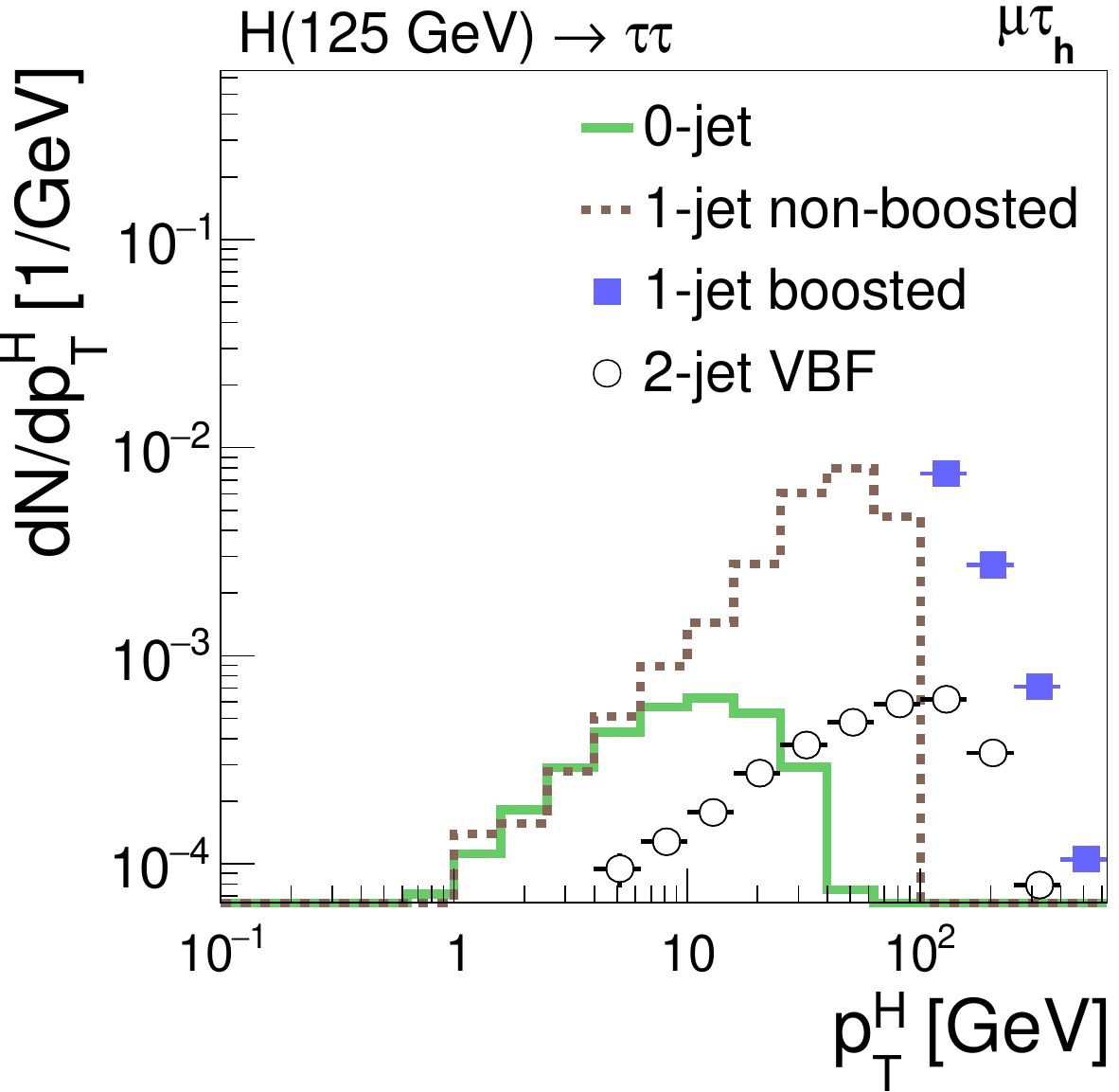}}}
\put(80.0, 81.5){\mbox{\includegraphics*[height=72mm]
  {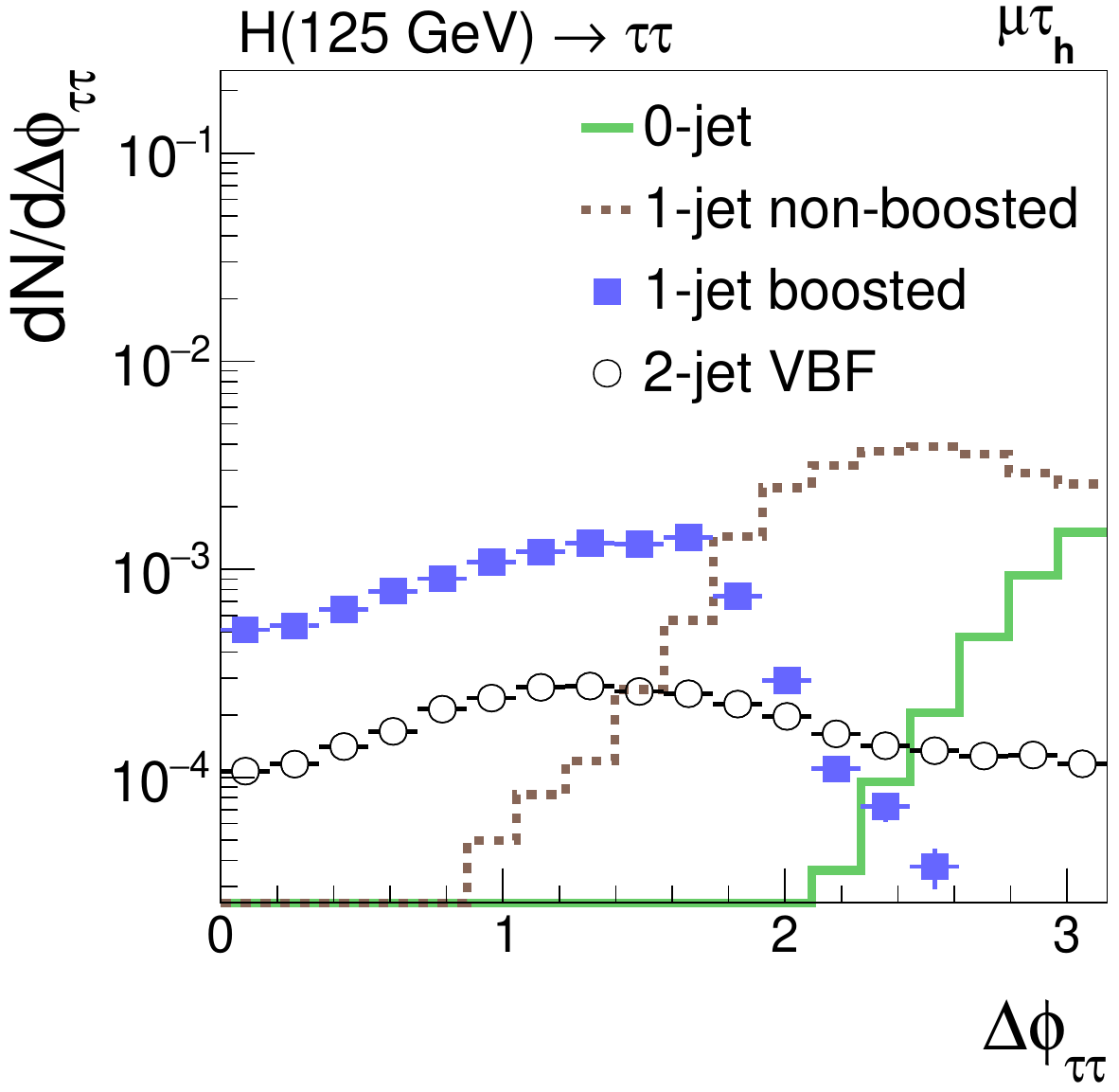}}}
\put(-1.5, 0.0){\mbox{\includegraphics*[height=72mm]
  {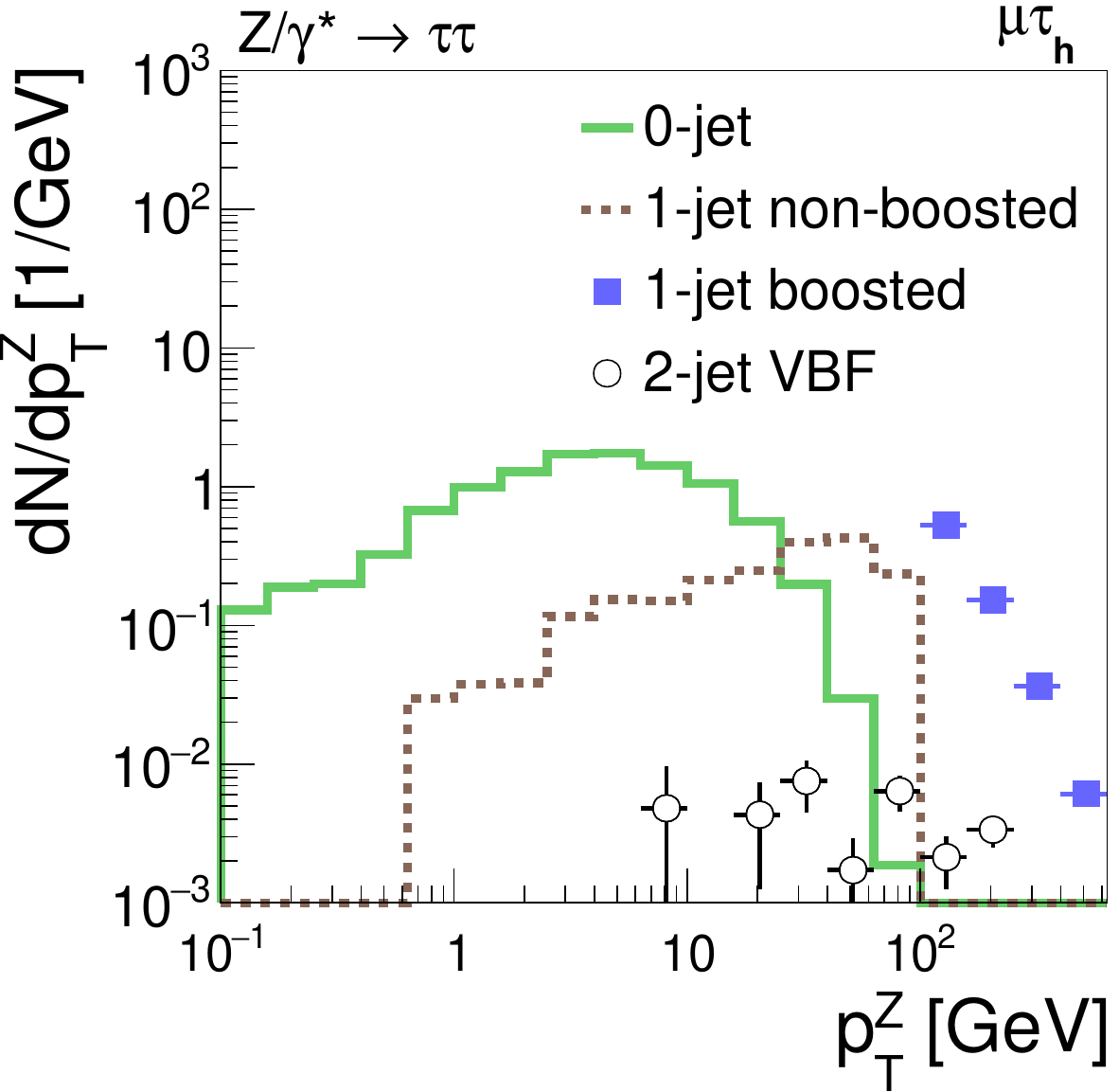}}}
\put(80.0, 0.0){\mbox{\includegraphics*[height=72mm]
  {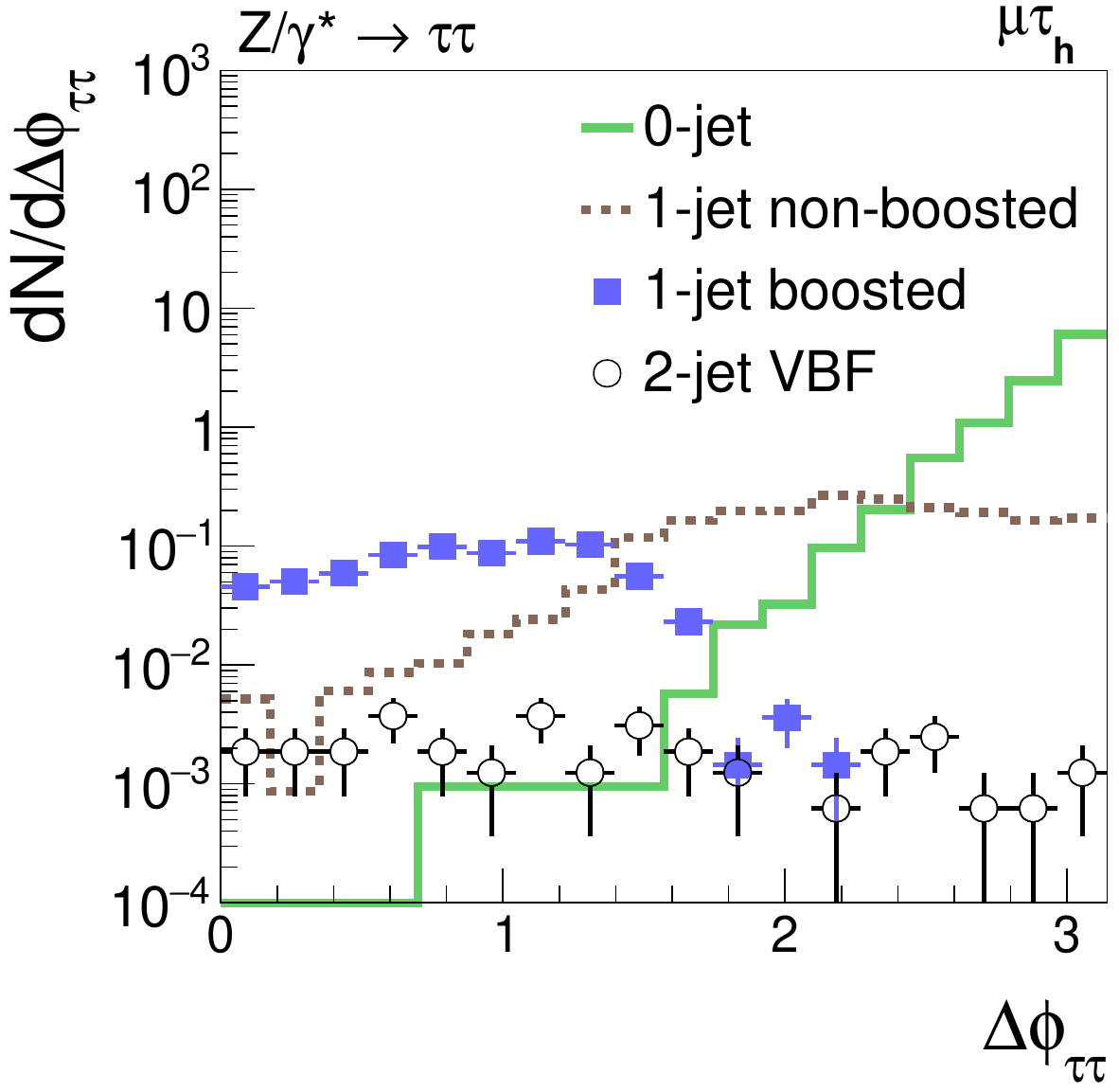}}}
\end{picture}
\end{center}
\caption{
  Distributions in $\pT$ of the $\PHiggs$ respectively $\PZ$ boson (a,c) and in the
  angle $\Delta\phi_{\Pgt\Pgt}$ (b,d) for SM $\PHiggs \to \Pgt\Pgt$ signal (a,b)
  and $\PZ/\Pggx \to \Pgt\Pgt$ background (c,d) events,
  selected in the $\Pgm\tauh$ decay channel.
  The signal events shown in the $0$-jet, $1$-jet non-boosted and
  $1$-jet boosted categories (in the $2$-jet VBF category) are produced via the $\Pg\Pg \to \PHiggs$
  ($\Pquark\APquark \to \PHiggs$) production process.   
}
\label{fig:ditau_pT_and_dphi}
\end{figure}

The resolution on $m_{\Pgt\Pgt}$ is significantly higher 
in events selected in the $1$-jet and $2$-jet VBF categories,
in which the $\Pgt$ lepton pair typically recoils against high $\pT$ jets.
As the $\pT$ of the $\PHiggs$ respectively $\PZ$ boson increases,
the angle $\Delta\phi_{\Pgt\Pgt}$ between the $\Pgt$ leptons decreases, due to the Lorentz boost in direction of the $\PHiggs$ or $\PZ$ boson.
The momenta of the neutrinos produced in the $\Pgt$ lepton decays add constructively in this case,
with the effect that the mass of the $\Pgt$ lepton pair is constrained by the measured value of $\MET$.
The correlation between the $\pT$ of the $\Pgt$ lepton pair and the angle $\Delta\phi_{\Pgt\Pgt}$ is visualized in Fig.~\ref{fig:ditau_pT_and_dphi}
for SM $\PHiggs \to \Pgt\Pgt$ signal and for $\PZ/\Pggx \to \Pgt\Pgt$ background events.
The events are selected in the $\Pgm\tauh$ decay channel.
The distributions for events selected in the $\tauh\tauh$ and $\Pe\Pgm$ decay channels are similar.

%
%

The variation of the resolution on $m_{\Pgt\Pgt}$ across event categories is most pronounced in the $\Pe\Pgm$ and least pronounced in the $\tauh\tauh$ decay channel.
This is because the fraction of $\Pgt$ lepton energy carried by the visible $\Pgt$ decay products is typically high for hadronic $\Pgt$ decays and typically low for leptonic $\Pgt$ decays,
\cf Fig.~\ref{fig:tauDecay_z}.
The consequence is that for events in the $\tauh\tauh$ decay channel,
regardless of the angle $\Delta\phi_{\Pgt\Pgt}$ between the $\Pgt$ leptons,
the best estimate for the mass of the $\Pgt$ lepton pair is typically not much higher than $m_{\vis}$,
as the energies of the neutrinos produced in hadronic $\Pgt$ decays are known to be most likely small.
For events in the $\Pe\Pgm$ decay channel on the other hand,
$\Pgt$ lepton decays with high energetic neutrinos are known to be likely and,
provided they are compatible with the measured value of $\MET$,
mass hypotheses of low $m_{\PHiggs}^{\textrm{test}(i)} \approx m_{\vis}$, corresponding to the case $z_{(1)} \approx 1$ and $z_{(2)} \approx 1$,
and of high $m_{\PHiggs}^{\textrm{test}(i)} \gg m_{\vis}$, corresponding to the case $z_{(1)} \ll 1$ and $z_{(2)} \ll 1$,
are often degenerate in terms of the probability density $\mathcal{P}$.

Numerical values for the resolution in $m_{\Pgt\Pgt}$ achieved by the different algorithms are given in Tables~\ref{tab:resolutions_sm_tautau} to~\ref{tab:resolutions_sm_emu}.
The mass of the visible $\Pgt$ decay products, $m_{\vis}$, is given in the tables for comparison.
The resolution is quantified separately for the low mass and for the high mass tail of the mass distribution reconstructed by a given algorithm.
The symbol $\sigma_{l}/\textrm{M}$ (the symbol $\sigma_{h}/\textrm{M}$) denotes the 
ratio of the difference between the median and the $16\%$ quantile (between the $84\%$ quantile and the median) relative to the median
of the mass distribution.
The capability of each algorithm to separate the SM $\PHiggs \to \Pgt\Pgt$ signal from the irreducible $\PZ/\Pggx \to \Pgt\Pgt$ background 
is additionally quantified in terms of $\textrm{S}/(\textrm{S} + \textrm{B})$, the ratio of the number of SM $\PHiggs \to \Pgt\Pgt$ signal events to the sum of signal plus $\PZ/\Pggx \to \Pgt\Pgt$ background events.
The yields $\textrm{S}$ of signal and $\textrm{B}$ of background events are computed within a mass window containing $68\%$ of signal events,
with $16\%$ of signal events on either side of the mass window, and for SM signal and background cross sections.

%
%
\begin{table}
\begin{center}
\begin{tabular}{|l|cccc|cccc|}
\hline
\multicolumn{9}{|c|}{$\tauh\tauh$ decay channel} \\
\hline
\hline
\multirow{2}{17mm}{Sample} & \multicolumn{4}{c|}{$m_{\vis}$} & \multicolumn{4}{c|}{SVfitSA} \\
\cline{2-9}
 & $\textrm{M}$~[\GeV\unskip] & $\sigma_{l}/\textrm{M}$ & $\sigma_{h}/\textrm{M}$ & $\tfrac{S}{S+B}$ & $\textrm{M}$~[\GeV\unskip] & $\sigma_{l}/\textrm{M}$ & $\sigma_{h}/\textrm{M}$ & $\tfrac{S}{S+B}$ \\
\hline
$\PZ \to \Pgt\Pgt$: & & & & & & & & \\ 
 $\quad$ $0$-jet & $100.0$ & $0.072$ & $0.101$ & $-$ & $117.9$ & $0.078$ & $0.090$ & $-$ \\
 $\quad$ $1$-jet non--boosted & $86.3$ & $0.107$ & $0.140$ & $-$ & $101.6$ & $0.100$ & $0.148$ & $-$ \\
 $\quad$ $1$-jet boosted & $70.1$ & $0.195$ & $0.170$ & $-$ & $90.4$ & $0.092$ & $0.100$ & $-$ \\
 $\quad$ $2$-jet VBF & $66.7$ & $0.190$ & $0.230$ & $-$ & $90.7$ & $0.090$ & $0.080$ & $-$ \\
SM $\Pg\Pg \to \PHiggs$, $\PHiggs \to \Pgt\Pgt$: & & & & & & & & \\ 
 $\quad$ $0$-jet & $109.2$ & $0.080$ & $0.088$ & $0.144$ & $130.0$ & $0.079$ & $0.095$ & $0.161$ \\
 $\quad$ $1$-jet non--boosted & $107.3$ & $0.108$ & $0.099$ & $0.177$ & $129.1$ & $0.087$ & $0.099$ & $0.251$ \\
 $\quad$ $1$-jet boosted & $92.4$ & $0.204$ & $0.173$ & $0.038$ & $122.8$ & $0.085$ & $0.081$ & $0.335$ \\
SM $\Pquark\APquark \to \PHiggs$, $\PHiggs \to \Pgt\Pgt$: & & & & & & & & \\ 
 $\quad$ $2$-jet VBF & $94.6$ & $0.198$ & $0.171$ & $0.213$ & $123.4$ & $0.088$ & $0.088$ & $0.846$ \\
\hline
\end{tabular}

\begin{tabular}{|l|cccc|cccc|}
\hline
\multirow{2}{17mm}{Sample} & \multicolumn{4}{c|}{cSVfit, $\kappa=0$} & \multicolumn{4}{c|}{cSVfit, $\kappa=5$} \\
\cline{2-9}
 & $\textrm{M}$~[\GeV\unskip] & $\sigma_{l}/\textrm{M}$ & $\sigma_{h}/\textrm{M}$ & $\tfrac{S}{S+B}$ & $\textrm{M}$~[\GeV\unskip] & $\sigma_{l}/\textrm{M}$ & $\sigma_{h}/\textrm{M}$ & $\tfrac{S}{S+B}$ \\
\hline
$\PZ \to \Pgt\Pgt$: & & & & & & & & \\ 
 $\quad$ $0$-jet & $128.8$ & $0.107$ & $0.169$ & $-$ & $115.6$ & $0.081$ & $0.103$ & $-$ \\
 $\quad$ $1$-jet non--boosted & $104.7$ & $0.107$ & $0.157$ & $-$ & $99.8$ & $0.099$ & $0.146$ & $-$ \\
 $\quad$ $1$-jet boosted & $91.9$ & $0.092$ & $0.100$ & $-$ & $89.0$ & $0.093$ & $0.099$ & $-$ \\
 $\quad$ $2$-jet VBF & $91.9$ & $0.091$ & $0.090$ & $-$ & $89.4$ & $0.088$ & $0.098$ & $-$ \\
SM $\Pg\Pg \to \PHiggs$, $\PHiggs \to \Pgt\Pgt$: & & & & & & & & \\ 
 $\quad$ $0$-jet & $143.6$ & $0.097$ & $0.153$ & $0.161$ & $127.6$ & $0.079$ & $0.093$ & $0.164$ \\
 $\quad$ $1$-jet non--boosted & $136.8$ & $0.093$ & $0.126$ & $0.255$ & $126.5$ & $0.086$ & $0.099$ & $0.246$ \\
 $\quad$ $1$-jet boosted & $124.7$ & $0.083$ & $0.081$ & $0.338$ & $121.2$ & $0.086$ & $0.082$ & $0.348$ \\
SM $\Pquark\APquark \to \PHiggs$, $\PHiggs \to \Pgt\Pgt$: & & & & & & & & \\ 
 $\quad$ $2$-jet VBF & $125.7$ & $0.088$ & $0.096$ & $0.734$ & $121.6$ & $0.088$ & $0.087$ & $0.807$ \\
\hline
\end{tabular}

\begin{tabular}{|l|cccc|cccc|}
\hline
\multirow{2}{17mm}{Sample} & \multicolumn{4}{c|}{SVfitMEM, $\kappa=0$} & \multicolumn{4}{c|}{SVfitMEM, $\kappa=5$} \\
\cline{2-9}
 & $\textrm{M}$~[\GeV\unskip] & $\sigma_{l}/\textrm{M}$ & $\sigma_{h}/\textrm{M}$ & $\tfrac{S}{S+B}$ & $\textrm{M}$~[\GeV\unskip] & $\sigma_{l}/\textrm{M}$ & $\sigma_{h}/\textrm{M}$ & $\tfrac{S}{S+B}$ \\
\hline
$\PZ \to \Pgt\Pgt$: & & & & & & & & \\ 
 $\quad$ $0$-jet & $128.1$ & $0.113$ & $0.156$ & $-$ & $115.6$ & $0.078$ & $0.107$ & $-$ \\
 $\quad$ $1$-jet non--boosted & $104.8$ & $0.106$ & $0.157$ & $-$ & $99.8$ & $0.107$ & $0.147$ & $-$ \\
 $\quad$ $1$-jet boosted & $91.8$ & $0.090$ & $0.100$ & $-$ & $89.1$ & $0.092$ & $0.100$ & $-$ \\
 $\quad$ $2$-jet VBF & $91.9$ & $0.096$ & $0.087$ & $-$ & $89.1$ & $0.098$ & $0.094$ & $-$ \\
SM $\Pg\Pg \to \PHiggs$, $\PHiggs \to \Pgt\Pgt$: & & & & & & & & \\ 
 $\quad$ $0$-jet & $142.2$ & $0.098$ & $0.140$ & $0.150$ & $127.9$ & $0.085$ & $0.098$ & $0.165$ \\
 $\quad$ $1$-jet non--boosted & $136.5$ & $0.097$ & $0.120$ & $0.248$ & $127.0$ & $0.090$ & $0.102$ & $0.234$ \\
 $\quad$ $1$-jet boosted & $124.8$ & $0.083$ & $0.082$ & $0.349$ & $121.3$ & $0.088$ & $0.081$ & $0.339$ \\
SM $\Pquark\APquark \to \PHiggs$, $\PHiggs \to \Pgt\Pgt$: & & & & & & & & \\ 
 $\quad$ $2$-jet VBF & $125.6$ & $0.086$ & $0.095$ & $0.847$ & $121.9$ & $0.091$ & $0.086$ & $0.768$ \\
\hline
\end{tabular}
\end{center}
\caption{
  Median $\textrm{M}$ and resolutions $\sigma_{l}/\textrm{M}$ and $\sigma_{h}/\textrm{M}$
  of the distributions in $m_{\vis}$ 
  and in $m_{\Pgt\Pgt}$ reconstructed by different versions of SVfit algorithm
  in simulated SM $\PHiggs \to \Pgt\Pgt$ signal (S) and $\PZ/\Pggx \to \Pgt\Pgt$ background (B) events, 
  selected in different event categories in the $\tauh\tauh$ decay channel.
  The improvement in signal-to-background separation is quantified by the ratio $\textrm{S}/(\textrm{S} + \textrm{B})$,
  computed within a mass window containing $68\%$ of signal events.
  The computation of the ratio $\textrm{S}/(\textrm{S} + \textrm{B})$ and the difference between the two resolutions $\sigma_{l}/\textrm{M}$ and $\sigma_{h}/\textrm{M}$
  is explained in the text.
}
\label{tab:resolutions_sm_tautau}
\end{table}

\begin{table}
\begin{center}
\begin{tabular}{|l|cccc|cccc|}
\hline
\multicolumn{9}{|c|}{$\Pgm\tauh$ decay channel} \\
\hline
\hline
\multirow{2}{17mm}{Sample} & \multicolumn{4}{c|}{$m_{\vis}$} & \multicolumn{4}{c|}{SVfitSA} \\
\cline{2-9}
 & $\textrm{M}$~[\GeV\unskip] & $\sigma_{l}/\textrm{M}$ & $\sigma_{h}/\textrm{M}$ & $\tfrac{S}{S+B}$ & $\textrm{M}$~[\GeV\unskip] & $\sigma_{l}/\textrm{M}$ & $\sigma_{h}/\textrm{M}$ & $\tfrac{S}{S+B}$ \\
\hline
$\PZ \to \Pgt\Pgt$: & & & & & & & & \\ 
 $\quad$ $0$-jet & $66.8$ & $0.123$ & $0.146$ & $-$ & $95.4$ & $0.117$ & $0.134$ & $-$ \\
 $\quad$ $1$-jet non--boosted & $64.7$ & $0.147$ & $0.170$ & $-$ & $94.5$ & $0.117$ & $0.133$ & $-$ \\
 $\quad$ $1$-jet boosted & $53.3$ & $0.242$ & $0.278$ & $-$ & $90.4$ & $0.096$ & $0.096$ & $-$ \\
 $\quad$ $2$-jet VBF & $56.0$ & $0.336$ & $0.295$ & $-$ & $92.5$ & $0.119$ & $0.152$ & $-$ \\
SM $\Pg\Pg \to \PHiggs$, $\PHiggs \to \Pgt\Pgt$: & & & & & & & & \\ 
 $\quad$ $0$-jet & $80.6$ & $0.172$ & $0.200$ & $0.008$ & $119.4$ & $0.161$ & $0.157$ & $0.012$ \\
 $\quad$ $1$-jet non--boosted & $79.8$ & $0.190$ & $0.217$ & $0.016$ & $122.8$ & $0.137$ & $0.137$ & $0.046$ \\
 $\quad$ $1$-jet boosted & $71.4$ & $0.261$ & $0.291$ & $0.020$ & $122.9$ & $0.092$ & $0.089$ & $0.280$ \\
SM $\Pquark\APquark \to \PHiggs$, $\PHiggs \to \Pgt\Pgt$: & & & & & & & & \\ 
 $\quad$ $2$-jet VBF & $74.2$ & $0.249$ & $0.263$ & $0.126$ & $122.9$ & $0.103$ & $0.101$ & $0.575$ \\
\hline
\end{tabular}

\begin{tabular}{|l|cccc|cccc|}
\hline
\multirow{2}{17mm}{Sample} & \multicolumn{4}{c|}{cSVfit, $\kappa=0$} & \multicolumn{4}{c|}{cSVfit, $\kappa=4$} \\
\cline{2-9}
 & $\textrm{M}$~[\GeV\unskip] & $\sigma_{l}/\textrm{M}$ & $\sigma_{h}/\textrm{M}$ & $\tfrac{S}{S+B}$ & $\textrm{M}$~[\GeV\unskip] & $\sigma_{l}/\textrm{M}$ & $\sigma_{h}/\textrm{M}$ & $\tfrac{S}{S+B}$ \\
\hline
$\PZ \to \Pgt\Pgt$: & & & & & & & & \\ 
 $\quad$ $0$-jet & $121.6$ & $0.178$ & $0.434$ & $-$ & $94.8$ & $0.122$ & $0.136$ & $-$ \\
 $\quad$ $1$-jet non--boosted & $103.5$ & $0.130$ & $0.205$ & $-$ & $93.7$ & $0.119$ & $0.134$ & $-$ \\
 $\quad$ $1$-jet boosted & $92.1$ & $0.088$ & $0.101$ & $-$ & $89.8$ & $0.099$ & $0.092$ & $-$ \\
 $\quad$ $2$-jet VBF & $94.5$ & $0.127$ & $0.185$ & $-$ & $91.8$ & $0.106$ & $0.139$ & $-$ \\
SM $\Pg\Pg \to \PHiggs$, $\PHiggs \to \Pgt\Pgt$: & & & & & & & & \\ 
 $\quad$ $0$-jet & $150.6$ & $0.176$ & $0.403$ & $0.011$ & $119.2$ & $0.166$ & $0.159$ & $0.011$ \\
 $\quad$ $1$-jet non--boosted & $136.4$ & $0.134$ & $0.218$ & $0.041$ & $122.3$ & $0.136$ & $0.135$ & $0.047$ \\
 $\quad$ $1$-jet boosted & $125.4$ & $0.087$ & $0.091$ & $0.265$ & $122.1$ & $0.093$ & $0.088$ & $0.288$ \\
SM $\Pquark\APquark \to \PHiggs$, $\PHiggs \to \Pgt\Pgt$: & & & & & & & & \\ 
 $\quad$ $2$-jet VBF & $127.3$ & $0.094$ & $0.130$ & $0.404$ & $122.0$ & $0.103$ & $0.102$ & $0.579$ \\
\hline
\end{tabular}

\begin{tabular}{|l|cccc|cccc|}
\hline
\multirow{2}{17mm}{Sample} & \multicolumn{4}{c|}{SVfitMEM, $\kappa=0$} & \multicolumn{4}{c|}{SVfitMEM, $\kappa=4$} \\
\cline{2-9}
 & $\textrm{M}$~[\GeV\unskip] & $\sigma_{l}/\textrm{M}$ & $\sigma_{h}/\textrm{M}$ & $\tfrac{S}{S+B}$ & $\textrm{M}$~[\GeV\unskip] & $\sigma_{l}/\textrm{M}$ & $\sigma_{h}/\textrm{M}$ & $\tfrac{S}{S+B}$ \\
\hline
$\PZ \to \Pgt\Pgt$: & & & & & & & & \\ 
 $\quad$ $0$-jet & $116.4$ & $0.161$ & $0.292$ & $-$ & $94.0$ & $0.124$ & $0.139$ & $-$ \\
 $\quad$ $1$-jet non--boosted & $102.8$ & $0.131$ & $0.195$ & $-$ & $92.8$ & $0.118$ & $0.137$ & $-$ \\
 $\quad$ $1$-jet boosted & $92.5$ & $0.093$ & $0.097$ & $-$ & $89.5$ & $0.098$ & $0.100$ & $-$ \\
 $\quad$ $2$-jet VBF & $94.0$ & $0.107$ & $0.186$ & $-$ & $91.5$ & $0.126$ & $0.137$ & $-$ \\
SM $\Pg\Pg \to \PHiggs$, $\PHiggs \to \Pgt\Pgt$: & & & & & & & & \\ 
 $\quad$ $0$-jet & $145.3$ & $0.168$ & $0.270$ & $0.011$ & $118.4$ & $0.168$ & $0.159$ & $0.011$ \\
 $\quad$ $1$-jet non--boosted & $135.4$ & $0.134$ & $0.193$ & $0.041$ & $121.6$ & $0.139$ & $0.140$ & $0.046$ \\
 $\quad$ $1$-jet boosted & $125.5$ & $0.088$ & $0.092$ & $0.273$ & $122.1$ & $0.092$ & $0.090$ & $0.290$ \\
SM $\Pquark\APquark \to \PHiggs$, $\PHiggs \to \Pgt\Pgt$: & & & & & & & & \\ 
 $\quad$ $2$-jet VBF & $127.5$ & $0.095$ & $0.124$ & $0.449$ & $121.9$ & $0.104$ & $0.102$ & $0.575$ \\
\hline
\end{tabular}
\end{center}
\caption{
  Median $\textrm{M}$ and resolutions $\sigma_{l}/\textrm{M}$ and $\sigma_{h}/\textrm{M}$
  of the distributions in $m_{\vis}$ 
  and in $m_{\Pgt\Pgt}$ reconstructed by different versions of SVfit algorithm
  in simulated SM $\PHiggs \to \Pgt\Pgt$ signal (S) and $\PZ/\Pggx \to \Pgt\Pgt$ background (B) events, 
  selected in different event categories in the $\Pgm\tauh$ decay channel.
  The improvement in signal-to-background separation is quantified by the ratio $\textrm{S}/(\textrm{S} + \textrm{B})$,
  computed within a mass window containing $68\%$ of signal events.
  The computation of the ratio $\textrm{S}/(\textrm{S} + \textrm{B})$ and the difference between the two resolutions $\sigma_{l}/\textrm{M}$ and $\sigma_{h}/\textrm{M}$
  is explained in the text.
}
\label{tab:resolutions_sm_mutau}
\end{table}

\begin{table}
\begin{center}
\begin{tabular}{|l|cccc|cccc|}
\hline
\multicolumn{9}{|c|}{$\Pe\Pgm$ decay channel} \\
\hline
\hline
\multirow{2}{17mm}{Sample} & \multicolumn{4}{c|}{$m_{\vis}$} & \multicolumn{4}{c|}{SVfitSA} \\
\cline{2-9}
 & $\textrm{M}$~[\GeV\unskip] & $\sigma_{l}/\textrm{M}$ & $\sigma_{h}/\textrm{M}$ & $\tfrac{S}{S+B}$ & $\textrm{M}$~[\GeV\unskip] & $\sigma_{l}/\textrm{M}$ & $\sigma_{h}/\textrm{M}$ & $\tfrac{S}{S+B}$ \\
\hline
$\PZ \to \Pgt\Pgt$: & & & & & & & & \\ 
 $\quad$ $0$-jet & $48.1$ & $0.196$ & $0.274$ & $-$ & $88.2$ & $0.200$ & $0.228$ & $-$ \\
 $\quad$ $1$-jet non--boosted & $47.0$ & $0.216$ & $0.265$ & $-$ & $89.5$ & $0.170$ & $0.192$ & $-$ \\
 $\quad$ $1$-jet boosted & $38.2$ & $0.304$ & $0.456$ & $-$ & $90.0$ & $0.116$ & $0.123$ & $-$ \\
 $\quad$ $2$-jet VBF & $41.0$ & $0.174$ & $0.349$ & $-$ & $89.2$ & $0.167$ & $0.096$ & $-$ \\
SM $\Pg\Pg \to \PHiggs$, $\PHiggs \to \Pgt\Pgt$: & & & & & & & & \\ 
 $\quad$ $0$-jet & $55.6$ & $0.240$ & $0.324$ & $0.003$ & $108.4$ & $0.244$ & $0.263$ & $0.003$ \\
 $\quad$ $1$-jet non--boosted & $55.2$ & $0.250$ & $0.334$ & $0.006$ & $117.3$ & $0.209$ & $0.191$ & $0.010$ \\
 $\quad$ $1$-jet boosted & $49.7$ & $0.303$ & $0.429$ & $0.013$ & $123.3$ & $0.100$ & $0.099$ & $0.157$ \\
SM $\Pquark\APquark \to \PHiggs$, $\PHiggs \to \Pgt\Pgt$: & & & & & & & & \\ 
 $\quad$ $2$-jet VBF & $51.2$ & $0.291$ & $0.383$ & $0.056$ & $122.1$ & $0.142$ & $0.116$ & $0.463$ \\
\hline
\end{tabular}

\begin{tabular}{|l|cccc|cccc|}
\hline
\multirow{2}{17mm}{Sample} & \multicolumn{4}{c|}{cSVfit, $\kappa=0$} & \multicolumn{4}{c|}{cSVfit, $\kappa=3$} \\
\cline{2-9}
 & $\textrm{M}$~[\GeV\unskip] & $\sigma_{l}/\textrm{M}$ & $\sigma_{h}/\textrm{M}$ & $\tfrac{S}{S+B}$ & $\textrm{M}$~[\GeV\unskip] & $\sigma_{l}/\textrm{M}$ & $\sigma_{h}/\textrm{M}$ & $\tfrac{S}{S+B}$ \\
\hline
$\PZ \to \Pgt\Pgt$: & & & & & & & & \\ 
 $\quad$ $0$-jet & $159.8$ & $0.307$ & $0.827$ & $-$ & $92.8$ & $0.202$ & $0.246$ & $-$ \\
 $\quad$ $1$-jet non--boosted & $106.9$ & $0.174$ & $0.408$ & $-$ & $92.0$ & $0.166$ & $0.199$ & $-$ \\
 $\quad$ $1$-jet boosted & $92.6$ & $0.099$ & $0.138$ & $-$ & $90.0$ & $0.114$ & $0.120$ & $-$ \\
 $\quad$ $2$-jet VBF & $95.5$ & $0.103$ & $0.169$ & $-$ & $89.8$ & $0.138$ & $0.104$ & $-$ \\
SM $\Pg\Pg \to \PHiggs$, $\PHiggs \to \Pgt\Pgt$: & & & & & & & & \\ 
 $\quad$ $0$-jet & $174.4$ & $0.258$ & $0.622$ & $0.004$ & $114.7$ & $0.243$ & $0.268$ & $0.003$ \\
 $\quad$ $1$-jet non--boosted & $140.2$ & $0.172$ & $0.320$ & $0.014$ & $120.6$ & $0.196$ & $0.191$ & $0.011$ \\
 $\quad$ $1$-jet boosted & $127.1$ & $0.089$ & $0.116$ & $0.144$ & $123.2$ & $0.098$ & $0.104$ & $0.157$ \\
SM $\Pquark\APquark \to \PHiggs$, $\PHiggs \to \Pgt\Pgt$: & & & & & & & & \\ 
 $\quad$ $2$-jet VBF & $129.2$ & $0.106$ & $0.187$ & $0.464$ & $122.7$ & $0.131$ & $0.122$ & $0.463$ \\
\hline
\end{tabular}

\begin{tabular}{|l|cccc|cccc|}
\hline
\multirow{2}{17mm}{Sample} & \multicolumn{4}{c|}{SVfitMEM, $\kappa=0$} & \multicolumn{4}{c|}{SVfitMEM, $\kappa=3$} \\
\cline{2-9}
 & $\textrm{M}$~[\GeV\unskip] & $\sigma_{l}/\textrm{M}$ & $\sigma_{h}/\textrm{M}$ & $\tfrac{S}{S+B}$ & $\textrm{M}$~[\GeV\unskip] & $\sigma_{l}/\textrm{M}$ & $\sigma_{h}/\textrm{M}$ & $\tfrac{S}{S+B}$ \\
\hline
$\PZ \to \Pgt\Pgt$: & & & & & & & & \\ 
 $\quad$ $0$-jet & $140.4$ & $0.258$ & $0.333$ & $-$ & $89.0$ & $0.209$ & $0.237$ & $-$ \\
 $\quad$ $1$-jet non--boosted & $105.2$ & $0.171$ & $0.356$ & $-$ & $90.2$ & $0.176$ & $0.197$ & $-$ \\
 $\quad$ $1$-jet boosted & $93.0$ & $0.103$ & $0.139$ & $-$ & $90.0$ & $0.118$ & $0.123$ & $-$ \\
 $\quad$ $2$-jet VBF & $95.0$ & $0.111$ & $0.159$ & $-$ & $89.5$ & $0.175$ & $0.110$ & $-$ \\
SM $\Pg\Pg \to \PHiggs$, $\PHiggs \to \Pgt\Pgt$: & & & & & & & & \\ 
 $\quad$ $0$-jet & $161.2$ & $0.233$ & $0.321$ & $0.004$ & $110.4$ & $0.255$ & $0.271$ & $0.003$ \\
 $\quad$ $1$-jet non--boosted & $138.6$ & $0.168$ & $0.269$ & $0.014$ & $118.8$ & $0.212$ & $0.194$ & $0.010$ \\
 $\quad$ $1$-jet boosted & $127.9$ & $0.088$ & $0.119$ & $0.148$ & $123.7$ & $0.099$ & $0.101$ & $0.164$ \\
SM $\Pquark\APquark \to \PHiggs$, $\PHiggs \to \Pgt\Pgt$: & & & & & & & & \\ 
 $\quad$ $2$-jet VBF & $129.6$ & $0.107$ & $0.175$ & $0.364$ & $122.6$ & $0.138$ & $0.118$ & $0.463$ \\
\hline
\end{tabular}
\end{center}
\caption{
  Median $\textrm{M}$ and resolutions $\sigma_{l}/\textrm{M}$ and $\sigma_{h}/\textrm{M}$
  of the distributions in $m_{\vis}$ 
  and in $m_{\Pgt\Pgt}$ reconstructed by different versions of SVfit algorithm
  in simulated SM $\PHiggs \to \Pgt\Pgt$ signal (S) and $\PZ/\Pggx \to \Pgt\Pgt$ background (B) events, 
  selected in different event categories in the $\Pe\Pgm$ decay channel.
  The improvement in signal-to-background separation is quantified by the ratio $\textrm{S}/(\textrm{S} + \textrm{B})$,
  computed within a mass window containing $68\%$ of signal events.
  The computation of the ratio $\textrm{S}/(\textrm{S} + \textrm{B})$ and the difference between the two resolutions $\sigma_{l}/\textrm{M}$ and $\sigma_{h}/\textrm{M}$
  is explained in the text.
}
\label{tab:resolutions_sm_emu}
\end{table}

The choice of quantifying the resolution in terms of the ratios $\sigma_{l}/\textrm{M}$ and $\sigma_{h}/\textrm{M}$ is motivated 
by the fact that the distributions in $m_{\Pgt\Pgt}$ and in $m_{\vis}$ may be shifted with respect to $m_{\Pgt\Pgt}^{\true}$.
While the distributions in $m_{\Pgt\Pgt}$ reconstructed by the different versions of the SVfit algorithm,
as well as by the CA method, peak close to $m_{\Pgt\Pgt}^{\true}$,
the distributions in $m_{\vis}$ exhibit significant shifts towards lower mass.
The shift of the $m_{\vis}$ distribution is in general highest for events in the $\Pe\Pgm$ decay channel and lowest for events in the $\tauh\tauh$ decay channel,
reflecting the fact that the visible $\Pgt$ decay products typically carry a lower fraction of $\Pgt$ lepton energy in leptonic compared to hadronic $\Pgt$ decays.
The shift is more pronounced in events selected in the $1$-jet and $2$-jet VBF categories and less pronounced in the $0$-jet category.
In particular in the $\tauh\tauh$ channel,
the $\pT$ cuts that are applied on the $\tauh$ remove most of the $\PZ/\Pggx \to \Pgt\Pgt$ events in the $0$-jet category,
except for a few events with large $m_{\vis}$.
The events selected in the $1$-jet and $2$-jet VBF categories typically have smaller $m_{\vis}$, as $m_{\vis}$ decreases proportional to the cosine of the angle between the $\Pgt$ leptons:
$m_{\vis} \approx \pT^{\vis(1)} \, \cosh\eta_{\vis(1)} \cdot \pT^{\vis(2)} \, \cosh\eta_{\vis(2)} \cdot \left( 1 - \cos\sphericalangle(\Pgt^{(1)},\Pgt^{(2)}) \right)$.
All algorithms can be trivially calibrated such that the median of each mass distribution coincides with the true mass of the $\Pgt$ lepton pair,
by scaling the output of the algorithm by a suitably chosen constant.
The advantage of quantifying the resolution in terms of the ratios $\sigma_{l}/\textrm{M}$ and $\sigma_{h}/\textrm{M}$ is that these ratios
are invariant under such scaling.

The ratio $\textrm{S}/(\textrm{S} + \textrm{B})$
of the number of SM $\PHiggs \to \Pgt\Pgt$ signal to the sum of signal plus $\PZ/\Pggx \to \Pgt\Pgt$ background events in general increases with jet multiplicity
and with the $\pT$ of the $\Pgt$ lepton pair ($\pT^{\PHiggs}$ respectively $\pT^{\PZ}$).
The categories with the highest signal-to-background ratio, the $1$-jet boosted and the $2$-jet VBF category,
in fact provide most of the sensitivity of the SM $\PHiggs \to \Pgt\Pgt$ analysis.
The improvement in mass resolution provided by the SVfit algorithm is largest in these categories,
enhancing the separation of the SM $\PHiggs \to \Pgt\Pgt$ signal from the $\PZ/\Pggx \to \Pgt\Pgt$ background where it matters most.
In the SM $\PHiggs \to \Pgt\Pgt$ analysis performed by the CMS collaboration during LHC Run $1$,
the use of $m_{\Pgt\Pgt}$ reconstructed by the SVfit algorithm has increased the expected significance for observing a signal by $\approx 40\%$ compared to $m_{\vis}$~\cite{HIG-13-004},
corresponding to a gain of a factor two in luminosity.

The reconstruction of $m_{\Pgt\Pgt}$ by the CA method performs suboptimal compared to SVfit.
In particular the pronounced high mass tails in the $m_{\Pgt\Pgt}$ distribution, arising from resolution effects,
significantly reduce the sensitivity for observing a signal,
as they cause a sizeable fraction of $\PZ/\Pggx \to \Pgt\Pgt$ background events
to be reconstructed near the signal region $m_{\Pgt\Pgt} \approx 125$~\GeV.
A further disadvantage of the CA method is that it fails to yield a physical solution for approximately half of the events,
whereas the SVfit algorithm provides a physical solution for every event. 

\begin{figure}
\setlength{\unitlength}{1mm}
\begin{center}
\begin{picture}(180,182)(0,0)
\put(-2.5, -2.0){\mbox{\includegraphics*[height=184mm]
{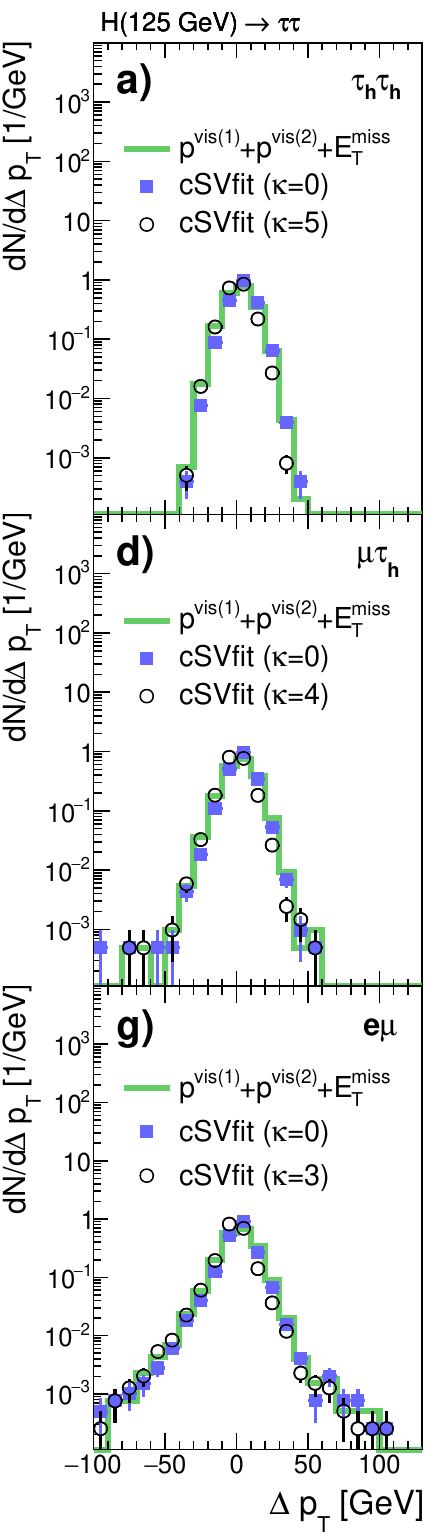}}}
\put(50.5, -2.0){\mbox{\includegraphics*[height=184mm]
{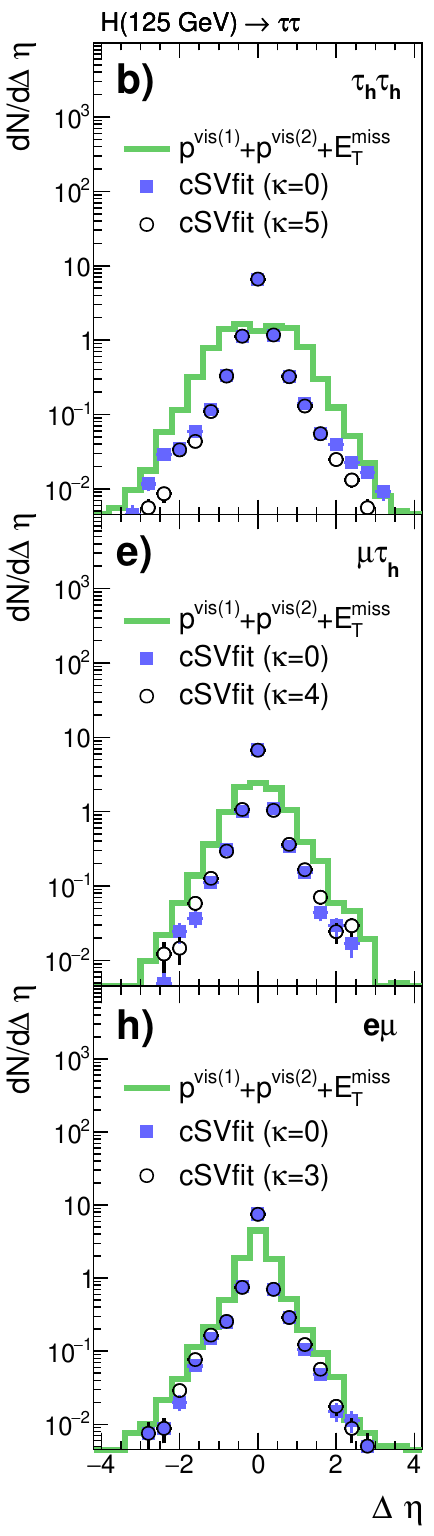}}}
\put(103.5, -2.0){\mbox{\includegraphics*[height=184mm]
{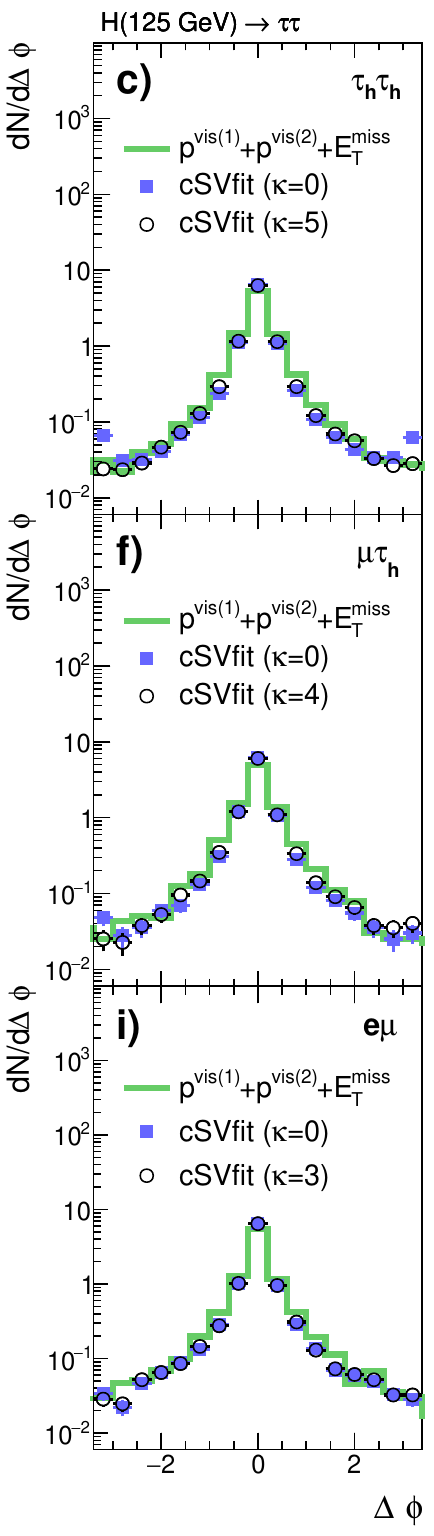}}}
\end{picture}
\end{center}
\caption{
  Resolution on $\pT$ (a,d,g), $\eta$ (b,e,h), and $\phi$ (c,f,i) of the $\Pgt$ lepton
  pair in SM $\PHiggs \to \Pgt\Pgt$ signal and $\PZ/\Pggx \to
  \Pgt\Pgt$ background events,
  separately for the decay channels $\tauh\tauh$ (a,b,c), $\Pgm\tauh$ (d,e,f),
  and $\Pe\Pgm$ (g,h,i).
  The SM $\PHiggs \to \Pgt\Pgt$ signal events are produced via the
  gluon fusion process.
  The resolutions on $\pT$ and $\phi$ achieved by computing the sum of
  the momenta of the visible $\Pgt$ decay products and $\MET$ are shown for comparison.
}
\label{fig:resolutions_sm_pT_eta_and_phi}
\end{figure}

The performance of the cSVfit algorithm to reconstruct the $\pT$,
$\eta$, and $\phi$ of the $\Pgt$ lepton pair is studied in SM $\PHiggs
\to \Pgt\Pgt$ signal events produced via the gluon fusion process,
separately for the decay channels $\tauh\tauh$, $\Pgm\tauh$, and $\Pe\Pgm$.
Distributions of the difference between reconstructed and true $\pT$,
$\eta$, and $\phi$ of the $\PHiggs$ boson are shown in
Fig.~\ref{fig:resolutions_sm_pT_eta_and_phi}.
Compared to the case that the $\PHiggs$ boson $\pT$ and $\phi$ are
reconstructed by taking the vectorial sum of the momenta of the visible $\Pgt$
decay products and of $\MET$, the cSVfit algorithm improves the
resolution by $10$--$20\%$. The pseudo-rapidity $\eta$ of the
$\PHiggs$ boson can only be reconstructed with the cSVfit algorithm.
Typical resolutions on the $\pT$, $\eta$, and $\phi$ of the $\PHiggs$ boson,
reconstructed by the cSVfit algorithm, amount to $10$~\GeV, $0.4$~\rad, and $0.8$~\rad,
respectively.

%
%

\begin{figure}
\setlength{\unitlength}{1mm}
\begin{center}
\begin{picture}(160,212)(0,0)
\put(-4.5, -2.0){\mbox{\includegraphics*[height=214mm]
  {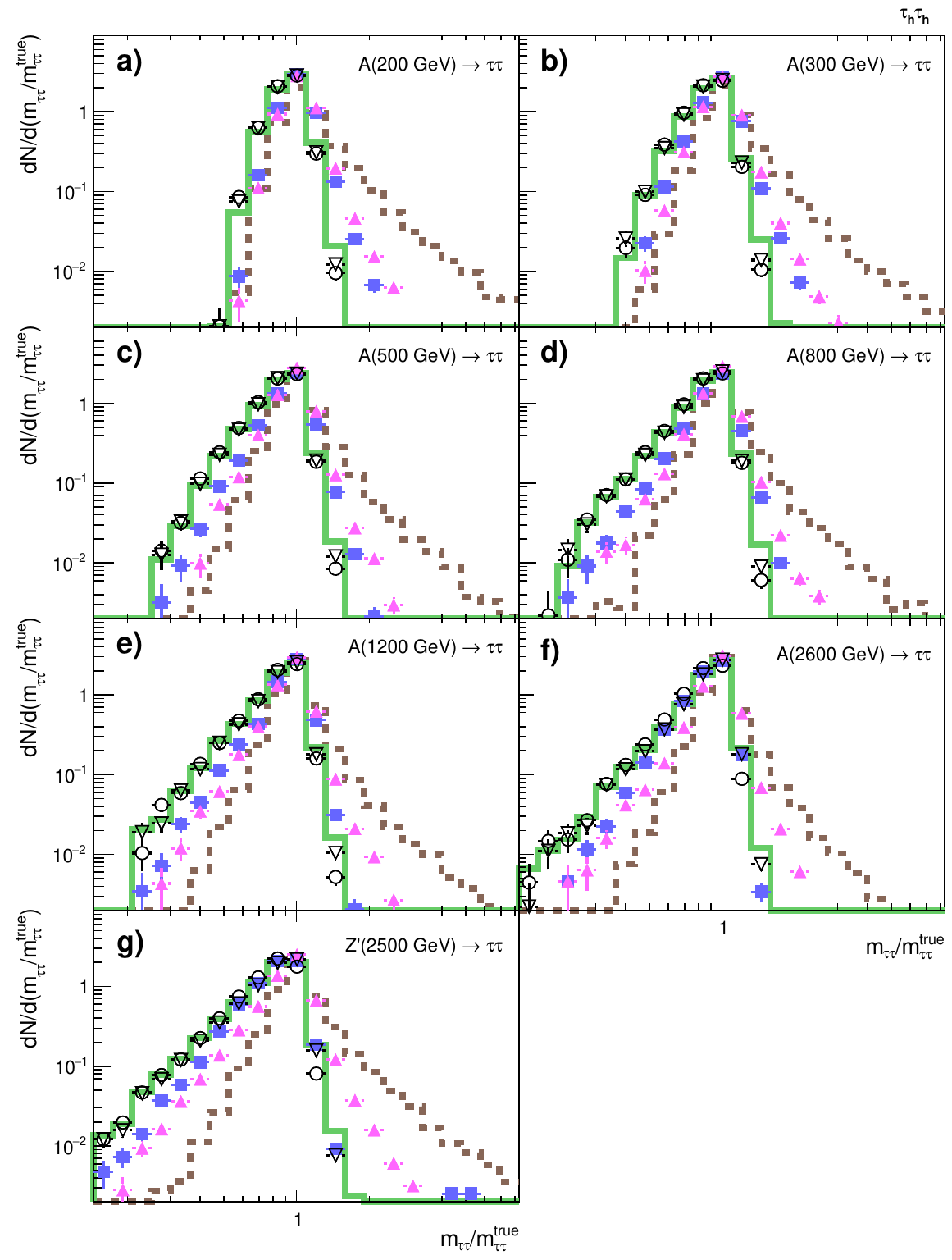}}}
\put(107.0, 9.0){\mbox{\includegraphics*[height=36mm]
  {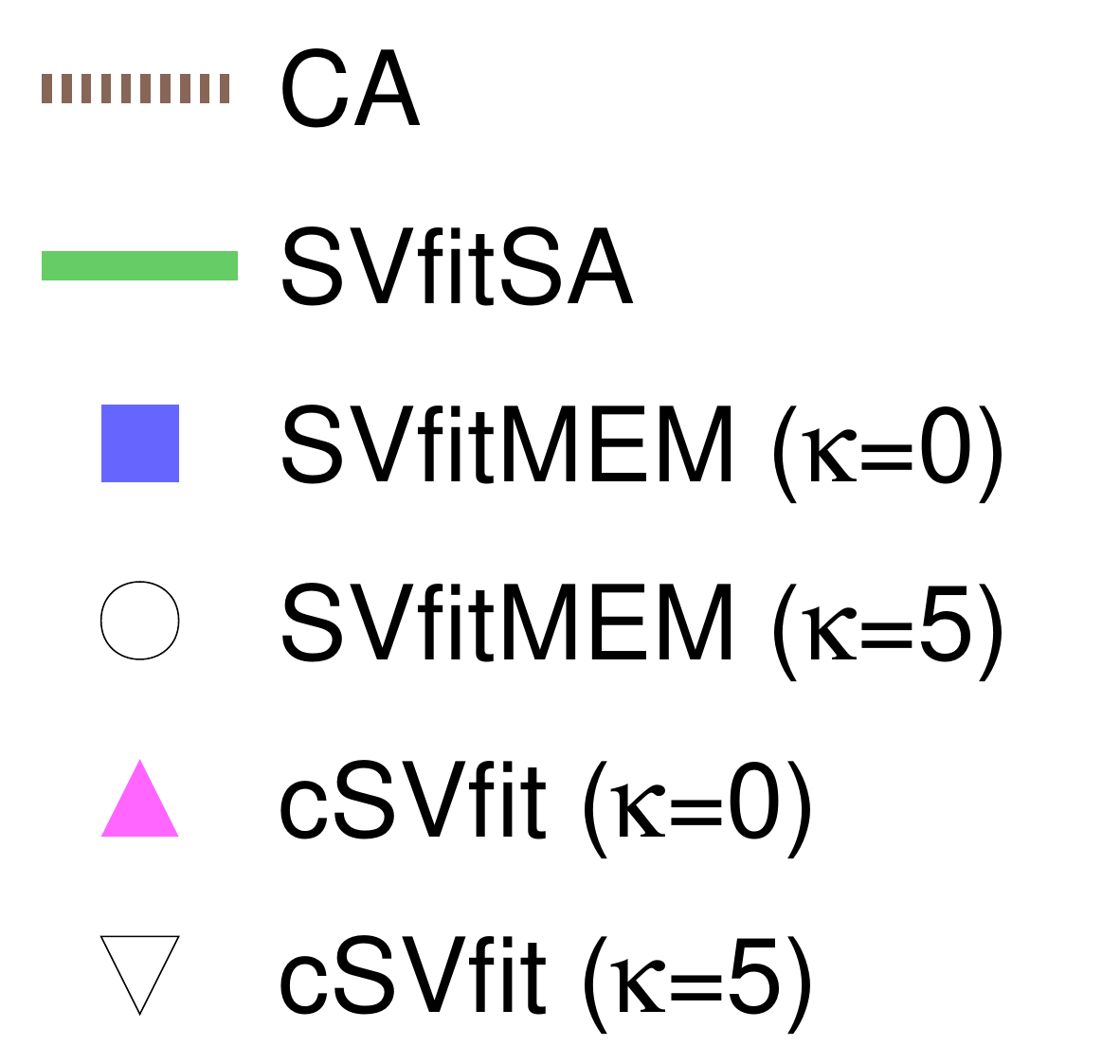}}}
\end{picture}
\end{center}
\caption{
  Distributions in $m_{\Pgt\Pgt}$ reconstructed by the CA method and different versions of the SVfit algorithm in simulated $\PHiggsps \to \Pgt\Pgt$ signal events of different mass $m_{\PHiggsps}$:
  $200$~\GeV (a), $300$~\GeV (b), $500$~\GeV (c), $800$~\GeV (d), $1200$~\GeV (e), and $2600$~\GeV (f), as well as in $\PZ' \to \Pgt\Pgt$ events of mass $m_{\PZ'} = 2500$~\GeV (g).
  The events are selected in the $\tauh\tauh$ decay channel.
  The axis of abscissae ranges from $0.15$ to $8$.
}
\label{fig:massDistributions_mssm_tautau}
\end{figure}

\begin{figure}
\setlength{\unitlength}{1mm}
\begin{center}
\begin{picture}(160,212)(0,0)
\put(-4.5, -2.0){\mbox{\includegraphics*[height=214mm]
  {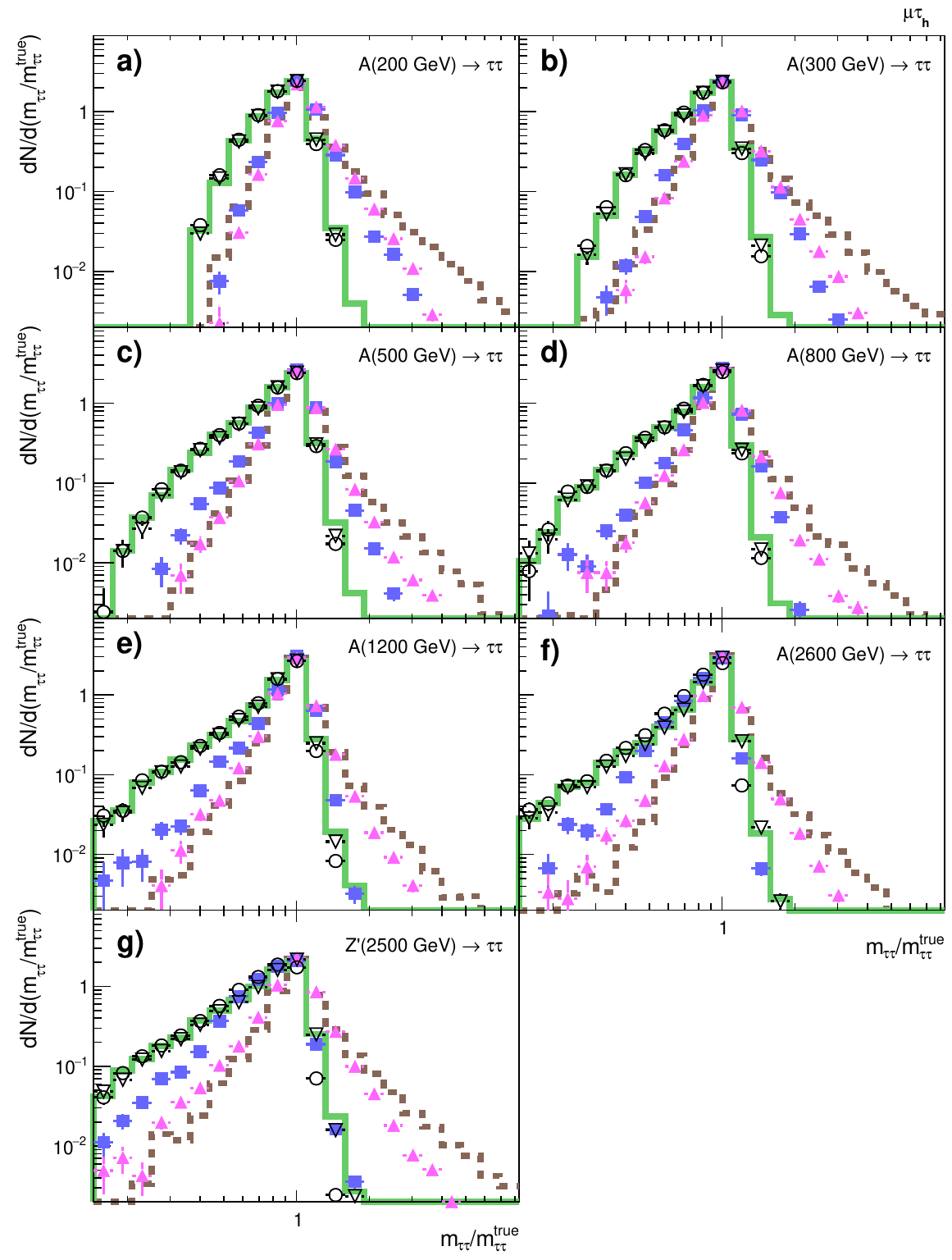}}}
\put(107.0, 9.0){\mbox{\includegraphics*[height=36mm]
  {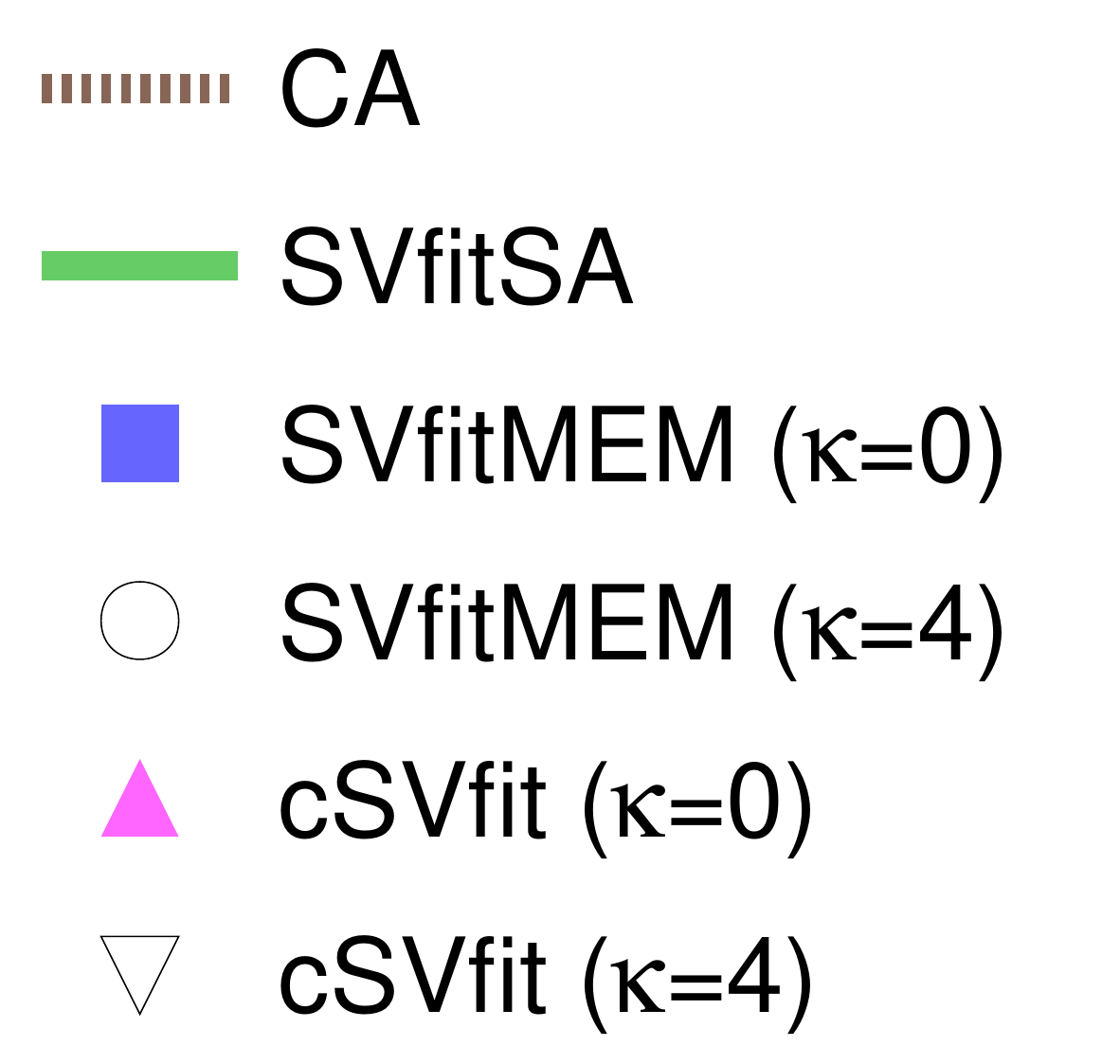}}}
\end{picture}
\end{center}
\caption{
  Distributions in $m_{\Pgt\Pgt}$ reconstructed by the CA method and different versions of the SVfit algorithm in simulated $\PHiggsps \to \Pgt\Pgt$ signal events of different mass $m_{\PHiggsps}$:
  $200$~\GeV (a), $300$~\GeV (b), $500$~\GeV (c), $800$~\GeV (d), $1200$~\GeV (e), and $2600$~\GeV (f), as well as in $\PZ' \to \Pgt\Pgt$ events of mass $m_{\PZ'} = 2500$~\GeV (g).
  The events are selected in the $\Pgm\tauh$ decay channel.
  The axis of abscissae ranges from $0.15$ to $8$.
}
\label{fig:massDistributions_mssm_mutau}
\end{figure}

\begin{figure}
\setlength{\unitlength}{1mm}
\begin{center}
\begin{picture}(160,212)(0,0)
\put(-4.5, -2.0){\mbox{\includegraphics*[height=214mm]
  {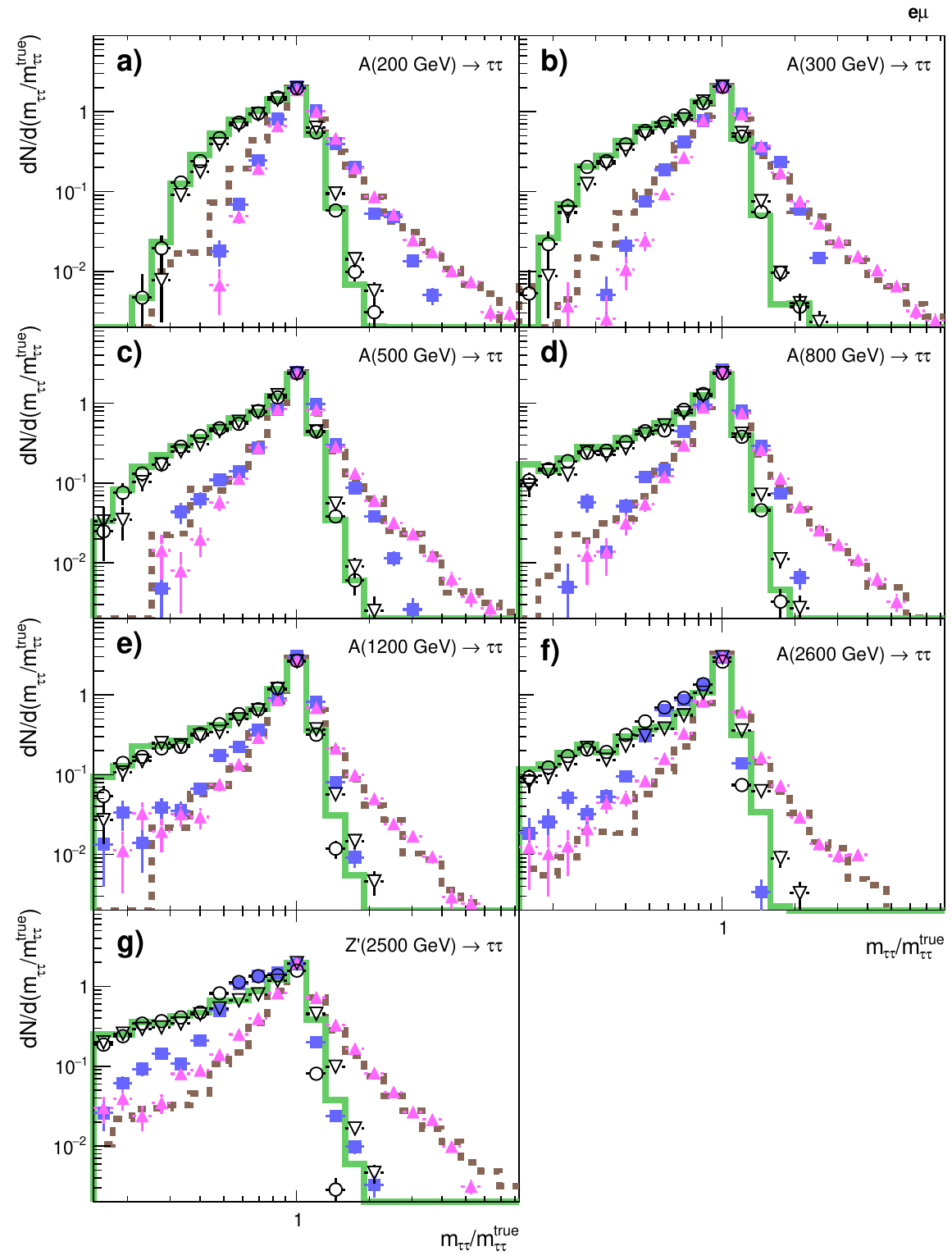}}}
\put(107.0, 9.0){\mbox{\includegraphics*[height=36mm]
  {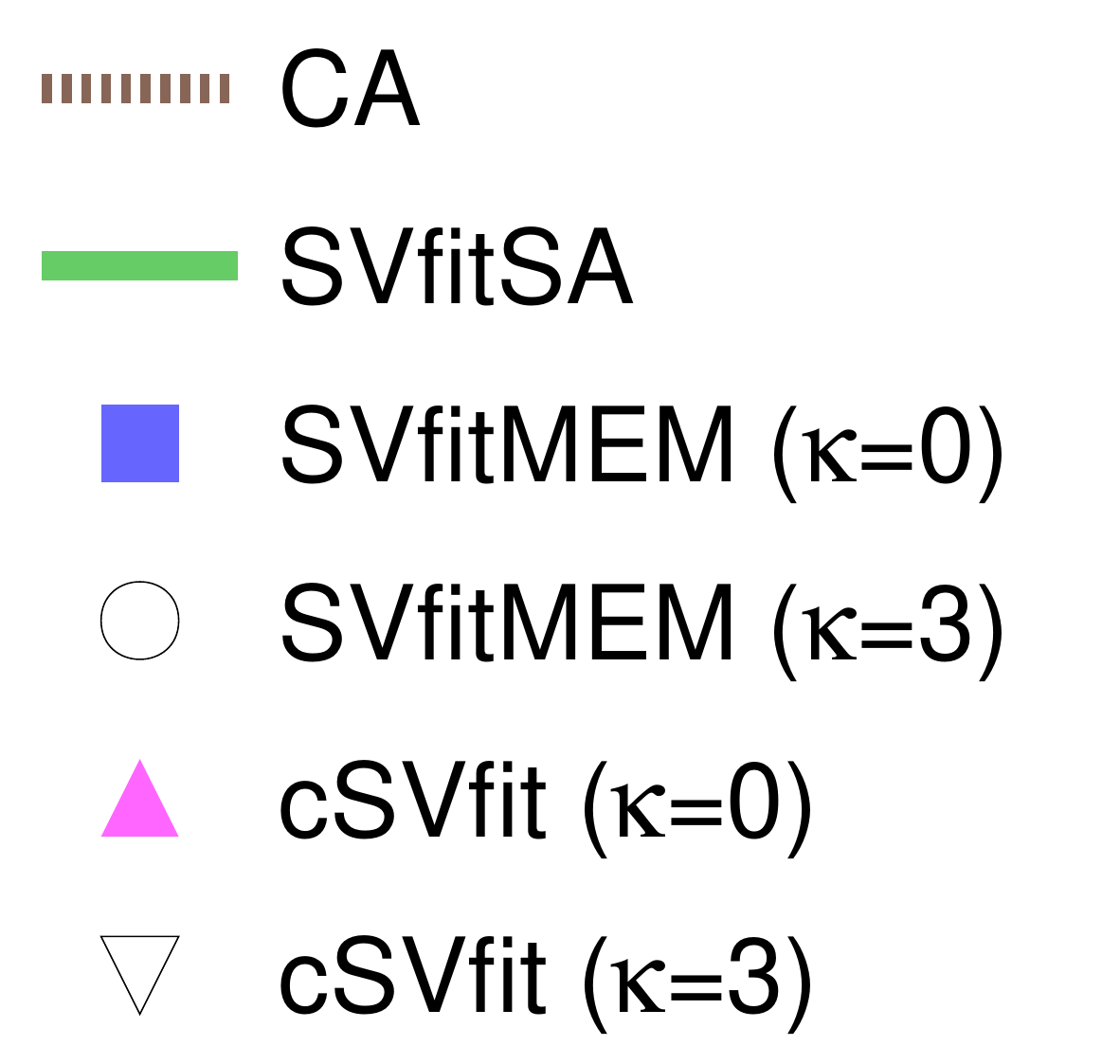}}}
\end{picture}
\end{center}
\caption{
  Distributions in $m_{\Pgt\Pgt}$ reconstructed by the CA method and different versions of the SVfit algorithm in simulated $\PHiggsps \to \Pgt\Pgt$ signal events of different mass $m_{\PHiggsps}$:
  $200$~\GeV (a), $300$~\GeV (b), $500$~\GeV (c), $800$~\GeV (d), $1200$~\GeV (e), and $2600$~\GeV (f), as well as in $\PZ' \to \Pgt\Pgt$ events of mass $m_{\PZ'} = 2500$~\GeV (g).
  The events are selected in the $\Pe\Pgm$ decay channel.
  The axis of abscissae ranges from $0.15$ to $8$.
}
\label{fig:massDistributions_mssm_emu}
\end{figure}

Distributions in $m_{\Pgt\Pgt}$ reconstructed by different versions of
the SVfit algorithm and by the CA method, as well as in $m_{\vis}$, in events containing hypothetical heavy
pseudoscalar Higgs bosons and heavy spin $1$ resonances are shown in
Figs.~\ref{fig:massDistributions_mssm_tautau}
to~\ref{fig:massDistributions_mssm_emu}.
The distributions in $m_{\Pgt\Pgt}$ reconstructed by the cSVfit and
SVfitMEM algorithms are shown for the case that the artificial regularization
term described in Section~\ref{sec:mem_logM} is used and for the case that it is not used. 

%
%

The SVfit algorithm significantly improves the separation of 
the heavy $\PHiggsps$ and $\PZ'$ boson signals from the irreducible $\PZ/\Pggx \to \Pgt\Pgt$
background in all three decay channels.
Numerical values for the median $\textrm{M}$ as well as for the resolutions $\sigma_{l}/\textrm{M}$ and $\sigma_{h}/\textrm{M}$ 
are given in Tables~\ref{tab:resolutions_mssm_tautau}
to~\ref{tab:resolutions_mssm_emu}.

%
%
\begin{table}
\begin{center}
\begin{tabular}{|l|ccc|ccc|}
\hline
\multicolumn{7}{|c|}{$\tauh\tauh$ decay channel} \\
\hline
\hline
\multirow{2}{17mm}{Sample} & \multicolumn{3}{c|}{$m_{\vis}$} & \multicolumn{3}{c|}{SVfitSA} \\
\cline{2-7}
 & $\textrm{M}$~[\GeV\unskip] & $\sigma_{l}/\textrm{M}$ & $\sigma_{h}/\textrm{M}$ & $\textrm{M}$~[\GeV\unskip] & $\sigma_{l}/\textrm{M}$ & $\sigma_{h}/\textrm{M}$ \\
\hline
$\PHiggsps \to \Pgt\Pgt$: & & & & & & \\ 
 $\quad$ $m_{\PHiggsps} = 200$~\GeV & $148$ & $0.172$ & $0.173$ & $191$ & $0.137$ & $0.111$ \\
 $\quad$ $m_{\PHiggsps} = 300$~\GeV & $202$ & $0.215$ & $0.233$ & $278$ & $0.176$ & $0.118$ \\
 $\quad$ $m_{\PHiggsps} = 500$~\GeV & $318$ & $0.271$ & $0.272$ & $455$ & $0.205$ & $0.128$ \\
 $\quad$ $m_{\PHiggsps} = 800$~\GeV & $487$ & $0.302$ & $0.305$ & $733$ & $0.205$ & $0.118$ \\
 $\quad$ $m_{\PHiggsps} = 1200$~\GeV & $724$ & $0.318$ & $0.312$ & $1101$ & $0.206$ & $0.116$ \\
 $\quad$ $m_{\PHiggsps} = 1800$~\GeV & $1078$ & $0.329$ & $0.314$ & $1662$ & $0.196$ & $0.110$ \\
 $\quad$ $m_{\PHiggsps} = 2600$~\GeV & $1553$ & $0.339$ & $0.316$ & $2414$ & $0.198$ & $0.103$ \\
$\PZ' \to \Pgt\Pgt$: & & & & & & \\ 
 $\quad$ $m_{\PZ'} = 2500$~\GeV & $1486$ & $0.340$ & $0.325$ & $2196$ & $0.245$ & $0.147$ \\
\hline
\end{tabular}

\begin{tabular}{|l|ccc|ccc|}
\hline
\multirow{2}{17mm}{Sample} & \multicolumn{3}{c|}{cSVfit, $\kappa=0$} & \multicolumn{3}{c|}{cSVfit, $\kappa=5$} \\
\cline{2-7}
 & $\textrm{M}$~[\GeV\unskip] & $\sigma_{l}/\textrm{M}$ & $\sigma_{h}/\textrm{M}$ & $\textrm{M}$~[\GeV\unskip] & $\sigma_{l}/\textrm{M}$ & $\sigma_{h}/\textrm{M}$ \\
\hline
$\PHiggsps \to \Pgt\Pgt$: & & & & & & \\ 
 $\quad$ $m_{\PHiggsps} = 200$~\GeV & $207$ & $0.116$ & $0.155$ & $188$ & $0.140$ & $0.113$ \\
 $\quad$ $m_{\PHiggsps} = 300$~\GeV & $303$ & $0.135$ & $0.158$ & $275$ & $0.178$ & $0.118$ \\
 $\quad$ $m_{\PHiggsps} = 500$~\GeV & $497$ & $0.140$ & $0.148$ & $453$ & $0.202$ & $0.127$ \\
 $\quad$ $m_{\PHiggsps} = 800$~\GeV & $788$ & $0.138$ & $0.135$ & $731$ & $0.205$ & $0.116$ \\
 $\quad$ $m_{\PHiggsps} = 1200$~\GeV & $1179$ & $0.141$ & $0.125$ & $1101$ & $0.201$ & $0.113$ \\
 $\quad$ $m_{\PHiggsps} = 1800$~\GeV & $1769$ & $0.134$ & $0.117$ & $1661$ & $0.193$ & $0.107$ \\
 $\quad$ $m_{\PHiggsps} = 2600$~\GeV & $2560$ & $0.139$ & $0.113$ & $2413$ & $0.192$ & $0.102$ \\
$\PZ' \to \Pgt\Pgt$: & & & & & & \\ 
 $\quad$ $m_{\PZ'} = 2500$~\GeV & $2445$ & $0.182$ & $0.154$ & $2219$ & $0.242$ & $0.140$ \\
\hline
\end{tabular}

\begin{tabular}{|l|ccc|ccc|}
\hline
\multirow{2}{17mm}{Sample} & \multicolumn{3}{c|}{SVfitMEM, $\kappa=0$} & \multicolumn{3}{c|}{SVfitMEM, $\kappa=5$} \\
\cline{2-7}
 & $\textrm{M}$~[\GeV\unskip] & $\sigma_{l}/\textrm{M}$ & $\sigma_{h}/\textrm{M}$ & $\textrm{M}$~[\GeV\unskip] & $\sigma_{l}/\textrm{M}$ & $\sigma_{h}/\textrm{M}$ \\
\hline
$\PHiggsps \to \Pgt\Pgt$: & & & & & & \\ 
 $\quad$ $m_{\PHiggsps} = 200$~\GeV & $203$ & $0.118$ & $0.138$ & $188$ & $0.139$ & $0.112$ \\
 $\quad$ $m_{\PHiggsps} = 300$~\GeV & $298$ & $0.146$ & $0.144$ & $273$ & $0.179$ & $0.124$ \\
 $\quad$ $m_{\PHiggsps} = 500$~\GeV & $495$ & $0.168$ & $0.234$ & $451$ & $0.211$ & $0.132$ \\
 $\quad$ $m_{\PHiggsps} = 800$~\GeV & $791$ & $0.164$ & $0.419$ & $723$ & $0.207$ & $0.123$ \\
 $\quad$ $m_{\PHiggsps} = 1200$~\GeV & $1166$ & $0.151$ & $0.134$ & $1088$ & $0.204$ & $0.118$ \\
 $\quad$ $m_{\PHiggsps} = 1800$~\GeV & $1684$ & $0.152$ & $0.104$ & $1619$ & $0.193$ & $0.112$ \\
 $\quad$ $m_{\PHiggsps} = 2600$~\GeV & $2399$ & $0.183$ & $0.106$ & $2325$ & $0.205$ & $0.117$ \\
$\PZ' \to \Pgt\Pgt$: & & & & & & \\ 
 $\quad$ $m_{\PZ'} = 2500$~\GeV & $2222$ & $0.220$ & $0.142$ & $2131$ & $0.243$ & $0.152$ \\
\hline
\end{tabular}
\end{center}
\caption{
  Median $\textrm{M}$ and resolutions $\sigma_{l}/\textrm{M}$ and $\sigma_{h}/\textrm{M}$
  of the distributions in $m_{\vis}$ 
  and in $m_{\Pgt\Pgt}$ reconstructed by different versions of SVfit algorithm,
  in simulated signal events containing either heavy pseudoscalar Higgs
  bosons $\PHiggsps$ or heavy spin $1$ resonances $\PZ'$
  and in simulated $\PZ/\Pggx \to \Pgt\Pgt$ background events.
  The events are selected in the $\tauh\tauh$ decay channel.
  The difference between the two resolutions $\sigma_{l}/\textrm{M}$ and $\sigma_{h}/\textrm{M}$
  is explained in the text.
}
\label{tab:resolutions_mssm_tautau}
\end{table}

\begin{table}
\begin{center}
\begin{tabular}{|l|ccc|ccc|}
\hline
\multicolumn{7}{|c|}{$\Pgm\tauh$ decay channel} \\
\hline
\hline
\multirow{2}{17mm}{Sample} & \multicolumn{3}{c|}{$m_{\vis}$} & \multicolumn{3}{c|}{SVfitSA} \\
\cline{2-7}
 & $\textrm{M}$~[\GeV\unskip] & $\sigma_{l}/\textrm{M}$ & $\sigma_{h}/\textrm{M}$ & $\textrm{M}$~[\GeV\unskip] & $\sigma_{l}/\textrm{M}$ & $\sigma_{h}/\textrm{M}$ \\
\hline
$\PHiggsps \to \Pgt\Pgt$: & & & & & & \\ 
 $\quad$ $m_{\PHiggsps} = 200$~\GeV & $113$ & $0.252$ & $0.288$ & $187$ & $0.201$ & $0.139$ \\
 $\quad$ $m_{\PHiggsps} = 300$~\GeV & $153$ & $0.294$ & $0.352$ & $276$ & $0.237$ & $0.138$ \\
 $\quad$ $m_{\PHiggsps} = 500$~\GeV & $236$ & $0.346$ & $0.412$ & $460$ & $0.266$ & $0.134$ \\
 $\quad$ $m_{\PHiggsps} = 800$~\GeV & $361$ & $0.380$ & $0.453$ & $742$ & $0.259$ & $0.118$ \\
 $\quad$ $m_{\PHiggsps} = 1200$~\GeV & $529$ & $0.408$ & $0.477$ & $1121$ & $0.252$ & $0.108$ \\
 $\quad$ $m_{\PHiggsps} = 1800$~\GeV & $779$ & $0.434$ & $0.499$ & $1694$ & $0.244$ & $0.099$ \\
 $\quad$ $m_{\PHiggsps} = 2600$~\GeV & $1111$ & $0.454$ & $0.511$ & $2463$ & $0.228$ & $0.095$ \\
$\PZ' \to \Pgt\Pgt$: & & & & & & \\ 
 $\quad$ $m_{\PZ'} = 2500$~\GeV & $1077$ & $0.438$ & $0.485$ & $2212$ & $0.312$ & $0.155$ \\
\hline
\end{tabular}

\begin{tabular}{|l|ccc|ccc|}
\hline
\multirow{2}{17mm}{Sample} & \multicolumn{3}{c|}{cSVfit, $\kappa=0$} & \multicolumn{3}{c|}{cSVfit, $\kappa=4$} \\
\cline{2-7}
 & $\textrm{M}$~[\GeV\unskip] & $\sigma_{l}/\textrm{M}$ & $\sigma_{h}/\textrm{M}$ & $\textrm{M}$~[\GeV\unskip] & $\sigma_{l}/\textrm{M}$ & $\sigma_{h}/\textrm{M}$ \\
\hline
$\PHiggsps \to \Pgt\Pgt$: & & & & & & \\ 
 $\quad$ $m_{\PHiggsps} = 200$~\GeV & $214$ & $0.138$ & $0.292$ & $187$ & $0.200$ & $0.138$ \\
 $\quad$ $m_{\PHiggsps} = 300$~\GeV & $313$ & $0.136$ & $0.261$ & $276$ & $0.235$ & $0.137$ \\
 $\quad$ $m_{\PHiggsps} = 500$~\GeV & $510$ & $0.132$ & $0.226$ & $461$ & $0.261$ & $0.129$ \\
 $\quad$ $m_{\PHiggsps} = 800$~\GeV & $810$ & $0.129$ & $0.191$ & $742$ & $0.250$ & $0.116$ \\
 $\quad$ $m_{\PHiggsps} = 1200$~\GeV & $1204$ & $0.127$ & $0.165$ & $1121$ & $0.244$ & $0.105$ \\
 $\quad$ $m_{\PHiggsps} = 1800$~\GeV & $1802$ & $0.122$ & $0.152$ & $1695$ & $0.235$ & $0.098$ \\
 $\quad$ $m_{\PHiggsps} = 2600$~\GeV & $2603$ & $0.118$ & $0.147$ & $2467$ & $0.221$ & $0.094$ \\
$\PZ' \to \Pgt\Pgt$: & & & & & & \\ 
 $\quad$ $m_{\PZ'} = 2500$~\GeV & $2541$ & $0.165$ & $0.259$ & $2234$ & $0.303$ & $0.149$ \\
\hline
\end{tabular}

\begin{tabular}{|l|ccc|ccc|}
\hline
\multirow{2}{17mm}{Sample} & \multicolumn{3}{c|}{SVfitMEM, $\kappa=0$} & \multicolumn{3}{c|}{SVfitMEM, $\kappa=4$} \\
\cline{2-7}
 & $\textrm{M}$~[\GeV\unskip] & $\sigma_{l}/\textrm{M}$ & $\sigma_{h}/\textrm{M}$ & $\textrm{M}$~[\GeV\unskip] & $\sigma_{l}/\textrm{M}$ & $\sigma_{h}/\textrm{M}$ \\
\hline
$\PHiggsps \to \Pgt\Pgt$: & & & & & & \\ 
 $\quad$ $m_{\PHiggsps} = 200$~\GeV & $208$ & $0.136$ & $0.217$ & $186$ & $0.200$ & $0.135$ \\
 $\quad$ $m_{\PHiggsps} = 300$~\GeV & $306$ & $0.156$ & $0.208$ & $274$ & $0.241$ & $0.139$ \\
 $\quad$ $m_{\PHiggsps} = 500$~\GeV & $504$ & $0.161$ & $0.171$ & $458$ & $0.270$ & $0.131$ \\
 $\quad$ $m_{\PHiggsps} = 800$~\GeV & $795$ & $0.156$ & $0.155$ & $733$ & $0.256$ & $0.122$ \\
 $\quad$ $m_{\PHiggsps} = 1200$~\GeV & $1181$ & $0.152$ & $0.113$ & $1111$ & $0.260$ & $0.107$ \\
 $\quad$ $m_{\PHiggsps} = 1800$~\GeV & $1708$ & $0.167$ & $0.091$ & $1641$ & $0.239$ & $0.103$ \\
 $\quad$ $m_{\PHiggsps} = 2600$~\GeV & $2433$ & $0.213$ & $0.092$ & $2350$ & $0.255$ & $0.109$ \\
$\PZ' \to \Pgt\Pgt$: & & & & & & \\ 
 $\quad$ $m_{\PZ'} = 2500$~\GeV & $2207$ & $0.257$ & $0.149$ & $2081$ & $0.298$ & $0.178$ \\
\hline
\end{tabular}
\end{center}
\caption{
  Median $\textrm{M}$ and resolutions $\sigma_{l}/\textrm{M}$ and $\sigma_{h}/\textrm{M}$
  of the distributions in $m_{\vis}$ 
  and in $m_{\Pgt\Pgt}$ reconstructed by different versions of SVfit algorithm,
  in simulated signal events containing either heavy pseudoscalar Higgs
  bosons $\PHiggsps$ or heavy spin $1$ resonances $\PZ'$
  and in simulated $\PZ/\Pggx \to \Pgt\Pgt$ background events.
  The events are selected in the $\Pgm\tauh$ decay channel.
  The difference between the two resolutions $\sigma_{l}/\textrm{M}$ and $\sigma_{h}/\textrm{M}$
  is explained in the text.
}
\label{tab:resolutions_mssm_mutau}
\end{table}

\begin{table}
\begin{center}
\begin{tabular}{|l|ccc|ccc|}
\hline
\multicolumn{7}{|c|}{$\Pe\Pgm$ decay channel} \\
\hline
\hline
\multirow{2}{17mm}{Sample} & \multicolumn{3}{c|}{$m_{\vis}$} & \multicolumn{3}{c|}{SVfitSA} \\
\cline{2-7}
 & $\textrm{M}$~[\GeV\unskip] & $\sigma_{l}/\textrm{M}$ & $\sigma_{h}/\textrm{M}$ & $\textrm{M}$~[\GeV\unskip] & $\sigma_{l}/\textrm{M}$ & $\sigma_{h}/\textrm{M}$ \\
\hline
$\PHiggsps \to \Pgt\Pgt$: & & & & & & \\ 
 $\quad$ $m_{\PHiggsps} = 200$~\GeV & $77$ & $0.318$ & $0.426$ & $181$ & $0.303$ & $0.187$ \\
 $\quad$ $m_{\PHiggsps} = 300$~\GeV & $104$ & $0.372$ & $0.514$ & $272$ & $0.355$ & $0.176$ \\
 $\quad$ $m_{\PHiggsps} = 500$~\GeV & $162$ & $0.427$ & $0.564$ & $464$ & $0.351$ & $0.136$ \\
 $\quad$ $m_{\PHiggsps} = 800$~\GeV & $246$ & $0.463$ & $0.620$ & $744$ & $0.348$ & $0.133$ \\
 $\quad$ $m_{\PHiggsps} = 1200$~\GeV & $357$ & $0.487$ & $0.661$ & $1130$ & $0.353$ & $0.116$ \\
 $\quad$ $m_{\PHiggsps} = 1800$~\GeV & $538$ & $0.517$ & $0.659$ & $1727$ & $0.320$ & $0.094$ \\
 $\quad$ $m_{\PHiggsps} = 2600$~\GeV & $748$ & $0.528$ & $0.716$ & $2501$ & $0.312$ & $0.086$ \\
$\PZ' \to \Pgt\Pgt$: & & & & & & \\ 
 $\quad$ $m_{\PZ'} = 2500$~\GeV & $733$ & $0.522$ & $0.694$ & $2198$ & $0.438$ & $0.190$ \\
\hline
\end{tabular}

\begin{tabular}{|l|ccc|ccc|}
\hline
\multirow{2}{17mm}{Sample} & \multicolumn{3}{c|}{cSVfit, $\kappa=0$} & \multicolumn{3}{c|}{cSVfit, $\kappa=3$} \\
\cline{2-7}
 & $\textrm{M}$~[\GeV\unskip] & $\sigma_{l}/\textrm{M}$ & $\sigma_{h}/\textrm{M}$ & $\textrm{M}$~[\GeV\unskip] & $\sigma_{l}/\textrm{M}$ & $\sigma_{h}/\textrm{M}$ \\
\hline
$\PHiggsps \to \Pgt\Pgt$: & & & & & & \\ 
 $\quad$ $m_{\PHiggsps} = 200$~\GeV & $223$ & $0.162$ & $0.463$ & $186$ & $0.272$ & $0.191$ \\
 $\quad$ $m_{\PHiggsps} = 300$~\GeV & $321$ & $0.155$ & $0.444$ & $279$ & $0.325$ & $0.167$ \\
 $\quad$ $m_{\PHiggsps} = 500$~\GeV & $518$ & $0.136$ & $0.369$ & $469$ & $0.322$ & $0.133$ \\
 $\quad$ $m_{\PHiggsps} = 800$~\GeV & $815$ & $0.130$ & $0.310$ & $751$ & $0.319$ & $0.128$ \\
 $\quad$ $m_{\PHiggsps} = 1200$~\GeV & $1214$ & $0.126$ & $0.263$ & $1141$ & $0.317$ & $0.111$ \\
 $\quad$ $m_{\PHiggsps} = 1800$~\GeV & $1812$ & $0.115$ & $0.202$ & $1732$ & $0.267$ & $0.097$ \\
 $\quad$ $m_{\PHiggsps} = 2600$~\GeV & $2607$ & $0.119$ & $0.172$ & $2518$ & $0.268$ & $0.089$ \\
$\PZ' \to \Pgt\Pgt$: & & & & & & \\ 
 $\quad$ $m_{\PZ'} = 2500$~\GeV & $2578$ & $0.186$ & $0.495$ & $2268$ & $0.411$ & $0.181$ \\
\hline
\end{tabular}

\begin{tabular}{|l|ccc|ccc|}
\hline
\multirow{2}{17mm}{Sample} & \multicolumn{3}{c|}{SVfitMEM, $\kappa=0$} & \multicolumn{3}{c|}{SVfitMEM, $\kappa=3$} \\
\cline{2-7}
 & $\textrm{M}$~[\GeV\unskip] & $\sigma_{l}/\textrm{M}$ & $\sigma_{h}/\textrm{M}$ & $\textrm{M}$~[\GeV\unskip] & $\sigma_{l}/\textrm{M}$ & $\sigma_{h}/\textrm{M}$ \\
\hline
$\PHiggsps \to \Pgt\Pgt$: & & & & & & \\ 
 $\quad$ $m_{\PHiggsps} = 200$~\GeV & $214$ & $0.151$ & $0.397$ & $183$ & $0.288$ & $0.181$ \\
 $\quad$ $m_{\PHiggsps} = 300$~\GeV & $316$ & $0.172$ & $0.321$ & $275$ & $0.342$ & $0.171$ \\
 $\quad$ $m_{\PHiggsps} = 500$~\GeV & $517$ & $0.146$ & $0.227$ & $471$ & $0.336$ & $0.128$ \\
 $\quad$ $m_{\PHiggsps} = 800$~\GeV & $806$ & $0.156$ & $0.200$ & $745$ & $0.320$ & $0.129$ \\
 $\quad$ $m_{\PHiggsps} = 1200$~\GeV & $1200$ & $0.154$ & $0.122$ & $1131$ & $0.342$ & $0.106$ \\
 $\quad$ $m_{\PHiggsps} = 1800$~\GeV & $1737$ & $0.179$ & $0.081$ & $1663$ & $0.285$ & $0.101$ \\
 $\quad$ $m_{\PHiggsps} = 2600$~\GeV & $2456$ & $0.270$ & $0.082$ & $2364$ & $0.333$ & $0.108$ \\
$\PZ' \to \Pgt\Pgt$: & & & & & & \\ 
 $\quad$ $m_{\PZ'} = 2500$~\GeV & $2125$ & $0.290$ & $0.198$ & $1938$ & $0.367$ & $0.270$ \\
\hline
\end{tabular}
\end{center}
\caption{
  Median $\textrm{M}$ and resolutions $\sigma_{l}/\textrm{M}$ and $\sigma_{h}/\textrm{M}$
  of the distributions in $m_{\vis}$ 
  and in $m_{\Pgt\Pgt}$ reconstructed by different versions of SVfit algorithm,
  in simulated signal events containing either heavy pseudoscalar Higgs
  bosons $\PHiggsps$ or heavy spin $1$ resonances $\PZ'$
  and in simulated $\PZ/\Pggx \to \Pgt\Pgt$ background events.
  The events are selected in the $\Pe\Pgm$ decay channel.
  The difference between the two resolutions $\sigma_{l}/\textrm{M}$ and $\sigma_{h}/\textrm{M}$
  is explained in the text.
}
\label{tab:resolutions_mssm_emu}
\end{table}

We interpret the fact that the performance of the SVfitMEM algorithm is similar for an $\PHiggsps \to \Pgt\Pgt$ signal of mass $2600$~\GeV
and for a $\PZ' \to \Pgt\Pgt$ signal of mass $2500$~\GeV as evidence that the usage of a
LO ME for the gluon fusion process $\Pg\Pg \to \PHiggs$ in the SVfitMEM algorithm represents no limitation for using
the SVfitMEM algorithm in data analyses of $\Pgt$ lepton pair production other than studies of Higgs boson production.

\begin{figure}
\setlength{\unitlength}{1mm}
\begin{center}
\begin{picture}(160,214)(0,0)
\put(-2.5, 150.0){\mbox{\includegraphics*[height=70mm]
  {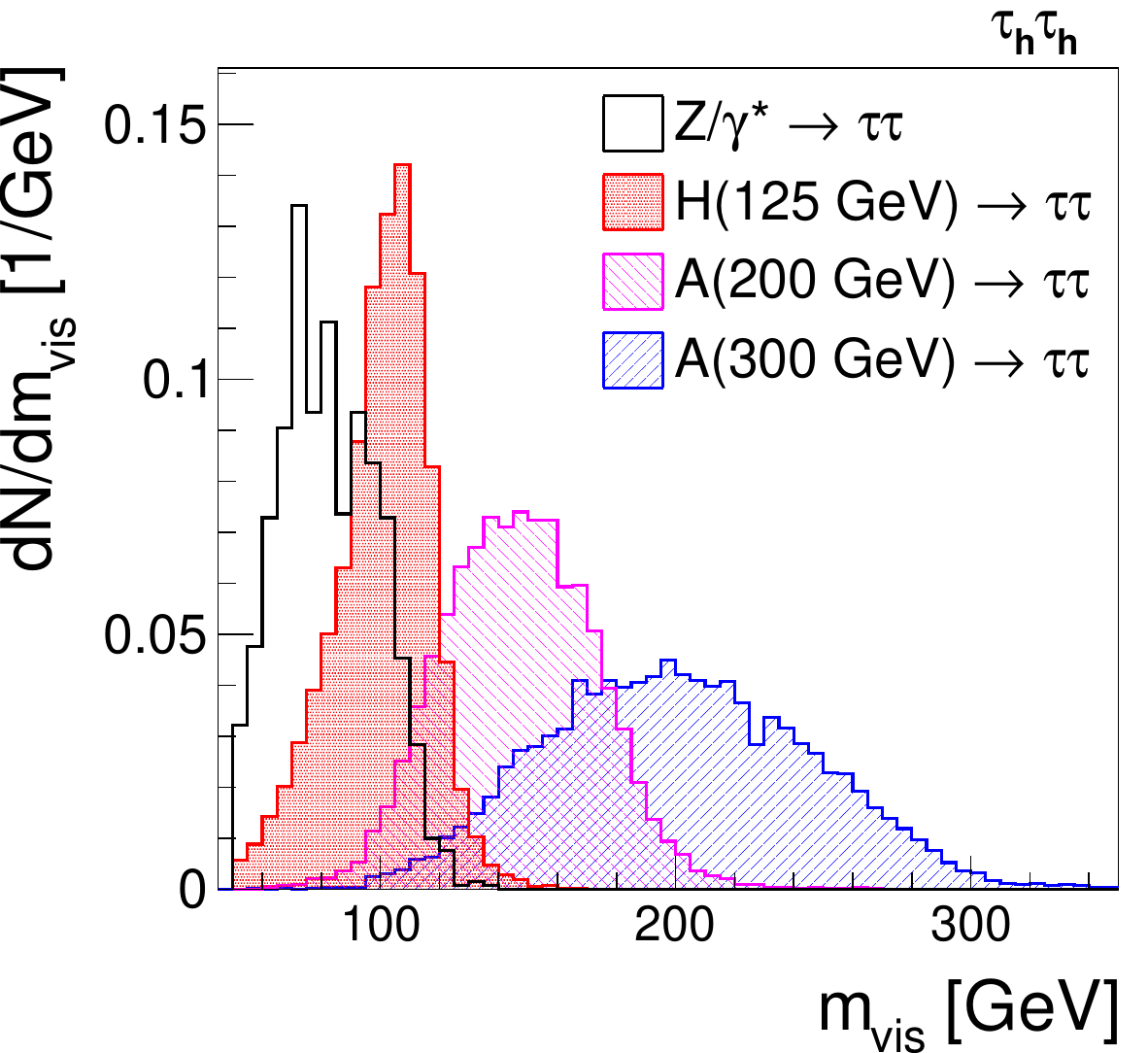}}}
\put(80.0, 150.0){\mbox{\includegraphics*[height=70mm]
  {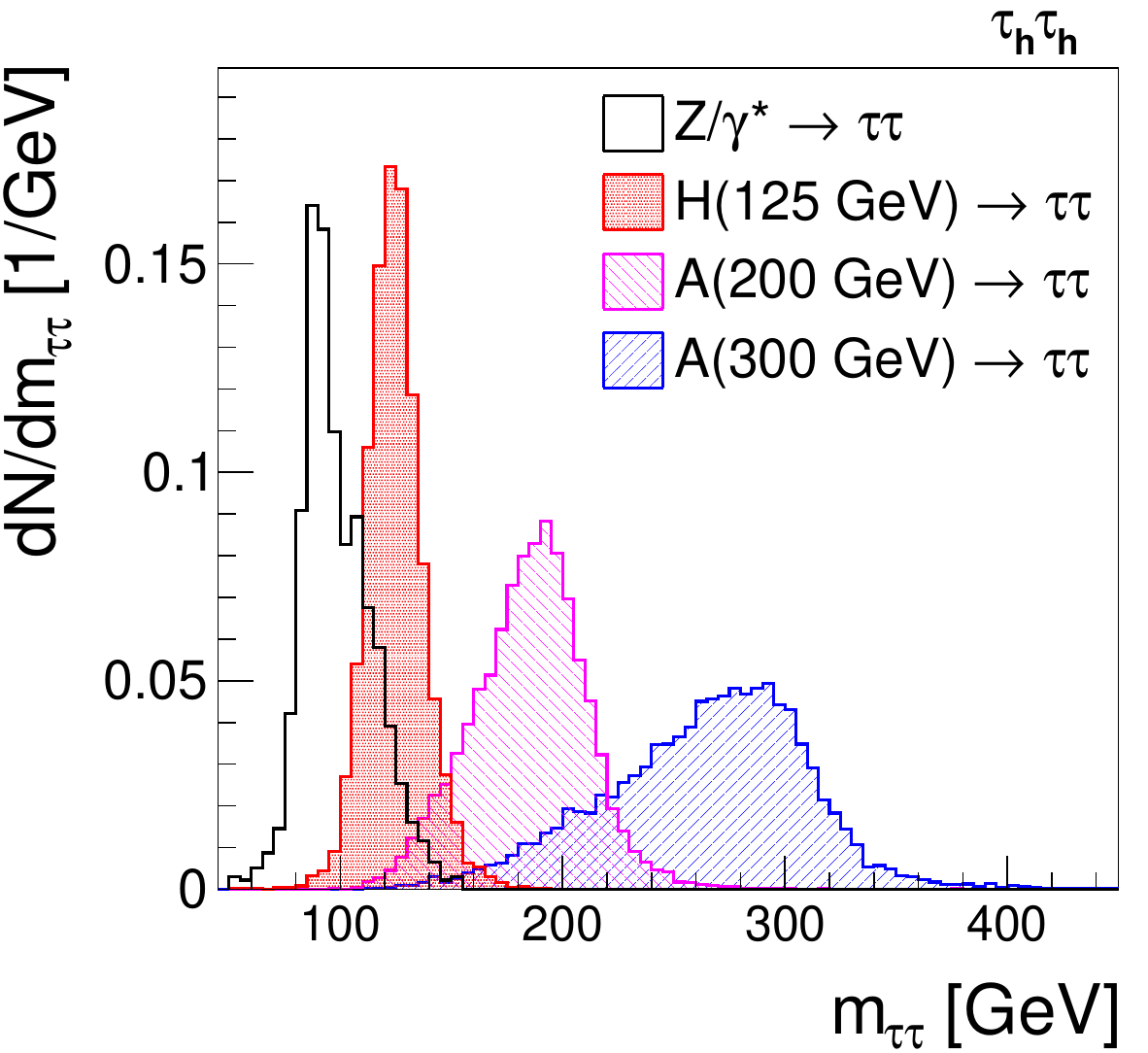}}}
\put(-2.5, 75.0){\mbox{\includegraphics*[height=70mm]
  {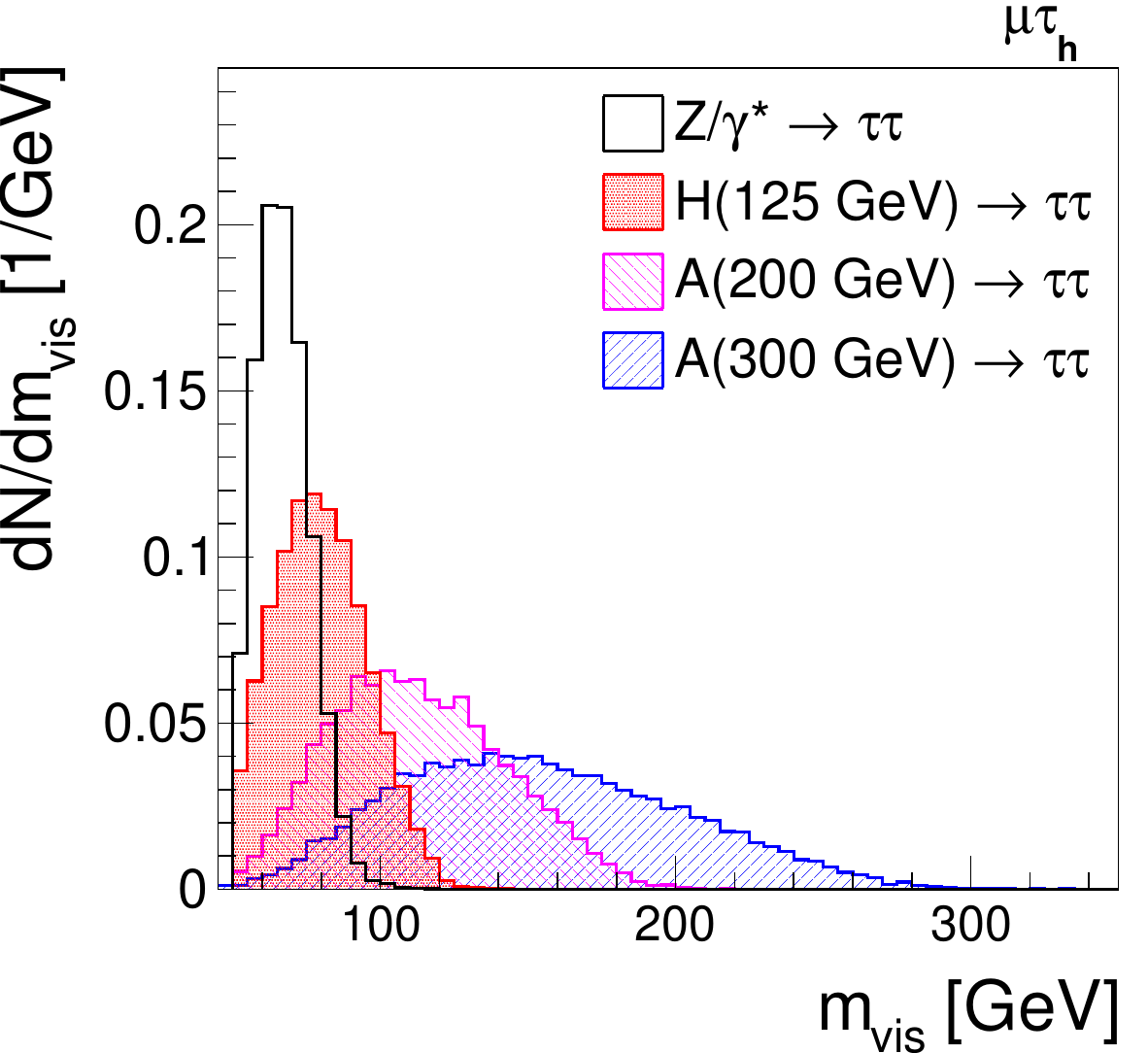}}}
\put(80.0, 75.0){\mbox{\includegraphics*[height=70mm]
  {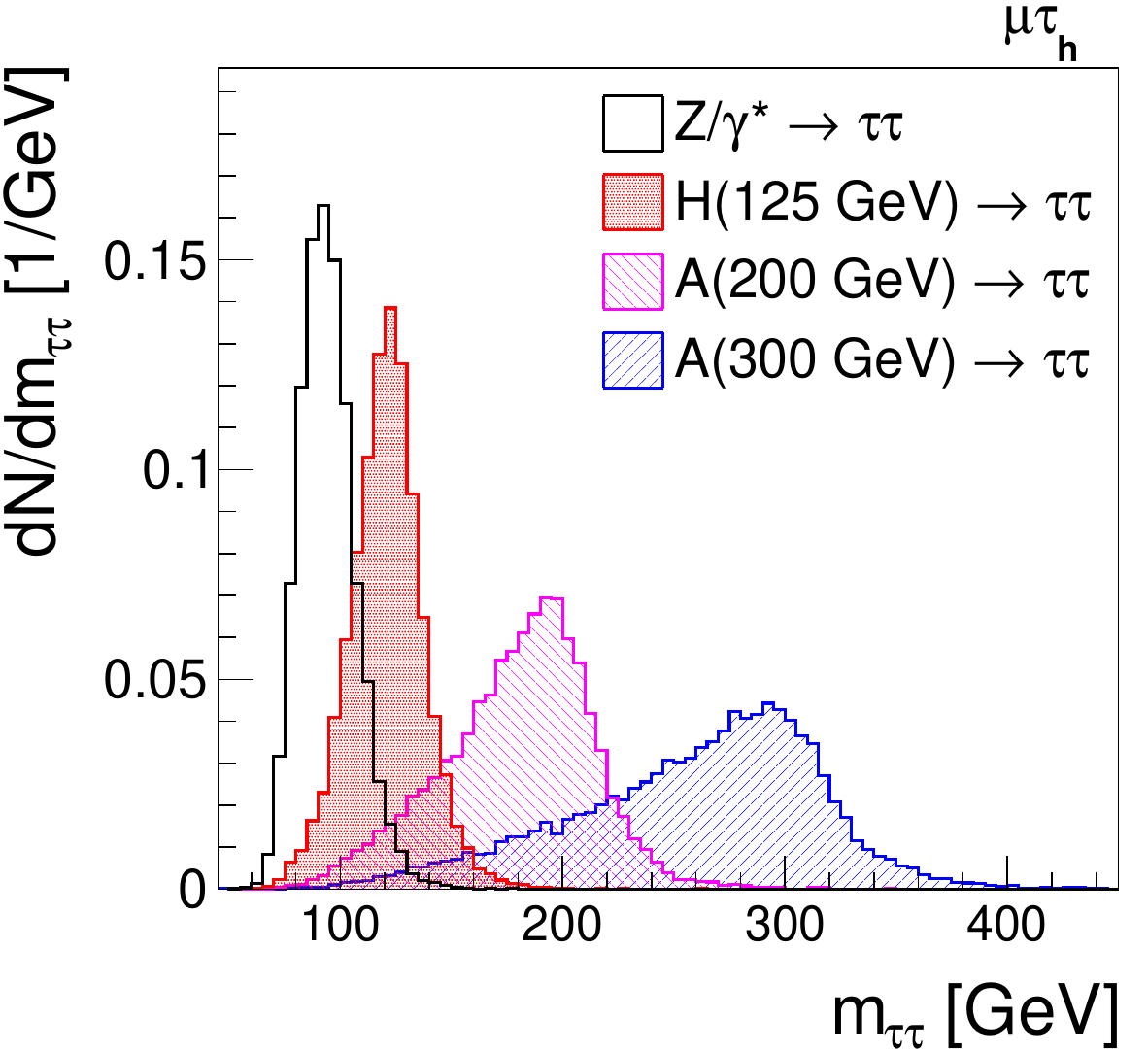}}}
\put(-2.5, 0.0){\mbox{\includegraphics*[height=70mm]
  {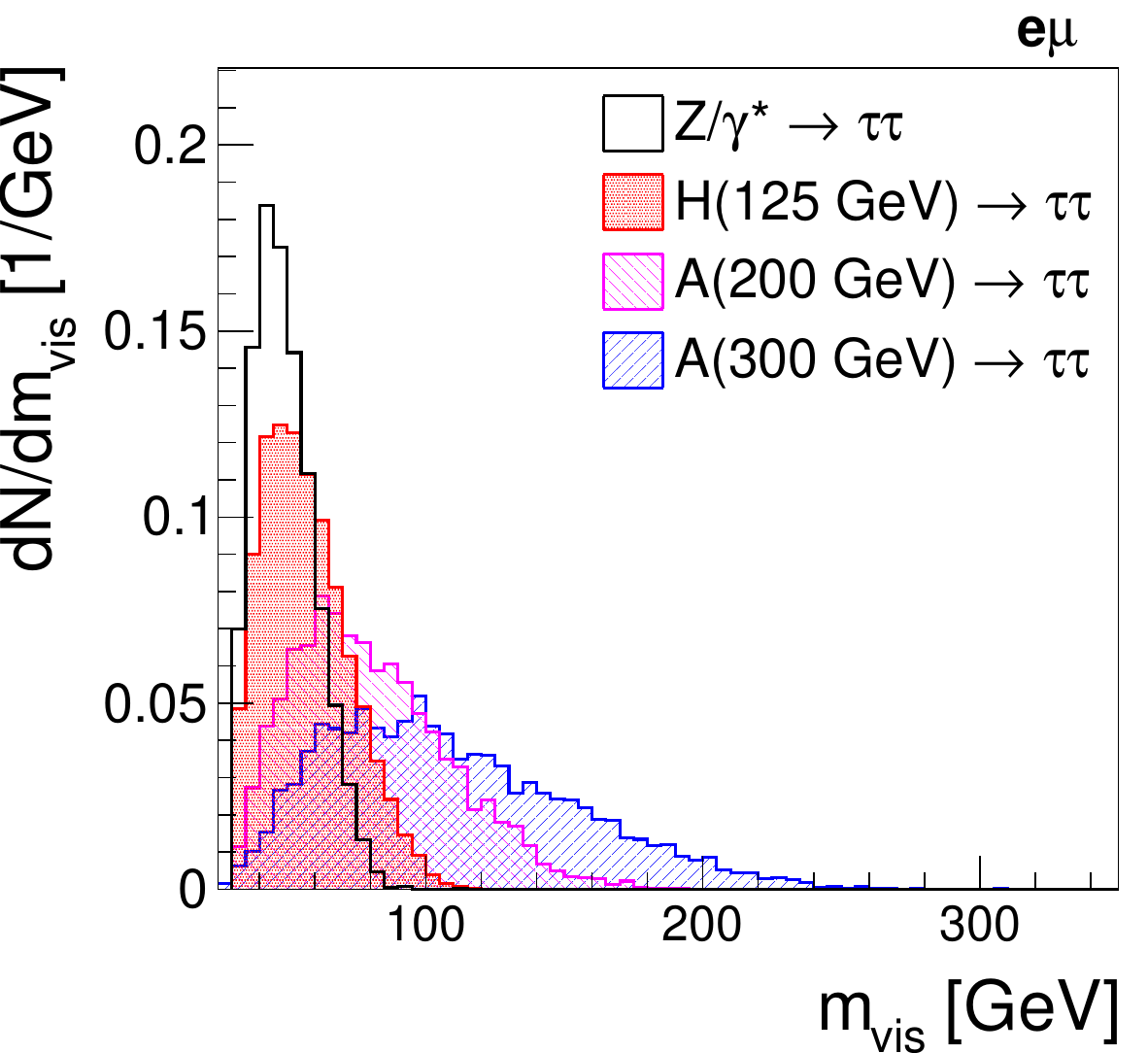}}}
\put(80.0, 0.0){\mbox{\includegraphics*[height=70mm]
  {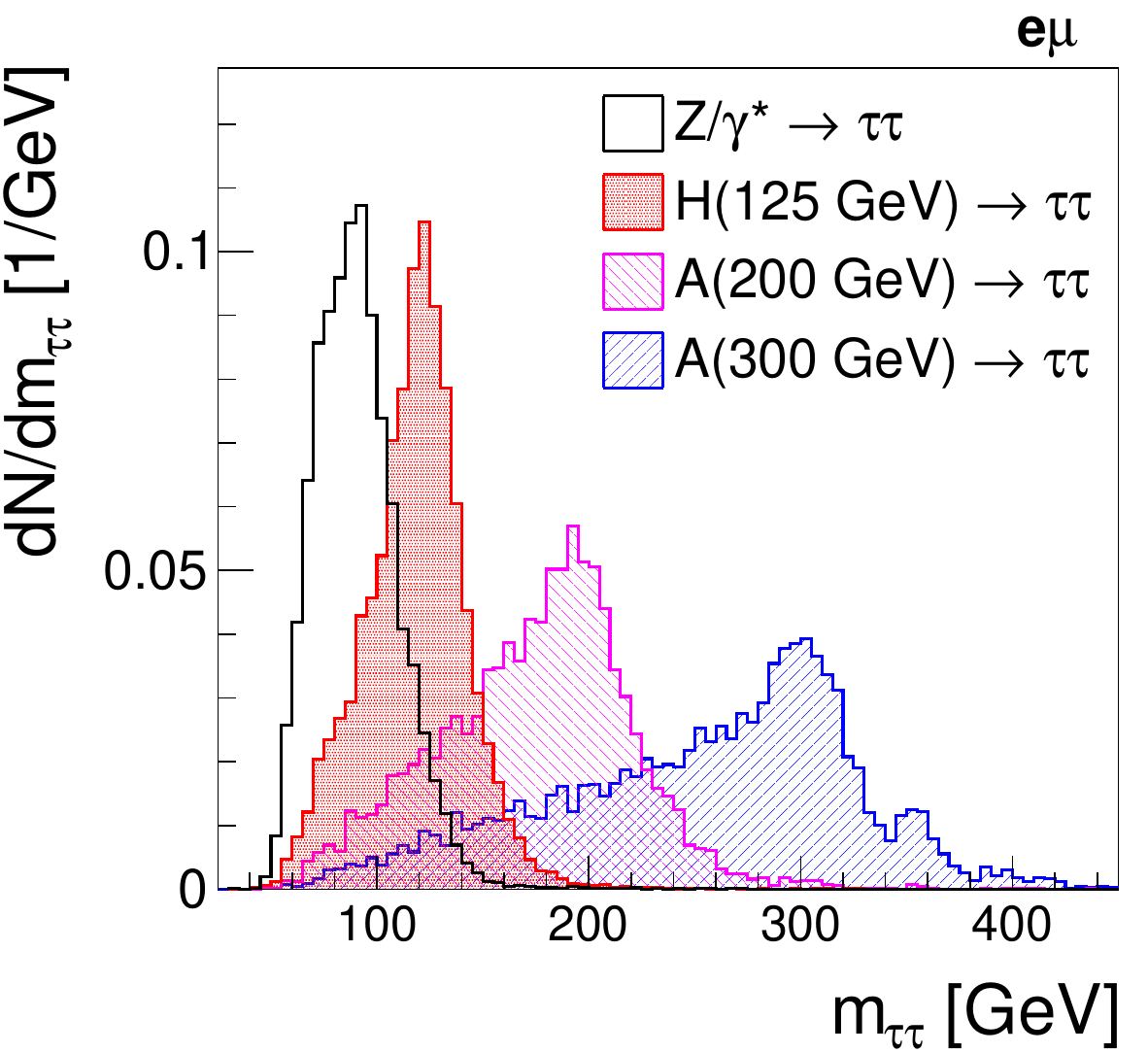}}}
\end{picture}
\end{center}
\caption{
  Distributions in $m_{\vis}$ (left) and in $m_{\Pgt\Pgt}$ reconstructed by the SVfitMEM algorithm with small positive $\kappa$ (right)
  in simulated $\PZ/\Pggx \to \Pgt\Pgt$ background events and SM $\PHiggs \to \Pgt\Pgt$ respectively $\PHiggsps \to \Pgt\Pgt$ signal events,
  selected in the decay channels $\tauh\tauh$ ($\kappa = 5$, top), $\Pgm\tauh$ ($\kappa = 4$, centre), and $\Pe\Pgm$ ($\kappa = 3$, bottom).
  Hypothetical signal events containing heavy pseudoscalar Higgs bosons $\PHiggsps$ are generated for masses $m_{\PHiggsps}$ of $200$ and $300$~\GeV. 
}
\label{fig:distributions_mVis_vs_SVfit}
\end{figure}

The improvement in signal-to-background separation provided by the
SVfit algorithm in searches for high mass resonances decaying to $\Pgt$ lepton pairs is illustrated in Fig.~\ref{fig:distributions_mVis_vs_SVfit}.

We conclude the discussion of the performance of the SVfit algorithm
with a comparison of the computing time requirements of the SVfitMEM,
cSVfit, and SVfitSA algorithms.
The computing time is dominated by the
evaluation of the integrand during the numeric integration of Eqs.~(\ref{eq:mem_with_hadRecoil}) and~(\ref{eq:cSVfit_with_hadRecoil}), and of Eq.~(2) of Ref.~\cite{SVfit}, respectively.
The computing time requirement is expected to be
highest for the SVfitMEM algorithm, due to the time needed for evaluation of the PDF and of the TF in Eq.~(\ref{eq:mem_with_hadRecoil}).

The time for reconstructing $m_{\Pgt\Pgt}$ in a single event scales approximately linearly
with the number of evaluations of the integrand.
The SVfitMEM (SVfitSA) algorithm performs $20\,000$ ($10\,000$)
evaluations of the integrand per mass hypothesis
$m_{\PHiggs}^{\textrm{test}(i)}$.
The number of evaluations of the integrand per mass hypothesis $m_{\PHiggs}^{\textrm{test}(i)}$ is chosen such that the
resolution on $m_{\Pgt\Pgt}$ is within $1\%$ compared to the resolution obtained in case an infinite (very large) number of evaluations of the
integrand is performed.
The number of evaluations is higher in case of the SVfitMEM algorithm, 
accounting for the fact that the additional PDF and TF factors raise the variation of the integrand, 
such that a higher number of evaluations is necessary in order to reach a given precision on the value of the integral.

The number of mass hypotheses tested for an event varies depending on the
conditions described in Section~\ref{sec:mem_numericalMaximization}.
For the series of $m_{\PHiggs}^{\textrm{test}(i)}$ values defined by Eq.~(\ref{eq:mTauTau_step_size}),
the number of mass hypotheses for which the integral $\mathcal{P}$ gets computed according to Eq.~(\ref{eq:mem_with_hadRecoil}) 
increases approximately logarithmically with the true mass of the $\Pgt$ lepton pair. 
In signal events, the integral $\mathcal{P}$
typically gets computed for a series of $20$ to $30$ mass hypotheses.
The average number of mass hypotheses varies with the decay channel.
The series is typically the longest in the $\Pe\Pgm$ and the shortest in the $\tauh\tauh$ channel,
as the value of $m_{\vis}$ that is used to initialize the series is farther away respectively less far away from $m_{\Pgt\Pgt}^{\true}$
at which the value of $\mathcal{P}$ is expected to reach its maximum $\mathcal{P}^{\textrm{max}}$.
A larger number of mass hypotheses may be required for background events,
in particular for backgrounds containing neutrinos that do not originate from $\Pgt$ lepton decays,
such as $\PW$ boson production and the production of top quark pairs,
as the values of $\mathcal{P}$ may be almost degenerate over a large range in $m_{\PHiggs}^{\textrm{test}(i)}$.

The numeric integration is performed by the VAMP algorithm in case of
the SVfitMEM algorithm (\cf Section~\ref{sec:mem_numericalMaximization}) and by the VEGAS algorithm
in case of the SVfitSA algorithm.
In case of the cSVfit algorithm, the numeric integration is performed
by a custom implementation of the Markov chain Monte Carlo method, as described in Section~\ref{sec:classicSVfit}.
In the latter case, the integrand is evaluated $100\,000$ times per event and there is no
iteration over a sequence of mass hypotheses, \cf Eq.~(\ref{eq:cSVfit_with_hadRecoil}).

\begin{figure}
\setlength{\unitlength}{1mm}
\begin{center}
\begin{picture}(160,210)(0,0)
\put(-2.5, -4.0){\mbox{\includegraphics*[height=214mm]
  {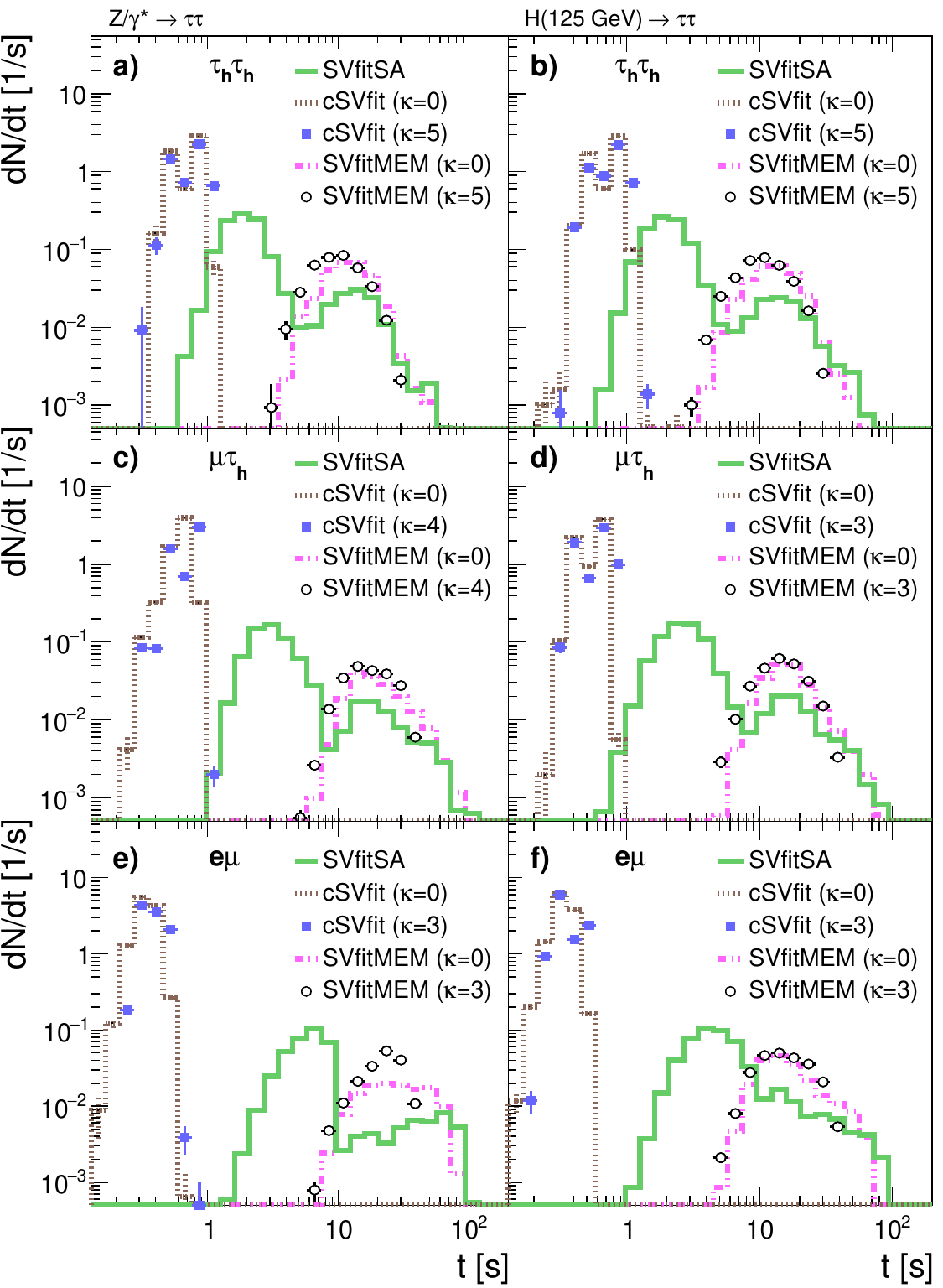}}}
\end{picture}
\end{center}
\caption{
  Distributions in CPU time, in seconds per event, needed to reconstruct $m_{\Pgt\Pgt}$ by
  the SVfitMEM, cSVfit, and SVfitSA algorithms
  in simulated $\PZ/\Pggx \to \Pgt\Pgt$ background (a,c,e)
  and SM $\PHiggs \to \Pgt\Pgt$ signal (b,d,f) events 
 in the decay channels $\tauh\tauh$ (a,b), $\Pgm\tauh$ (c,d),
  and $\Pe\Pgm$ (e,f).
}
\label{fig:computing_time}
\end{figure}

The computing time is measured by CPU time, 
using a single core of a $2.30$~GHz Intel\TReg~Xeon\TReg~E$5$-$2695$ v$3$ processor.
The distribution of the CPU time, in units of seconds per event, 
spent by the SVfitMEM, cSVfit and SVfitSA algorithms
to compute $m_{\Pgt\Pgt}$ in SM $\PHiggs \to \Pgt\Pgt$ signal events and in $\PZ/\Pggx \to \Pgt\Pgt$ background events
is shown in Fig.~\ref{fig:computing_time}.
Numerical values of the mean and RMS of the distributions are given in Table~\ref{tab:computing_time}.
The cSVfit algorithm requires about $0.5$s of CPU time per event to reconstruct the mass $m_{\Pgt\Pgt}$ of the $\Pgt$ lepton pair,
as well as its $\pT$, $\eta$, $\phi$, and transverse mass $m_{T\Pgt\Pgt}$.
The CPU time is reduced by a factor $5$ to $30$ compared to the SVfitSA algorithm, 
owing to the more efficient Markov chain Monte
Carlo integration method compared to
computing the probability density $\mathcal{P}$
for a series of mass hypotheses.
The computing time requirement of the SVfitMEM algorithm is higher by a factor $3$ to $4$ compared to the SVfitSA algorithm.

%
%

\begin{table}
\begin{center}
\begin{tabular}{|l|C{1.5cm}C{1.5cm}|C{1.5cm}C{1.5cm}|}
\hline
\multicolumn{5}{|c|}{$\tauh\tauh$ decay channel} \\
\hline
\hline
\multirow{2}{24mm}{Algorithm} & \multicolumn{2}{c|}{$\PZ/\Pggx \to \Pgt\Pgt$} & \multicolumn{2}{c|}{SM $\PHiggs \to \Pgt\Pgt$} \\
\cline{2-5}
& Mean~[s] & RMS~[s] & Mean~[s] & RMS~[s] \\
\hline
SVfitSA & $3.51$ & $5.24$ & $3.75$ & $5.87$ \\
cSVfit & & & & \\
$\quad \kappa=0$ & $0.72$ & $0.17$ & $0.73$ & $0.18$ \\
$\quad \kappa=5$ & $0.77$ & $0.20$ & $0.78$ & $0.20$ \\
SVfitMEM & & & & \\
$\quad \kappa=0$ & $13.07$ & $6.27$ & $14.03$ & $6.89$ \\
$\quad \kappa=5$ & $10.72$ & $4.72$ & $11.44$ & $5.02$ \\
\hline
\end{tabular}

\vspace{3.5mm}

\begin{tabular}{|l|C{1.5cm}C{1.5cm}|C{1.5cm}C{1.5cm}|}
\hline
\multicolumn{5}{|c|}{$\Pgm\tauh$ decay channel} \\
\hline
\hline
\multirow{2}{24mm}{Algorithm} & \multicolumn{2}{c|}{$\PZ/\Pggx \to \Pgt\Pgt$} & \multicolumn{2}{c|}{SM $\PHiggs \to \Pgt\Pgt$} \\
\cline{2-5}
& Mean~[s\unskip] & RMS~[s\unskip] & Mean~[s\unskip] & RMS~[s\unskip] \\
\hline
SVfitSA & $5.82$ & $9.22$ & $5.02$ & $7.91$ \\
cSVfit & & & & \\
$\quad \kappa=0$ & $0.62$ & $0.11$ & $0.56$ & $0.12$ \\
$\quad \kappa=4$ & $0.73$ & $0.17$ & $0.61$ & $0.16$ \\
SVfitMEM & & & & \\
$\quad \kappa=0$ & $22.64$ & $12.99$ & $18.45$ & $10.15$ \\
$\quad \kappa=4$ & $18.38$ & $7.49$ & $15.83$ & $6.68$ \\
\hline
\end{tabular}

\vspace{3.5mm}

\begin{tabular}{|l|C{1.5cm}C{1.5cm}|C{1.5cm}C{1.5cm}|}
\hline
\multicolumn{5}{|c|}{$\Pe\Pgm$ decay channel} \\
\hline
\hline
\multirow{2}{24mm}{Algorithm} & \multicolumn{2}{c|}{$\PZ/\Pggx \to \Pgt\Pgt$} & \multicolumn{2}{c|}{SM $\PHiggs \to \Pgt\Pgt$} \\
\cline{2-5}
& Mean~[s\unskip] & RMS~[s\unskip] & Mean~[s\unskip] & RMS~[s\unskip] \\
\hline
SVfitSA & $10.26$ & $14.65$ & $7.68$ & $10.35$ \\
cSVfit & & & & \\
$\quad \kappa=0$ & $0.34$ & $0.06$ & $0.34$ & $0.06$ \\
$\quad \kappa=3$ & $0.39$ & $0.08$ & $0.37$ & $0.09$ \\
SVfitMEM & & & & \\
$\quad \kappa=0$ & $30.92$ & $16.59$ & $19.37$ & $12.29$ \\
$\quad \kappa=3$ & $22.56$ & $7.70$ & $16.62$ & $7.43$ \\
\hline
\end{tabular}
\end{center}
\caption{
 CPU time, in seconds per event, needed to reconstruct $m_{\Pgt\Pgt}$ by
  the SVfitMEM, cSVfit, and SVfitSA algorithms
  in simulated $\PZ/\Pggx \to \Pgt\Pgt$ background
  and SM $\PHiggs \to \Pgt\Pgt$ signal events 
  in the decay channels $\tauh\tauh$, $\Pgm\tauh$,
  and $\Pe\Pgm$.
}
\label{tab:computing_time}
\end{table}

%% file: discussion.tex
\section{Discussion}
\label{sec:discussion}

Considering that the resolution on $m_{\Pgt\Pgt}$,
quantified by $\sigma_{l}/\textrm{M}$ and $\sigma_{h}/\textrm{M}$, achieved
by the SVfitMEM and cSVfit algorithms is almost identical, we find that the cSVfit
algorithm represents the best compromise between physics performance and computing time requirements in practical applications of the SVfit algorithm.

The optimal resolution is achieved in case an artificial regularization term of the type described in Section~\ref{sec:mem_logM}, with small positive $\kappa$,
is added to the likelihood function.
We expect the optimal choice of $\kappa$ to depend on the experimental resolution
as well as on the rates of signal and background processes,
and we recommend to perform a reoptimization of $\kappa$ in practical applications of our algorithm.

The merit of the SVfitMEM algorithm is that the 
formalism to treat $\Pgt$ lepton decays in the ME method, developed
for the SVfitMEM algorithm, can be used
in future applications of the ME method to data analyses with $\Pgt$
leptons in the final state.
An example for such an application is the analysis of SM $\PHiggs$ boson production in association with a pair of top quarks ($\Ptop\APtop\PHiggs$)
in final states with a $\Pgt$ lepton~\cite{HIG-17-003},
in which the existence of neutrinos from $\PW \to \Plepton\Pnu$ decays preclude the reconstruction of the $\PHiggs$ boson mass by the SVfit algorithm.

%% file: summary.tex
\section{Summary}
\label{sec:summary}

An algorithm for reconstruction of the $\PHiggs$ boson mass in events
in which the $\PHiggs$ boson decays into a pair of $\Pgt$ leptons has been
presented.
The relative resolution on the $\PHiggs$ boson mass amounts to typically
$15$--$20\%$.
The algorithm has been used in data analyses performed by the CMS
collaboration during LHC Run $1$.
It improves the sensitivity of the SM $\PHiggs \to \Pgt\Pgt$ analysis by $\approx 40\%$,
corresponding to a gain in luminosity by about a factor of two.

An improved version of the algorithm has been developed in preparation
for LHC Run $2$.
Two variants of the improved algorithm have been implemented.
The first variant, SVfitMEM, has been rigorously developed within the
formalism of the ME method. It is based on proper normalization of the probability density 
$\mathcal{P}$, given by Eq.~(\ref{eq:mem_with_hadRecoil}).
The second variant of the improved algorithm uses a likelihood
function of arbitrary normalization.
It allows to compute, on an event-by-event basis, any kinematic
function of the two $\Pgt$ leptons and provides a substantial reduction in computing time requirements.
A further improvement concerns the modelling of the experimental
resolution on the $\pT$ of $\tauh$ via TF, described in
Section~\ref{sec:mem_TF_tauToHadDecays}.

The performance of the algorithm has been studied in simulated SM
$\PHiggs \to \Pgt\Pgt$ signal and $\PZ/\Pggx \to \Pgt\Pgt$ background
events, as well as in simulated samples of heavy pseudoscalar Higgs
bosons and heavy spin $1$ resonances.
The SVfit algorithm is found to perform well in all event
categories and over the full range in true mass of the $\Pgt$ lepton
pair that we studied.

The development of the formalism to handle $\Pgt$ lepton decays
in the ME method constitutes an important result of this paper.
The formalism allows one to extend the matrix elements generated by automatized tools such as
CompHEP or MadGraph by the capability to handle hadronic as well as leptonic $\Pgt$ decays.
We expect that the formalism will be very useful for future
applications of the ME method to data analyses with $\Pgt$ leptons in
the final state.

%% file: acknowledgements.tex
\section*{Acknowledgements}

JC and CV wish to acknowledge the collaboration with our former colleague Evan Friis,
who made vital contributions to the early stage of this project. LM acknowledges the Estonian Research Council for supporting his work with the grant PUTJD110. 

%% file: appendix.tex
\section{Appendix}
\label{sec:appendix}

We derive here the relations for the product of the squared moduli of the ME and the phase space elements
$d\Phi^{(i)}_{\tauhnu}$ and $d\Phi^{(i)}_{\ellnunu}$ for,
respectively, the decays $\Pgt \to \textrm{hadrons} + \Pnut$ and $\Pgt \to \ellnunu$, given by Eq.~(\ref{eq:PSint}).
We start with the simpler case of hadronic $\Pgt$ decays in
Section~\ref{sec:appendix_tauToHadDecays} and turn to the more complex
case of leptonic $\Pgt$ decays in
Section~\ref{sec:appendix_tauToLepDecays}.
For clarity of notation, we omit the hat symbol in this section and
use the convention that all symbols refer to true values in the laboratory frame, unless indicated
explicitely otherwise.

%% file: appendix_tauToHadDecays.tex
\subsection{The decay $\Pgt \to \textrm{hadrons} + \Pnut$}
\label{sec:appendix_tauToHadDecays}

We treat hadronic $\Pgt$ decays as a two-body decay into a hadronic
system $\tauh$ plus a $\Pnut$,
as explained in Section~\ref{sec:mem_ME},
and take the squared modulus of the ME to be a constant,
which we denote by $\vert\mathcal{M}^{\eff}_{\Pgt \to \tauhnu}\vert^{2}$.
We further denote the momentum of the neutrino produced in the $\Pgt$ decay by
$\bm{p}^{\inv}$ and its energy by $E_{\inv}$ (\cf Section~\ref{sec:mem_PSintegration}).
For reasons that will become clear later, we allow the neutrino to
have non-zero mass $m_{\inv}$.

The product of the squared modulus of the ME and of the phase space
element $d\Phi^{(i)}_{\tauhnu}$ reads:
\begin{align}
 & \, \vert \BW_{\Pgt} \vert^{2} \cdot \vert\mathcal{M}_{\Pgt\to\cdots}(\bm{\tilde{p}})\vert^{2} \,
 d\Phi_{\tauhnu} = \vert \BW_{\Pgt} \vert^{2} \cdot \vert \mathcal{M}^{(i)}_{\Pgt\to\cdots}(\bm{\tilde{p}})
\vert^{2} \, \frac{d^{3}\bm{p}^{\vis}}{(2\pi)^{3} \, 2
   E_{\vis}} \, \frac{d^{3}\bm{p}^{\inv}}{(2\pi)^{3} \, 2 E_{\inv}}
 \cdot \nonumber \\
= & \, (2\pi)^{3} \, \int \, \frac{\pi}{m_{\Pgt} \, \Gamma_{\Pgt}} \,
\delta ( q_{\Pgt}^{2} - m_{\Pgt}^{2} ) \cdot \vert\mathcal{M}^{\eff}_{\Pgt \to
  \tauh\Pnut}\vert^{2} \, \delta \left( E_{\Pgt} - E_{\vis} -
  E_{\inv} \right) \, \delta^{3} \left( \bm{p}^{\Pgt} - \bm{p}^{\vis}
  - \bm{p}^{\inv} \right) \nonumber \\
& \qquad \frac{d^{3}\bm{p}^{\Pgt}}{(2\pi)^{3} \, 2 E_{\Pgt}} \, 
  \frac{d^{3}\bm{p}^{\vis}}{(2\pi)^{3} \, 2E_{\vis}} \, \frac{d^{3}\bm{p}^{\inv}}{(2\pi)^{3} \, 2 E_{\inv}} \, dq^{2}_{\Pgt} \nonumber \\
= & \frac{8\pi^{4}}{m_{\Pgt} \, \Gamma_{\Pgt}} \, \vert\mathcal{M}^{\eff}_{\Pgt \to
  \tauh\Pnut}\vert^{2} \, \delta \left( E_{\Pgt} - E_{\vis} -
  E_{\inv} \right) \, \delta^{3} \left( \bm{p}^{\Pgt} - \bm{p}^{\vis}
  - \bm{p}^{\inv} \right) \nonumber \\
& \qquad \frac{d^{3}\bm{p}^{\Pgt}}{(2\pi)^{3} \, 2 E_{\Pgt}} \, 
  \frac{d^{3}\bm{p}^{\vis}}{(2\pi)^{3} \, 2 E_{\vis}} \, \frac{d^{3}\bm{p}^{\inv}}{(2\pi)^{3} \, 2
    E_{\inv}} \nonumber \\
= & \, \frac{\pi}{m_{\Pgt} \, \Gamma_{\Pgt}} \, \frac{\vert\mathcal{M}^{\eff}_{\Pgt \to
  \tauh\Pnut}\vert^{2}}{(2\pi)^{6}} 
 \cdot \frac{1}{2 E_{\Pgt}(\bm{p}^{\vis}, \bm{p}^{\inv})} \, \delta
 \left( E_{\Pgt}(\bm{p}^{\vis}, \bm{p}^{\inv}) - E_{\vis} - E_{\inv}
 \right) \cdot \nonumber \\
& \qquad
  \frac{d^{3}\bm{p}^{\vis}}{2 E_{\vis}} \, \frac{\vert\bm{p}^{\inv}\vert}{2} \, dE_{\inv} \, d\cos\theta_{\inv} \, d\phi_{\inv} \, ,
\label{eq:hadTauDecaysPSint}
\end{align}
where we have used the formula for recursive phase space generation,
given by Eq.~(46.12) in Ref.~\cite{PDG}, for transforming the first line into the second
and the identity:
\begin{equation} 
d^{3}\bm{p}^{\inv} = \vert\bm{p}^{\inv}\vert^{2} \,
dp^{inv} \, d\cos\theta_{\inv} \, d\phi_{\inv} =
\vert\bm{p}^{\inv}\vert \, E_{\inv} \, dE_{\inv} \, d\cos\theta_{\inv}
\, d\phi_{\inv}
\end{equation} 
for rewriting the third line by the fourth.
The factor $\vert \BW_{\Pgt} \vert^{2} = \frac{\pi}{m_{\Pgt} \,
  \Gamma_{\Pgt}} \, \delta ( q^{2}_{\Pgt} - m^{2}_{\Pgt} )$ removes
the integration over $dq^{2}_{\Pgt}$, enforcing the $\Pgt$ lepton
energy and momentum to be related by $E_{\Pgt} =
\sqrt{\vert\bm{p}^{\Pgt}\vert^{2} + m_{\Pgt}^{2}}$.
The symbol $E_{\Pgt}(\bm{p}^{\vis}, \bm{p}^{\inv})$
indicates that $E_{\Pgt}$ is a function of $\bm{p}^{\vis}$
and $\bm{p}^{\inv}$, as is necessary to satisfy the $
\delta$-function $\delta^{3} ( \bm{p}^{\Pgt} - \bm{p}^{\vis} - \bm{p}^{\inv} )$.

We define $z = E_{\vis}/E_{\Pgt}$ according to Eq.~(\ref{eq:def_z}) and replace the integration over $dE_{\inv}$ by an integration over $z$.
The Jacobi factor related to this transformation is:
\begin{equation}
E_{\inv} = E_{\Pgt} - E_{\vis} = \left( 1 - z \right)
\, E_{\Pgt} = \frac{1 - z}{z} \, E_{\vis}
  \quad \Longleftrightarrow \quad dE_{\inv} = \Big\lvert \frac{\partial E_{\inv}}{\partial z} \Big\rvert \, dz = \frac{E_{\vis}}{z^{2}} \, dz \, .
\label{eq:hadTauDecaysJacobi}
\end{equation}

We then perform the integration over $d\cos\theta_{\inv}$.
Following the convention that we introduced in Section~\ref{sec:mem_PSintegration}, we choose the coordinate system such that
$\theta_{\inv}$ is equal to the angle between the $\bm{p}^{\vis}$ and $\bm{p}^{\inv}$ vectors.
The $\delta$-function $\delta \left( E_{\Pgt}(\bm{p}^{\vis}, \bm{p}^{\inv}) - E_{\vis} - E_{\inv} \right)$ depends on $\cos\theta_{\inv}$ via:
\begin{align}
& \, E_{\Pgt}(\bm{p}^{\vis}, \bm{p}^{\inv}) 
= \sqrt{\vert\bm{p}^{\Pgt}\vert^{2} + m^{2}_{\Pgt}} = \sqrt{\left( \bm{p}^{\vis} + \bm{p}^{\inv} \right)^2 + m^{2}_{\Pgt}} \nonumber \\
& \quad = \sqrt{\vert\bm{p}^{\vis}\vert^{2} +
  \vert\bm{p}^{\inv}\vert^{2} + 2 \, \bm{p}^{\vis}
  \cdot \bm{p}^{\inv} + m^{2}_{\Pgt}} \nonumber \\
& \quad = \sqrt{\vert\bm{p}^{\vis}\vert^{2} +
  \vert\bm{p}^{\inv}\vert^{2} + 2 \, \vert\bm{p}^{\vis}\vert \, \vert\bm{p}^{\inv}\vert \, \cos\theta_{\inv} + m^{2}_{\Pgt}}.
\end{align}
The $\delta$-function argument vanishes if $E_{\Pgt}(\bm{p}^{\vis}, \bm{p}^{\inv}) - E_{\vis} - E_{\inv} = 0$. 
This yields:
\begin{equation}
\cos\theta_{\inv} 
  = \frac{E_{\vis} E_{\inv} - \frac{1}{2} \left(m^{2}_{\Pgt} - \left(
        m^{2}_{\vis} + m^{2}_{\inv} \right)
    \right)}{\vert\bm{p}^{\vis}\vert \, \vert\bm{p}^{\inv}\vert} \, .
\label{eq:hadTauDecaysCosTheta}
\end{equation}

When substituting the expressions of Eqs.~(\ref{eq:hadTauDecaysJacobi}) and~(\ref{eq:hadTauDecaysCosTheta}) into Eq.~(\ref{eq:hadTauDecaysPSint}),
we need to account for the $\delta$-function rule:
\begin{equation} 
\delta \left( g(x) \right) = \sum_{k} \frac{\delta \left( x - x_{k}
  \right)}{\vert g'(x_{k}) \vert} \, .
\label{eq:deltaFuncRule}
\end{equation}
We identify:
\begin{equation} 
g(\cos\theta_{\inv}) = \sqrt{\vert\bm{p}^{\vis}\vert^{2} + \vert\bm{p}^{\inv}\vert^{2}
  + 2 \vert\bm{p}^{\vis}\vert \, \vert\bm{p}^{\inv}\vert \,
  \cos\theta_{\inv} + m^{2}_{\Pgt}} - E_{\vis} - E_{\inv}
\end{equation}
and obtain $\vert g'(x_{0}) \vert = \vert\bm{p}^{\vis}\vert \,
\vert\bm{p}^{\inv}\vert / \left( E_{\vis} + E_{\inv} \right) = \vert\bm{p}^{\vis}\vert \,
\vert\bm{p}^{\inv}\vert / E_{\Pgt}$.

This yields:
\begin{equation}
\vert \BW_{\Pgt} \vert^{2} \cdot \vert\mathcal{M}_{\Pgt\to\cdots}(\bm{\tilde{p}})\vert^{2} \,
 d\Phi^{(i)}_{\tauhnu} = \frac{\vert\mathcal{M}^{\eff}_{\Pgt \to
  \tauh\Pnut}\vert^{2}}{256\pi^{5}\, m_{\Pgt} \, \Gamma_{\Pgt}} \cdot 
    \frac{E_{\vis}}{\vert\bm{p}^{\vis}\vert \, z^{2}} \, 
    \frac{d^{3}\bm{p}^{\vis}}{2 E_{\vis}} \, dz \, d\phi_{\inv} \, .
\label{eq:hadTauDecaysResult}
\end{equation}

We define:
\begin{equation}
f_{h}\left(\bm{p}^{\vis}, m_{\vis}, \bm{p}^{\inv}\right) = 
  \frac{\vert\mathcal{M}^{\eff}_{\Pgt \to
  \tauh\Pnut}\vert^{2}}{256\pi^{6}} \cdot \frac{E_{\vis}}{\vert\bm{p}^{\vis}\vert \, z^{2}} 
\label{eq:hadTauDecays_f}
\end{equation}
to obtain:
\begin{equation}
\vert \BW_{\Pgt} \vert^{2} \cdot \vert\mathcal{M}_{\Pgt\to\cdots}(\bm{\tilde{p}})\vert^{2} \,
 d\Phi^{(i)}_{\tauhnu} = \frac{\pi}{m_{\Pgt} \, \Gamma_{\Pgt}} \,
 f_{h}(\bm{p}^{\vis}, m_{\vis}, \bm{p}^{\inv}) \, \frac{d^{3}\bm{p}^{\vis}}{2 E_{\vis}} \, dz \, d\phi_{\inv}
 \, ,
\end{equation}
which is the result that we quote in Eq.~(\ref{eq:PSint}).

%% file: appendix_tauToLepDecays.tex
\subsection{The decays $\Pgt \to \enunu$ and $\Pgt \to \mununu$}
\label{sec:appendix_tauToLepDecays}

We treat leptonic $\Pgt$ decays as three-body decays and do
account for the ME. Assuming the taus to be unpolarized,
the squared modulus of the ME is given by~\cite{Barger:1987nn}:
\begin{equation}
\vert\mathcal{M}_{\Pgt \to \ellnunu} \vert^{2} = 64 \, G^{2}_{F} \,
\left( E_{\Pgt} E_{\APnu} - \bm{p}^{\Pgt} \cdot \bm{p}^{\APnu} \right)
\, \left( E_{\Plepton} E_{\Pnu} - \bm{p}^{\Plepton} \cdot \bm{p}^{\Pnu} \right) \, , 
\label{eq:lepTauDecaysME}
\end{equation}
where $G_{F}$ denotes the Fermi constant, given by Eq.~(\ref{eq:def_G_F}).

The product of the squared modulus of the ME and the phase space
element $d\Phi_{\ellnunu}$ reads:
\begin{align}
& \, \vert \BW_{\Pgt} \vert^{2} \cdot
\underbrace{\vert\mathcal{M}_{\Pgt\to\cdots}(\bm{\tilde{p}})\vert^{2}}_{=
\vert\mathcal{M}_{\Pgt\to\cdots}(\bm{p})\vert^{2}} \,
 d\Phi_{\ellnunu} = \vert \BW_{\Pgt} \vert^{2} \cdot \vert \mathcal{M}^{(i)}_{\Pgt\to\cdots}(\bm{p})
\vert^{2} \, 
  \frac{d^{3}\bm{p}^{\vis}}{(2\pi)^{3} \, 2 E_{\vis}} \, 
  \frac{d^{3}\bm{p}^{\APnu}}{(2\pi)^{3} \, 2 E_{\APnu}} \,
  \frac{d^{3}\bm{p}^{\Pnu}}{(2\pi)^{3} \, 2 E_{\Pnu}} \nonumber \\
= & \, (2\pi)^{3} \, \int \, \frac{\pi}{m_{\Pgt} \, \Gamma_{\Pgt}} \,
\delta ( q_{\Pgt}^{2} - m_{\Pgt}^{2} ) \cdot
\vert\mathcal{M}_{\Pgt \to \ellnunu}(\bm{p})\vert^{2} \, \delta \left(
  E_{\Pgt} - E_{\vis} - E_{\APnu} - E_{\Pnu} \right) \cdot \nonumber \\
& \qquad
\delta^{3} \left( \bm{p}^{\Pgt} - \bm{p}^{\vis} - \bm{p}^{\Pnu} - \bm{p}^{\APnu} \right) \, \frac{d^{3}\bm{p}^{\Pgt}}{(2\pi)^{3} \, 2 E_{\Pgt}} \,
  \frac{d^{3}\bm{p}^{\vis}}{(2\pi)^{3} \, 2 E_{\vis}} \, 
  \frac{d^{3}\bm{p}^{\APnu}}{(2\pi)^{3} \, 2 E_{\APnu}} \, 
  \frac{d^{3}\bm{p}^{\Pnu}}{(2\pi)^{3} \, 2 E_{\Pnu}} \, dq^{2}_{\Pgt} \nonumber \\
= & \frac{8\pi^{4}}{m_{\Pgt} \, \Gamma_{\Pgt}} \, \vert\mathcal{M}_{\Pgt \to
  \ellnunu}\vert^{2} \, \delta \left( E_{\Pgt} - E_{\vis} - E_{\APnu} - E_{\Pnu} \right) \, \delta^{3} \left( \bm{p}^{\Pgt} -
  \bm{p}^{\vis} - \bm{p}^{\APnu} - \bm{p}^{\Pnu} \right)  \cdot \nonumber \\
& \qquad
  \frac{d^{3}\bm{p}^{\Pgt}}{(2\pi)^{3} \, 2 E_{\Pgt}} \,
  \frac{d^{3}\bm{p}^{\vis}}{(2\pi)^{3} \, 2 E_{\vis}} \,
  \frac{dE_{\APnu} \, d^{3}\bm{p}^{\APnu}}{(2\pi)^{3}} \,
  \theta(E_{\APnu}) \, \delta \left( E_{\APnu}^{2} - \vert\bm{p}^{\APnu}\vert^{2} \right) \,
  \frac{dE_{\Pnu} \, d^{3}\bm{p}^{\Pnu}}{(2\pi)^{3}} \,
  \theta(E_{\Pnu}) \, \delta \left( E_{\Pnu}^{2} - \vert\bm{p}^{\Pnu}\vert^{2} \right) \, .
\end{align}
The equality
$\vert\mathcal{M}_{\Pgt\to\cdots}(\bm{\tilde{p}})\vert^{2} =
\vert\mathcal{M}_{\Pgt\to\cdots}(\bm{p})\vert^{2}$
follows from the fact that the matrix element is a Lorentz invariant
quantity and as such has the same value in any frame.
We have used the identity:
\begin{equation}
\int \, dE \, d^{3}\bm{p} \, \theta(E) \, \delta \left( E^{2} -
  \vert\bm{p}\vert^{2} - m^{2} \right) = \int \, \frac{d^{3}\bm{p}}{2
  \, E} \, ,
\label{eq:PSintFourDim}
\end{equation}
for expressing the second line by the third.
Eq.~(\ref{eq:PSintFourDim}) follows from the $\delta$-function rule Eq.~(\ref{eq:deltaFuncRule}).
We assume that the mass of the $\APnu$ as well as the mass of the $\Pnu$ is zero.

We perform a variable transformation from $(E_{\Pnu}, \bm{p}^{\Pnu})$
and $(E_{\APnu}, \bm{p}^{\APnu})$ to:
\begin{align}
(u_{0}, \bm{u}) = & \frac{1}{\sqrt{2}} \, (E_{\APnu} + E_{\Pnu}, \bm{p}^{\APnu} +
\bm{p}^{\Pnu}) \, , \qquad (v_{0}, \bm{v}) = \frac{1}{\sqrt{2}} (
  E_{\APnu} - E_{\Pnu}, \bm{p}^{\APnu} - \bm{p}^{\Pnu} ) \, . \nonumber 
\end{align}
The variables $u_{0}$ and $\bm{u}$ represent the energy and momentum of the neutrino pair.
The magnitude of the determinant of the Jacobi matrix for the transformation from
$(E_{\APnu}, \bm{p}^{\APnu}; E_{\Pnu}, \bm{p}^{\Pnu})$
to $(u_{0}, \bm{u}; v_{0}, \bm{v})$ equals unity.
Expressed in the new variables, the energy and momentum of the
$\Pnu$ and of the $\APnu$ produced in the tau decay are given by:
\begin{align}
( E_{\APnu}, \bm{p}^{\APnu} ) = \frac{1}{\sqrt{2}} (
u_{0} + v_{0}, \bm{u} + \bm{v} ) \, \mbox{ and } \,
( E_{\Pnu}, \bm{p}^{\Pnu} ) = & \, \frac{1}{\sqrt{2}} ( u_{0} - v_{0}, \bm{u}
- \bm{v} ) \, . \nonumber 
\end{align}

The product of the squared modules of the ME and the phase space
element can then be expressed by:
\begin{align}
& \, \vert \BW_{\Pgt} \vert^{2} \cdot \vert\mathcal{M}_{\Pgt\to\cdots}\vert^{2} \,
 d\Phi_{\ellnunu} = \frac{8\pi^{4}}{m_{\Pgt} \, \Gamma_{\Pgt}} \,
 \delta \left( E_{\Pgt} - E_{\vis} - E_{\APnu} -
  E_{\Pnu} \right) \, \delta^{3} \left( \bm{p}^{\Pgt} -
  \bm{p}^{\vis} - \bm{p}^{\APnu} - \bm{p}^{\Pnu} \right) \nonumber \\
 & \qquad
\frac{d^{3}\bm{p}^{\Pgt}}{(2\pi)^{3} \, 2 E_{\Pgt}} \,
\frac{d^{3}\bm{p}^{\vis}}{(2\pi)^{3} \, 2 E_{\vis}} \,
\frac{du_{0} \, d^{3}\bm{u}}{(2\pi)^{3}} \nonumber \\
 & \qquad
  \vert\mathcal{M}_{\Pgt \to
  \ellnunu}\vert^{2} \, \theta(u_{0} + v_{0}) \, \delta
\left( \frac{u_{0}^{2} - \vert\bm{u}\vert^{2}}{2} + \frac{v_{0}^{2} - \vert\bm{v}\vert^{2}}{2} +
  u_{0} \, v_{0} - \bm{u} \cdot \bm{v} \right) \, 
  \frac{dv_{0} \, d^{3}\bm{v}}{(2\pi)^{3}} \nonumber \\
 & \qquad
  \theta(u_{0}
  - v_{0}) \, \delta \left( \frac{u_{0}^{2} - \vert\bm{u}\vert^{2}}{2} +
    \frac{v_{0}^{2} - \vert\bm{v}\vert^{2}}{2} - u_{0} \, v_{0} + \bm{u}
    \cdot \bm{v} \right) \, .
\label{eq:lepTauDecaysPSint}
\end{align}

We define:
\begin{align}
& \, I_{\inv} = \vert\mathcal{M}_{\Pgt \to
  \ellnunu}\vert^{2} \, \theta(u_{0} + v_{0}) \, \delta
\left( \frac{u_{0}^{2} - \vert\bm{u}\vert^{2}}{2} + \frac{v_{0}^{2} - \vert\bm{v}\vert^{2}}{2} +
  u_{0} \, v_{0} - \bm{u} \cdot \bm{v} \right) \, 
  \frac{dv_{0} \, d^{3}\bm{v}}{(2\pi)^{3}} \cdot \nonumber \\
 & \qquad
  \theta(u_{0}
  - v_{0}) \, \delta \left( \frac{u_{0}^{2} - \vert\bm{u}\vert^{2}}{2} +
    \frac{v_{0}^{2} - \vert\bm{v}\vert^{2}}{2} - u_{0} \, v_{0} + \bm{u}
    \cdot \bm{v} \right) \, .
\label{eq:def_Iinv}
\end{align}
The quantity $I_{\inv}$ is a Lorentz invariant quantity. 
As such, it can be computed in any frame and will yield the same value as in the laboratory frame.
We choose to evaluate it in the rest frame of the neutrino pair.
In this frame, the energy is given by $u_{0} = m_{\inv}$ 
and the momentum by $\bm{u} = ( 0, 0, 0 )$, with $m_{\inv}$ denoting
the mass of the neutrino pair.
Hence $u_{0} \, v_{0} - \bm{u} \cdot \bm{v} = m_{\inv} \, v_{0} $ in this frame.
Performing the integration over $v_{0}$, we obtain:
\begin{align}
I_{\inv}
= & \, \vert\mathcal{M}_{\Pgt \to
  \ellnunu}\vert^{2} \, \frac{1}{2} \, \frac{dv_{0} \, d^{3}\bm{v}}{(2\pi)^{3}} \, \theta ( u_{0} + v_{0} ) \, 
    \delta \left( \underbrace{\frac{u_{0}^{2} - \vert\bm{u}\vert^{2}}{2}}_{=
        \frac{m^{2}_{\inv}}{2}} + \underbrace{\frac{v_{0}^{2} -
        \vert\bm{v}\vert^{2}}{2}}_{= -\frac{1}{2} \, \vert\bm{v}\vert^{2}} \right)
    \cdot \nonumber \\
& \qquad
    \theta ( u_{0} - v_{0} ) \, \underbrace{\delta \left(
        u_{0} \, v_{0} - \bm{u} \cdot \bm{v} \right)}_{= \frac{1}{m_{\inv}} \, \delta ( v_{0} )} \nonumber \\
= & \, \frac{1}{2 m_{\inv}} \, \theta ( u_{0} ) \, \int \, \frac{d^{3}\bm{v}}{(2\pi)^{3}} \, 
  \vert\mathcal{M}_{\Pgt \to
  \ellnunu}\vert^{2} \, \underbrace{\delta \left( \frac{m^{2}_{\inv}}{2} - \frac{\vert\bm{v}\vert^{2}}{2} \right)}_{
    = \frac{1}{\vert\bm{v}\vert} \, \delta \left( \vert\bm{v}\vert - m_{\inv} \right)} \nonumber \\
= & \, \frac{1}{2 m_{\inv}} \, \theta ( u_{0} ) \, \int \, \frac{\vert\bm{v}\vert^2 d\vert\bm{v}\vert d\Omega_{v}}{(2\pi)^{3}} \, 
  \vert\mathcal{M}_{\Pgt \to
  \ellnunu}\vert^{2} \, \frac{1}{\vert\bm{v}\vert} \, \delta \left( \vert\bm{v}\vert - m_{\inv} \right) \nonumber \\
= & \, \frac{1}{2} \, \theta ( u_{0} ) \, \int \, \frac{d\Omega_{v}}{(2\pi)^{3}} \, \vert\mathcal{M}_{\Pgt \to
  \ellnunu}\vert^{2} \, . 
\label{eq:lepTauDecaysI}
\end{align}
We have used the relation $\delta(a + b) \, \delta(a - b) = \delta(2b) \, \delta(a - b) = \frac{1}{2} \, \delta(b) \, \delta(a)$ 
to express Eq.~(\ref{eq:def_Iinv}) by the first line in
Eq.~(\ref{eq:lepTauDecaysI}).
The identify $\frac{v_{0}^{2} - \vert\bm{v}\vert^{2}}{2} =
-\frac{1}{2} \, \vert\bm{v}\vert^{2}$ follows from the presence of the 
$\delta$-function $\delta ( v_{0} )$ in the first line.

In the rest frame of the neutrino pair:
\begin{align}
( E_{\Pgt}, \bm{p}^{\Pgt} ) = & \, ( E_{\Pgt}, 0, 0, \vert\bm{p}^{\Pgt}\vert ) \nonumber \\
( E_{\vis}, \bm{p}^{\vis} ) = & \, ( E_{\vis}, 0, 0, \vert\bm{p}^{\vis}\vert ) \nonumber \\
( E_{\Pnu}, \bm{p}^{\Pnu} ) = & \, \frac{m_{\inv}}{2} \, ( 1, 0, \sin\theta, \cos\theta ) \nonumber \\
( E_{\APnu}, \bm{p}^{\APnu} ) = & \, \frac{m_{\inv}}{2} \, ( 1, 0, \mbox{-}\sin\theta, \mbox{-}\cos\theta ) \, , \nonumber 
\end{align}
where we have chosen the polar axis such that it is parallel to $\bm{p}^{\vis}$.

The ME given by Eq.~(\ref{eq:lepTauDecaysME}) evaluates to:
\begin{align}
\vert\mathcal{M}_{\Pgt \to \ellnunu}\vert^{2} 
 = & \, 64 \, G^{2}_{F} \, 
  \underbrace{\left( E_{\Pgt} \, E_{\vis} - \bm{p}^{\Pgt} \cdot \bm{p}^{\APnu} \right)}_{= E_{\Pgt} \, \frac{m_{\inv}}{2} + \vert\bm{p}^{\Pgt}\vert \, \frac{m_{\inv}}{2} \cos\theta \quad} \,
  \underbrace{\left( E_{\vis} \, E_{\Pnu} - \bm{p}^{\vis} \cdot \bm{p}^{\Pnu} \right)}_{= E_{\vis} \, \frac{m_{\inv}}{2} - \vert\bm{p}^{\vis}\vert \, \frac{m_{\inv}}{2} \cos\theta} \nonumber \\
 = & \, 16 \, G^{2}_{F} \, m^{2}_{\inv} \, \left( E_{\Pgt} + \vert\bm{p}^{\Pgt}\vert \, \cos\theta \right)  \left( E_{\vis} - \vert\bm{p}^{\vis}\vert \, \cos\theta \right) 
\label{eq:lepTauDecaysMErf}
\end{align}
in the rest frame of the neutrino pair.

The energy of the $\Pgt$ lepton and of the electron or muon are
related by:
\begin{align}
m^{2}_{\vis} = & E_{\vis}^{2} - \vert\bm{p}^{\vis}\vert^{2} 
 = \left( E_{\Pgt} - u_{0} \right)^{2} - \left( \bm{p}^{\Pgt} - \bm{u} \right)^{2} \nonumber \\
& \qquad
 = m^{2}_{\Pgt} + m^{2}_{\inv} - 2 (E_{\Pgt} \, u_{0} - \bm{p}^{\Pgt} \cdot \bm{u}) 
 = m^{2}_{\Pgt} + m^{2}_{\inv} - 2 m_{\inv} E_{\Pgt} \, ,
\end{align}
from which it follows that:
\begin{equation}
E_{\Pgt} = \frac{m^{2}_{\Pgt} + m^{2}_{\inv} - m^{2}_{\vis}}{2
  m_{\inv}} \quad \mbox{ and } \quad E_{\vis} = E_{\Pgt} - m_{\inv} =
\frac{m^{2}_{\Pgt} - m^{2}_{\inv} - m^{2}_{\vis}}{2 m_{\inv}} \, .
\label{eq:lepTauDecaysEn}
\end{equation}

Substituting Eq.~(\ref{eq:lepTauDecaysMErf}) into Eq.~(\ref{eq:lepTauDecaysI}) yields:
\begin{align}
I_{\inv} 
= & \, \frac{1}{2} \, \theta ( u_{0} ) \, \int \, \frac{d\Omega_{v}}{(2\pi)^{3}} \, \vert\mathcal{M}_{\Pgt \to
  \ellnunu}\vert^{2} \nonumber \\
= & \, 8 \, G^{2}_{F} \, m^{2}_{\inv} \, \theta ( u_{0} ) \, \int \, \frac{d\cos\theta \, d\phi}{(2\pi)^{3}} \, 
  \left( E_{\Pgt} + \vert\bm{p}^{\Pgt}\vert \, \cos\theta \right)  \left( E_{\vis} - \vert\bm{p}^{\vis}\vert \, \cos\theta \right) \nonumber \\
= & \, \frac{G^{2}_{F}}{\pi^{3}} \, m^{2}_{\inv} \, \theta ( u_{0} ) \, \underbrace{\int_{0}^{2 \pi} \, d\phi}_{= 2 \pi} \, 
  \left( E_{\Pgt} \, E_{\vis} \, \underbrace{\int_{-1}^{+1}
      d\cos\theta}_{= 2} \right. \nonumber \\
& \qquad
     \left. + \left( \vert\bm{p}^{\Pgt}\vert \, E_{\vis} - E_{\Pgt} \,
       \vert\bm{p}^{\vis}\vert \right) \,
     \underbrace{\int_{-1}^{+1} \, d\cos\theta \, \cos\theta}_{= 0} 
  - \vert\bm{p}_{\Pgt}\vert \, \vert\bm{p}_{\vis}\vert
       \, \underbrace{\int_{-1}^{+1} \, d\cos\theta \, \cos^{2}\theta}_{= \frac{2}{3}} \right) \nonumber \\
= & \, \frac{2 \, G^{2}_{F}}{\pi^{2}} \, m^{2}_{\inv} \, \theta ( u_{0} ) \, 
  \left( 2 \, E_{\Pgt} \, E_{\vis} - \frac{2}{3} \, \sqrt{E_{\Pgt}^{2} - m^{2}_{\Pgt}} \, \sqrt{E_{\vis}^{2} - m^{2}_{\vis}} \right) \, ,
\label{eq:lepTauDecaysIrf}
\end{align}
with $E_{\Pgt}$ and $E_{\vis}$ given by Eq.~(\ref{eq:lepTauDecaysEn}).
Note that $I
_{\inv}$ solely depends on a single kinematic variable, $m_{\inv}$, 
as $m_{\Pgt}$ and $m_{\vis}$ are constants.

Before substituting Eq.~(\ref{eq:lepTauDecaysIrf}) into Eq.~(\ref{eq:lepTauDecaysPSint}),
we perform a variable transformation from $( u_{0}, u_{1}, u_{2}, u_{3} )$ to $( m^{2}_{\inv}, u_{1}, u_{2}, u_{3} ) = ( 2 \, ( {u_{0}}^{2} - {u_{1}}^{2} - {u_{2}}^{2} - {u_{3}}^{2}), u_{1}, u_{2}, u_{3} )$.
The magnitude of the determinant of the Jacobi matrix for this transformation is $\vert J
\vert = 4 \, u_{0}$,
from which it follows that:
\begin{equation}
du_{0} \, d^{3}\bm{u} = \frac{1}{\vert J \vert} \, dm^{2}_{\inv} \,
d^{3}\bm{u} = \frac{1}{4 u_{0}} \, dm^{2}_{\inv} \, d^{3}\bm{u} \, .
\label{eq:lepTauDecaysJacobi}
\end{equation}

Substituting Eqs.~(\ref{eq:lepTauDecaysIrf})
and~(\ref{eq:lepTauDecaysJacobi}) into
Eq.~(\ref{eq:lepTauDecaysPSint}), we then obtain:
\begin{align}
& \, \vert \BW_{\Pgt} \vert^{2} \cdot \vert\mathcal{M}_{\Pgt\to\cdots}\vert^{2} \,
 d\Phi_{\ellnunu} = \frac{8\pi^{4}}{m_{\Pgt} \, \Gamma_{\Pgt}} \,
 \delta \left( E_{\Pgt} - E_{\vis} - u_{0} \right)
 \, \delta^{3} \left( \bm{p}^{\Pgt} - \bm{p}^{\vis} -
  \bm{u} \right) \cdot \nonumber \\
& \qquad \frac{d^{3}\bm{p}^{\Pgt}}{(2\pi)^{3} \, 2 E_{\Pgt}} \,
  \frac{d^{3}\bm{p}^{\vis}}{(2\pi)^{3} \, 2 E_{\vis}} \, 
  \frac{d^{3}\bm{u}}{(2\pi)^{3} \, 2 u_{0}} \, \frac{I_{\inv}}{2} \,
  dm^{2}_{\inv} \, ,
\label{eq:finalLepTauDecaysPSint}
\end{align}
with $u_{0} \equiv E_{\inv}$ and $\bm{u} \equiv \bm{p}^{\inv}$.
The expression in Eq.~(\ref{eq:finalLepTauDecaysPSint}) is very similar in structure to the third line of Eq.~(\ref{eq:hadTauDecaysPSint}),
if we identify the integration over the momentum of the neutrino pair,
given by the phase space element $d^{3}\bm{u}$ in Eq.~(\ref{eq:finalLepTauDecaysPSint}), 
with the integration over the neutrino momentum $d^{3}\bm{p}^{\inv}$ in Eq.~(\ref{eq:hadTauDecaysPSint}).
The differences between the formulae for leptonic and hadronic $\Pgt$ decays
are the additional integration over $dm^{2}_{\inv}$ and the
factor $I_{\inv}/2$ in Eq.~(\ref{eq:finalLepTauDecaysPSint}), which replaces the
factor $\vert\mathcal{M}_{\Pgt \to
  \tauh\Pnut}\vert^{2}$ in Eq.~(\ref{eq:hadTauDecaysPSint}).
Note that Eq.~(\ref{eq:finalLepTauDecaysPSint}) as well as Eq.~(\ref{eq:hadTauDecaysPSint}) refer to the laboratory frame.
The rest frame of the neutrino pair was used only for the purpose of evaluating the Lorentz invariant integral $I_{\inv}$.

Since, according to Eq.~(\ref{eq:lepTauDecaysIrf}), $I_{\inv}$ 
depends solely on the integration variable $m_{\inv}$, 
$I_{\inv}$ can be treated as a constant when performing the integration over $d^{3}\bm{p}^{\Pgt}$ and $d^{3}\bm{u}$.
We can hence use Eq.~(\ref{eq:hadTauDecaysResult}) of Section~\ref{sec:appendix_tauToHadDecays} to express Eq.~(\ref{eq:finalLepTauDecaysPSint}) by:
\begin{align}
\vert \BW_{\Pgt} \vert^{2} \cdot \vert\mathcal{M}_{\Pgt\to\cdots}\vert^{2} \,
 d\Phi_{\ellnunu} = \frac{I_{\inv}}{512\pi^{5}\, m_{\Pgt} \, \Gamma_{\Pgt}} \cdot 
    \frac{E_{\vis}}{\vert\bm{p}^{\vis}\vert \, z^{2}} \, 
    \frac{d^{3}\bm{p}^{\vis}}{2 E_{\vis}} \, dz \, dm^{2}_{\inv} \,
    d\phi_{\inv} \, ,
\label{eq:lepTauDecaysResult}
\end{align}
where the angle $\phi_{\inv}$ specifies the orientation of the neutrino
pair momentum vector $\bm{p}^{\inv}$ with respect to the momentum vector $\bm{p}^{\vis}$
of the electron or muon.

The opening angle between the vector $\bm{p}^{\inv}$ and the direction
of the electron respectively muon is given in analogy to Eq.~(\ref{eq:hadTauDecaysCosTheta}) by:
\begin{equation}
\cos\theta_{\inv} = \frac{E_{\vis} E_{\inv} - \frac{1}{2} \left(m^{2}_{\Pgt} - \left( m^{2}_{\vis} + m^{2}_{\inv} \right) \right)}{\vert\bm{p}^{\vis}\vert \, 
  \vert\bm{p}^{\inv}\vert}.
\label{eq:lepTauDecaysCosTheta}
\end{equation}

We define:
\begin{equation}
f_{\Plepton}\left(\bm{p}^{\vis}, m_{\vis}, \bm{p}^{\inv}\right) = 
\frac{I_{\inv}}{512\pi^{6}} \cdot \frac{E_{\vis}}{\vert\bm{p}^{\vis}\vert \, z^{2}}
\label{eq:lepTauDecays_f}
\end{equation}
to obtain:
\begin{equation}
\vert \BW_{\Pgt} \vert^{2} \cdot \vert\mathcal{M}_{\Pgt\to\cdots}\vert^{2} \,
 d\Phi_{\ellnunu} = \frac{\pi}{m_{\Pgt} \, \Gamma_{\Pgt}} \,
 f_{\Plepton}(\bm{p}^{\vis}, m_{\vis}, \bm{p}^{\inv}) \, \frac{d^{3}\bm{p}^{\vis}}{2 E_{\vis}} \, dz \, dm^{2}_{\inv} \, d\phi_{\inv}
 \, ,
\end{equation}
the result that we quote in Eq.~(\ref{eq:PSint}).